\documentclass[reprint, superscriptaddress, amsmath, amssymb, aps, prl, floatfix,
]{revtex4-2}
\pdfoutput=1
\usepackage[T1]{fontenc} 
\usepackage{lmodern}
\usepackage{amsmath}
\usepackage{array}
\usepackage{dsfont}
\usepackage{mathtools}
\usepackage{graphicx}
\usepackage{dcolumn}
\usepackage{bm}
\usepackage{hyperref}
\usepackage{physics}
\usepackage{empheq}
\usepackage{color}
\usepackage{multirow}
\usepackage{afterpage}
\usepackage{mathrsfs}
\usepackage{tikz}
\usetikzlibrary{shapes}
\usetikzlibrary{tikzmark,calc}

\newcommand{\hexagon}{\mathord{\raisebox{0.6pt}{\tikz{\node[draw,scale=.65,regular polygon, regular polygon sides=6](){};}}}}

\allowdisplaybreaks

\begin{document}

\title{Spectroscopic Signatures of Fractionalization in Octupolar Quantum Spin Ice}

\author{F\'elix Desrochers}
\email{felix.desrochers@mail.utoronto.ca}
\affiliation{%
 Department of Physics, University of Toronto, Toronto, Ontario M5S 1A7, Canada
}%
\author{Yong Baek Kim}%
\email{ybkim@physics.utoronto.ca}
\affiliation{%
 Department of Physics, University of Toronto, Toronto, Ontario M5S 1A7, Canada
}%

\date{\today}

\begin{abstract}
Recent investigations on the dipolar-octupolar compounds Ce$_2$Zr$_2$O$_7$ and Ce$_2$Sn$_2$O$_7$ suggest that they may stabilize so-called $\pi$-flux octupolar quantum spin ice ($\pi$-O-QSI), a novel three-dimensional quantum spin liquid hosting emergent photons. Confirmation of such an exotic phase would require the prediction of a distinctive signature and its subsequent experimental observation. So far, however, theoretical predictions for any such sharp smoking-gun signatures are lacking. In this Letter, we thoroughly investigate O-QSI using an extension of gauge mean-field theory. This framework produces a phase diagram consistent with previous work and an energy-integrated neutron scattering signal with intensity-modulated rod motifs, as reported in experiments and numerical studies. We predict that the dynamical spin structure factor of $\pi$-O-QSI is characterized by a broad continuum with three distinctive peaks as a consequence of the two mostly flat spinon bands. These three peaks should be measurable by high-resolution inelastic neutron scattering. Such spectroscopic signatures would be clear evidence for the realization of $\pi$-flux quantum spin ice. 
\end{abstract}
\maketitle

\textit{Introduction}.--- Quantum spin liquids (QSLs) are quantum paramagnetic ground states of spin systems where competition between local interactions prevents conventional long-range order (LRO) and instead results in a long-range entangled (LRE) state supporting fractionalized excitations coupled to emergent gauge fields~\cite{savary2016quantumspinliquids, knolle2019field, zhou2017quantum, balents2010spin, broholm2020quantum, wen2004quantum, wen2002quantum, wen2017colloquium}. Decades after their initial proposal, the quest for an unequivocal experimental realization of a QSL remains a current endeavor --- a testimony to how formidable of a task identifying a QSL is. Indeed, even though the experimental observation of fractionalized quasiparticles or emergent gauge fields would be direct evidence for a QSL ground state, conventional probes do not usually offer such unambiguous detection. For instance, a widely used method to ``diagnose'' a QSL is through the lack of signatures indicating LRO and the presence of a broad continuum in inelastic neutron scattering. However, this is an unconvincing state of affairs considering that scattering continua can be indicative of much less exotic phenomena than a continuum of fractionalized quasiparticles~\cite{winter2017breakdown, harris1971dynamics, chernyshev2006magnon, zhitomirsky2013colloquium, murani1978neutron, lee1996spin}. Therefore, since the universal features of QSLs (i.e., LRE) are not easily reachable by currently available probes, one has to resort to a much more careful and systematic approach where the microscopic parameters of a candidate material are first estimated by fitting a plethora of experimental measurements (e.g., heat capacity, magnetization, neutron scattering). Specific predictions about the nature of the ground state and its distinctive signatures can then be made to be later confirmed by empirical studies. 

\begin{figure}[b]
\includegraphics[width=1.0\linewidth]{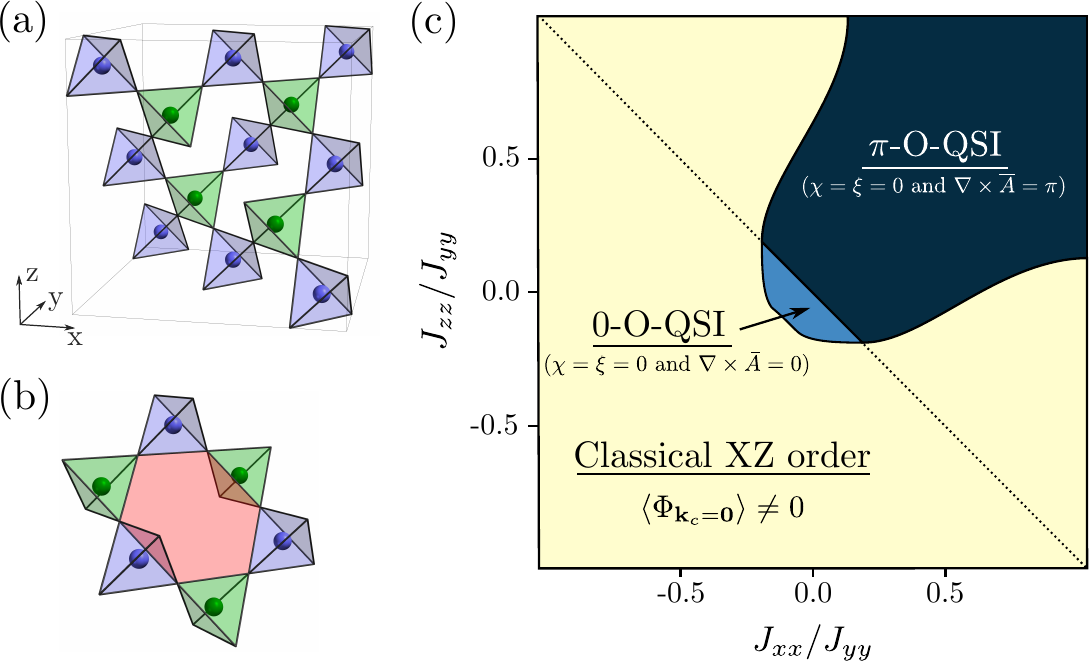}
\caption{
(a) The network of corner-sharing tetrahedra forming the pyrochlore lattice and the sites of its parent (premedial) diamond lattice. (b) Hexagonal plaquettes of the pyrochlore lattice. (c) GMFT phase diagram of DO systems in the octupolar dominant regime. The dotted line indicates the $J_{\pm}=0$  boundary. \label{fig: phase diagram}}
\end{figure}

Recent investigations on the dipolar-octupolar (DO) compounds Ce$_2$Zr$_2$O$_7$~\cite{gaudet2019quantum, gao2019experimental, gao2022magnetic, smith2022case, bhardwaj2022sleuthing,  smith2022reply, desrochers2022competing, Hosoi2022Uncovering}, Ce$_2$Sn$_2$O$_7$~\cite{sibille2015candidate, sibille2020quantum}, and Ce$_2$Hf$_2$O$_7$~\cite{Poiree2022, poree2023dipolar} have been particularly exciting in that regard. A large amount of experimental evidence indicates that they may be a realization of quantum spin ice (QSI); a QSL with an emergent compact $U(1)$ gauge structure that provides a lattice realization of quantum electrodynamics~\cite{hermele2004pyrochlore, benton2012seeing, castelnovo2012spin, lacroix2011introduction, udagawa2021spin}. In these DO compounds, the lowest lying doublet of the magnetically active ions forming a pyrochlore lattice can be described by pseudospin-1/2 with two components ($\mathrm{S}^{x}$, $\mathrm{S}^{z}$) that transform as dipoles and one ($\mathrm{S}^{y}$) as an octupolar moment~\cite{rau2019frustrated, huang2014quantum}. The most general Hamiltonian with nearest-neighbor couplings can be written as an $XYZ$ model in a rotated local basis $\mathcal{H} = \sum_{\langle i, j \rangle}\left[ J_{x x} \mathrm{S}_{i}^x \mathrm{S}_{j}^x + J_{yy} \mathrm{S}_{i}^{y} \mathrm{S}_{j}^{y} + J_{zz} \mathrm{S}_{i}^{z} \mathrm{S}_{j}^{z}\right]$. 

Combinations of experimental measurements and theoretical analyses have now strongly constrained the microscopic exchange parameters for Ce$_2$Zr$_2$O$_7$, Ce$_2$Hf$_2$O$_7$, and Ce$_2$Sn$_2$O$_7$. They indicate that the leading coupling is most likely associated with the octupolar component (i.e., $|J_{yy}|>|J_{xx}|,|J_{zz}|$)~\cite{smith2022case, bhardwaj2022sleuthing, sibille2020quantum, poree2023dipolar}, and in a region of parameter space that is predicted to realize the so-called $\pi$-flux octupolar quantum spin ice ($\pi$-O-QSI) phase, although conflicting results on Ce$_2$Sn$_2$O$_7$ have recently been reported~\cite{yahne2022dipolar}. In the $\pi$-O-QSI phase, the hexagonal plaquettes of the pyrochlore lattice are threaded by a static $\pi$ flux of the emergent gauge field (see Fig.~\ref{fig: phase diagram} (a) and (b))~\cite{benton2018quantum, lee2012generic, savary2021quantum, taillefumier2017competing, li2017symmetry, yao2020pyrochlore, chen2017spectral}. 

Even if the position of these compounds is solidly established in parameter space, there are still doubts regarding the nature of their ground state. The experimentally identified parameters are far from the perturbative Ising limit where the theoretical prediction for the $\pi$-flux QSI ground state is well established and disorder or interactions beyond the $XYZ$ model could potentially stabilize other competing states~\cite{desrochers2022competing}. These suspicions would be put aside if a prediction for a specific and distinct signature of the $\pi$-flux state could be experimentally observed. So far, no theoretical study of octupolar quantum spin ice (O-QSI) has been able to put forward such experimentally accessible smoking-gun signatures.

In this Letter, we present a comprehensive study of O-QSI using a recently introduced extension of gauge mean field theory (GMFT)~\cite{desrochers2023symmetry}. We first classify all symmetric GMFT Ans\"atze and study their stability to produce a phase diagram widely consistent with previous numerical investigations~\cite{patri2020distinguishing, benton2020ground, shannon2012quantum, huang2020extended, banerjee2008unusual}. We then compute the equal-time and dynamical spin structure factor for the $0$- and $\pi$-flux O-QSI states. It is crucially highlighted that, due to their two largely flat energy bands, the spinons' contribution to the dynamical spin structure factor of the $\pi$-flux state is formed of a broad continuum with three distinctive peaks, in contrast to only one peak in the 0-flux phase. This highly distinctive and unique feature should be accessible by neutron scattering on either powder or single crystal samples with a resolution of about an order of magnitude higher than the leading exchange coupling $J_{yy}$. The experimental identification of these structures would be cogent proof for the long sought-after experimental discovery of a three-dimensional QSL.


\textit{Model}.--- In this analysis, we examine the $XYZ$ model in the experimentally relevant octupolar dominant regime where $J_{yy}>0$ and $J_{yy}>|J_{xx}|,|J_{zz}|$. In the Ising limit (i.e., $J_{yy}>0$ and $J_{yy}\gg |J_{xx}|,|J_{zz}|$), the dominant term restricts the system to a subspace where the sum over the $y$-component of all spins is zero for every tetrahedron (i.e., 2-in-2-out). The transverse couplings lead to mixing between these states and promote the system from a classical spin liquid to a QSL whose low-energy behavior can be described by a compact $U(1)$ gauge theory of the form $\mathcal{H}_{\text {eff }}\! \sim\! (J_{xx} + J_{zz})^{3} / J_{yy}^{2} \sum_{\hexagon} \cos (\nabla \times \bar{A})$ on the parent diamond lattice~\cite{hermele2004pyrochlore, gingras2014quantum, henley2010coulomb, savary2021quantum}. From this perturbative argument, one expects a deconfined $U(1)$ QSL with 0 flux ($\pi$ flux) threading the hexagonal plaquettes for $J_{xx}+J_{zz}<0$ ($J_{xx}+J_{zz}>0$). The stability of the 0-flux state is now firmly established from quantum Monte Carlo (QMC) simulations~\cite{huang2020extended, banerjee2008unusual, huang2018dynamics, shannon2021quantum}.

To go beyond such a perturbative treatment, we employ GMFT where a bosonic matter field that conceptually corresponds to tetrahedra breaking the 2-in-2-out rule is introduced on the parent diamond lattice~\cite{savary2012coulombic, savary2013spin, savary2021quantum, lee2012generic, hao2014bosonic}. In such a framework, the pseudospins are expressed in terms of an emergent compact $U(1)$ gauge field $A$, its canonically conjugate electric field $E$ that takes on half-integer values, and the spinon operator $\Phi_{\mathbf{r}_{\alpha}}^{\dagger} = e^{i\varphi_{\mathbf{r}_{\alpha}}}$ that creates a gauge charge $Q_{\mathbf{r}_{\alpha}}$. Explicitly, $\mathrm{S}^+_{\mathbf{r}_{A}+ \mathbf{b}_{\mu}/2} = \Phi^{\dag}_{\mathbf{r}_A}  (e^{i A_{\mathbf{r}_{A},\mathbf{r}_{A} + \mathbf{b}_\mu}}/2)  \Phi_{\mathbf{r}_{A}+\mathbf{b}_\mu}$ and $\mathrm{S}^y_{\mathbf{r}_{\alpha}+\eta_{\alpha}\mathbf{b}_\mu/2} = \eta_{\alpha} E_{\mathbf{r}_{\alpha},\mathbf{r}_{\alpha}+\eta_{\alpha}\mathbf{b}_\mu}$ where $\mathrm{S}^{\pm}=\mathrm{S}^z \pm i \mathrm{S}^x$, $\mathbf{r}_{\alpha}$ labels the positions on the diamond lattice, $\eta_{\alpha}$=1($-1$) if the site belongs to the $\alpha$=$A(B)$ sublattice, and $\mathbf{b}_{\mu}$ ($\mu$=0,1,2,3) are vectors connecting the center of a tetrahedron to its four nearest-neighbor diamond lattice sites (see Supplemental Material \footnote{See Supplemental Material} for conventions). Such a mapping is exact if the discretized Gauss's law $ Q_{\mathbf{r}_{\alpha}} =  \sum_{\mu=0}^3 E_{\mathbf{r}_{\alpha},\mathbf{r}_{\alpha}+\eta_{\alpha}\mathbf{b}_\mu}$ is imposed on every tetrahedron.

Directly replacing the parton construction in the $XYZ$ Hamiltonian leads to an interacting quantum rotor model strongly coupled to a compact $U(1)$ gauge field. To get a tractable model, three successive approximations are carried out. (1) A mean-field (MF) decoupling is performed on the four bosons interaction arising from terms of the form $\mathrm{S}^{\pm}\mathrm{S}^{\pm}$ such that $\Phi_{i}^{\dagger} \Phi_{i}^{\dagger} \Phi_{j} \Phi_{k} \rightarrow \Phi_{i}^{\dagger} \Phi_{i}^{\dagger} \chi_{j,k} + \Phi_{j} \Phi_{k} \bar{\chi}^{0}_{i,i} +2 \Phi_{i}^{\dagger} \Phi_{j} \xi_{i, k}+2 \Phi_{i}^{\dagger} \Phi_{k} \xi_{i, j}$, where the inter-site pairing $\chi$, on-site pairing $\chi^{0}$, and inter-sublattice hopping $\xi$ MF parameters have been introduced. (2) The gauge field is fixed to a constant saddle point background $A\to \bar{A}$. This effectively decouples the matter and dynamical gauge field sectors. (3) Finally, a large-$N$ approximation is made by relaxing the constraint on the rotor length $|\Phi_{\mathbf{r}_{\alpha}}^{\dagger}\Phi_{\mathbf{r}_{\alpha}}|=1$ to an average one $\sum_{\mathbf{r}_{\alpha}}\expval{\Phi_{\mathbf{r}_{\alpha}}^{\dagger}\Phi_{\mathbf{r}_{\alpha}}}/N=\kappa$ for $\alpha=A,B$ imposed by the Lagrange multipliers $\lambda^{\alpha}$. We pick $\kappa=2$ since such a constraint recovers the correct spinon dispersion $\mathcal{E}_{\gamma}(\mathbf{k})=J_{yy}/2$ in the Ising limit (i.e., $J_{xx}/J_{yy}\to 0$ and $J_{zz}/J_{yy}\to 0$). It also reproduces the QMC results for the position of the phase transition from the 0-flux QSI to an ordered state and for the position of the lower and upper edges of the two-spinon continuum \cite{huang2020extended, huang2018dynamics, desrochers2023symmetry}. See Supplemental Material~\cite{Note1} for detailed discussion, which includes Refs.~\cite{wang2006spin, messio2013time, bieri2016projective, sachdev1992kagome, messio2010schwinger}. Following these prescriptions, we get the GMFT Hamiltonian
\begin{widetext}
\begin{align} \label{eq: GMFT Hamiltonian}
\mathcal{H}_{\mathrm{GMFT}}=&\frac{J_{y y}}{2} \sum_{\mathbf{r}_\alpha} Q_{\mathbf{r}_\alpha}^2 + \sum_{\mathbf{r}_\alpha} \lambda^{\alpha}\left(\Phi_{\mathbf{r}_\alpha}^{\dagger} \Phi_{\mathbf{r}_\alpha}-\kappa\right)  -\frac{J_{\pm}}{4} \sum_{\mathbf{r}_\alpha} \sum_{\mu, \nu \neq \mu} \Phi_{\mathbf{r}_\alpha+\eta_\alpha \mathbf{b}_\mu}^{\dagger} \Phi_{\mathbf{r}_\alpha+\eta_\alpha \mathbf{b}_\nu} e^{i \eta_\alpha\left(\overline{A}_{\mathbf{r}_\alpha, \mathbf{r}_\alpha+\eta_\alpha \mathbf{b}_\nu}-\overline{A}_{\mathbf{r}_\alpha, \mathbf{r}_\alpha+\eta_\alpha \mathbf{b}_\mu}\right)} \nonumber \\
&+\frac{J_{\pm \pm}}{8} \sum_{\mathbf{r}_\alpha} \sum_{\mu, \nu \neq \mu}\left[e^{i \eta_\alpha\left(\overline{A}_{\mathbf{r}_\alpha, \mathbf{r}_\alpha+\eta_\alpha \mathbf{b}_\nu}+\overline{A}_{\mathbf{r}_\alpha, \mathbf{r}_\alpha+\eta_\alpha \mathbf{b}_\mu}\right)}\right. \left(\Phi_{\mathbf{r}_\alpha}^{\dagger} \Phi_{\mathbf{r}_\alpha}^{\dagger} \chi_{\mathbf{r}_\alpha+\eta_\alpha \mathbf{b}_\mu, \mathbf{r}_\alpha+\eta_\alpha \mathbf{b}_\nu}+\bar{\chi}^{0}_{\mathbf{r}_\alpha, \mathbf{r}_\alpha} \Phi_{\mathbf{r}_\alpha+\eta_\alpha \mathbf{b}_\mu} \Phi_{\mathbf{r}_\alpha+\eta_\alpha \mathbf{b}_\nu}\right.  \nonumber \\
&\hspace{2.8cm}\left.\left. +2 \Phi_{\mathbf{r}_\alpha}^{\dagger} \Phi_{\mathbf{r}_\alpha+\eta_\alpha \mathbf{b}_\mu} \xi_{\mathbf{r}_\alpha, \mathbf{r}_\alpha+\eta_\alpha \mathbf{b}_\nu}+2 \Phi_{\mathbf{r}_\alpha}^{\dagger} \Phi_{\mathbf{r}_\alpha+\eta_\alpha \mathbf{b}_\nu} \xi_{\mathbf{r}_\alpha, \mathbf{r}_\alpha+\eta_\alpha \mathbf{b}_\mu}\right)+\text { H.c. }\right],
\end{align}
\end{widetext}
where $J_{\pm}=-(J_{xx}+J_{zz})/4$ and $J_{\pm\pm}=(J_{zz}-J_{xx})/4$. A detailed construction of $\mathcal{H}_{\text{GMFT}}$ is presented in the Supplemental Material~\cite{Note1}.

\begin{figure*}
    \centering
    \includegraphics[width=0.95\textwidth]{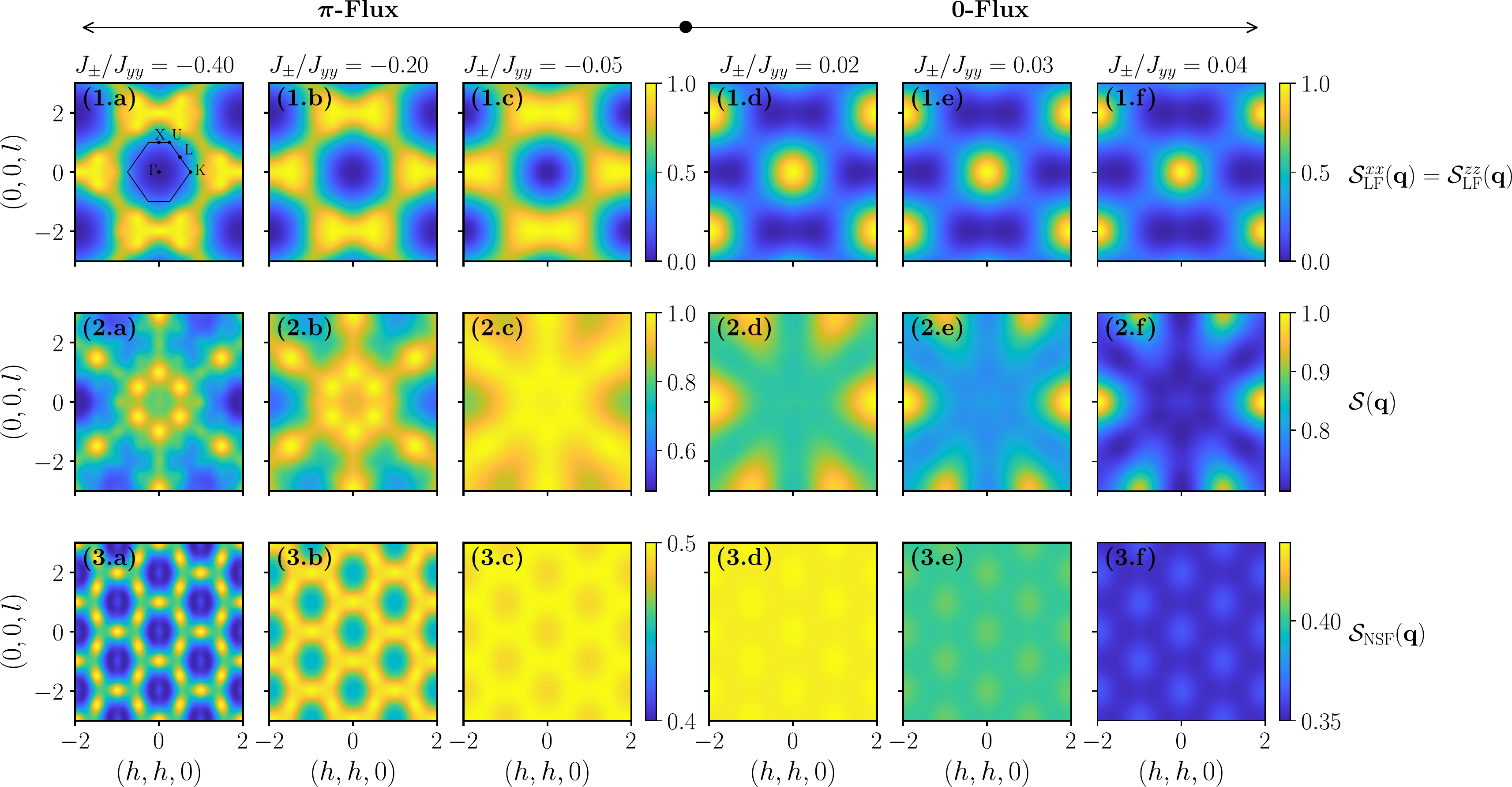}
    \caption{(1) Diagonal equal-time pseudospin correlations in the local frame for the transverse components, (2) total neutron scattering equal-time structure factor and its contribution to the (3) non-spin-flip channel in the [$hhl$] plane for $\pi$-O-QSI with $J_{\pm}/J_{yy}$ equal to (a) -0.40, (b) -0.20 and (c) -0.05 and $0$-O-QSI with $J_{\pm}/J_{yy}$ equal to (d) 0.02, (e) 0.03, and (e) 0.04.} 
    \label{fig: static correlations}
\end{figure*}


\textit{Phase diagram}.--- At this stage, one usually has to make an educated guess on the general form of the background gauge field and the other MF parameters \cite{savary2012coulombic, lee2012generic, savary2013spin}. Using a framework recently introduced by the authors~\cite{desrochers2023symmetry}, we make no such \textit{ad hoc} assumptions and classify all field configurations that yield QSI states symmetric under all space group operations of the diamond lattice. The non-trivial transformation properties of the DO pseudospin moments lead to a distinct classification than in the effective spin-1/2 case~\cite{desrochers2023symmetry}. For a chosen subset of inequivalent field configuration, we solve the self-consistency conditions and compute the ground state energy over the whole quadrupolar dominant quadrant to obtain the GMFT phase diagram presented in Fig.~\ref{fig: phase diagram} (c) (see Supplemental Material~\cite{Note1} for the classification). 

The GMFT phase diagram is consistent with previous works~\cite{patri2020distinguishing, benton2020ground}. We observe a large region where the spinon dispersion becomes gapless at the $\Gamma$ point, thus leading to condensation of the bosons $\expval{\Phi_{\mathbf{k}_c=\mathbf{0}}}\ne 0$. This region corresponds to X or Z all-in-all-out magnetic ordering, as expected by classical simulations, since it has ordering wave vector $\mathbf{k}_c=\mathbf{0}$ and condensation implies $\expval{\mathrm{S}^{\pm}}\sim e^{i\overline{A}}\expval{\Phi^{\dagger}}\expval{\Phi}\ne 0$. 

Most significantly, we find the $0$-O-QSI and $\pi$-O-QSI phases separated by the transition line $J_\pm=0$, as predicted by the perturbative argument outlined above. Along the $XXZ$ line (i.e., $J_{\pm\pm}=0$), we find a transition from 0-O-QSI to the ordered state at $J_{\pm}/J_{yy} \approx 0.048$, in spectacular agreement with QMC where the transition occurs at $J_{\pm}/J_{yy} \approx 0.05$~\cite{banerjee2008unusual, shannon2012quantum, kato2015numerical, huang2020extended}. In these deconfined QSI phases, the spinon spectrum is gapped, and all MF parameters vanish (i.e., $\chi=\chi^{0}=\xi=0$). When examining the GMFT Hamiltonian in Eq.~\eqref{eq: GMFT Hamiltonian}, we see that the disappearance of the MF terms implies that the $U(1)$ symmetry breaking term associated with the $J_{\pm\pm}$ coupling vanishes as well. Even though this emergent $U(1)$ symmetry within the QSI phases might \textit{a priori} seem like an artifact of GMFT, recent ED results corroborate its naturalness~\cite{Hosoi2022Uncovering}. ED calculations observed that for an anisotropic $XYZ$ model (i.e., $J_{xx}\ne J_{yy}\ne J_{zz}$) in a parameter regime where the $\pi$-O-QSI should be stable, the equal-time pseudospin correlations in the local frame (i.e., $\mathcal{S}_{\mathrm{LF}}^{a b}(\boldsymbol{q})=(1 / N) \sum_{i, j} e^{-i \boldsymbol{q} \cdot\left(\boldsymbol{R}_i-\boldsymbol{R}_j\right)}\left\langle \mathrm{S}_i^{a} \mathrm{S}_j^{b}\right\rangle$) satisfy $\mathcal{S}^{xx}_{\mathrm{LF}}=\mathcal{S}^{zz}_\mathrm{LF}$, whereas in classical simulations $\mathcal{S}^{xx}_{\mathrm{LF}}\ne \mathcal{S}^{zz}_\mathrm{LF}$. 


\begin{figure}[b]
\includegraphics[width=1.0\linewidth]{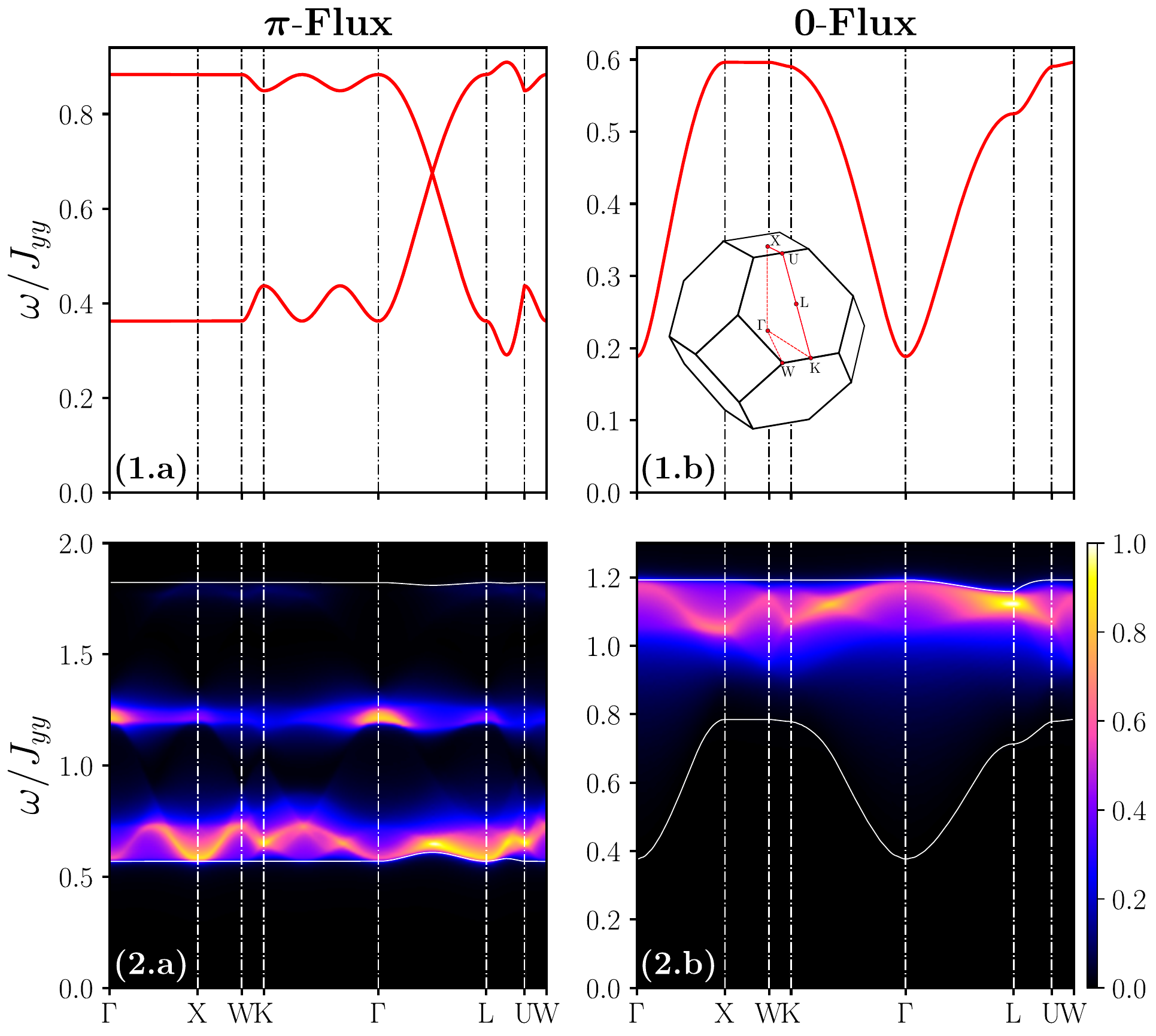}
\caption{(1) Spinon dispersion and (2) dynamical spin structure factor for (a) $\pi$-O-QSI with $J_{\pm}/J_{yy}=-0.1875$ and (b) $0$-O-QSI with $J_{\pm}/J_{yy}=0.04$ along high-symmetry lines of the pyrochlore lattice first Brillouin zone. The solid white lines denote the upper and lower edges of the two-spinon continuum. \label{fig: dssf path}} 
\end{figure}

\begin{figure*}
    \centering
    \includegraphics[width=1.0\textwidth]{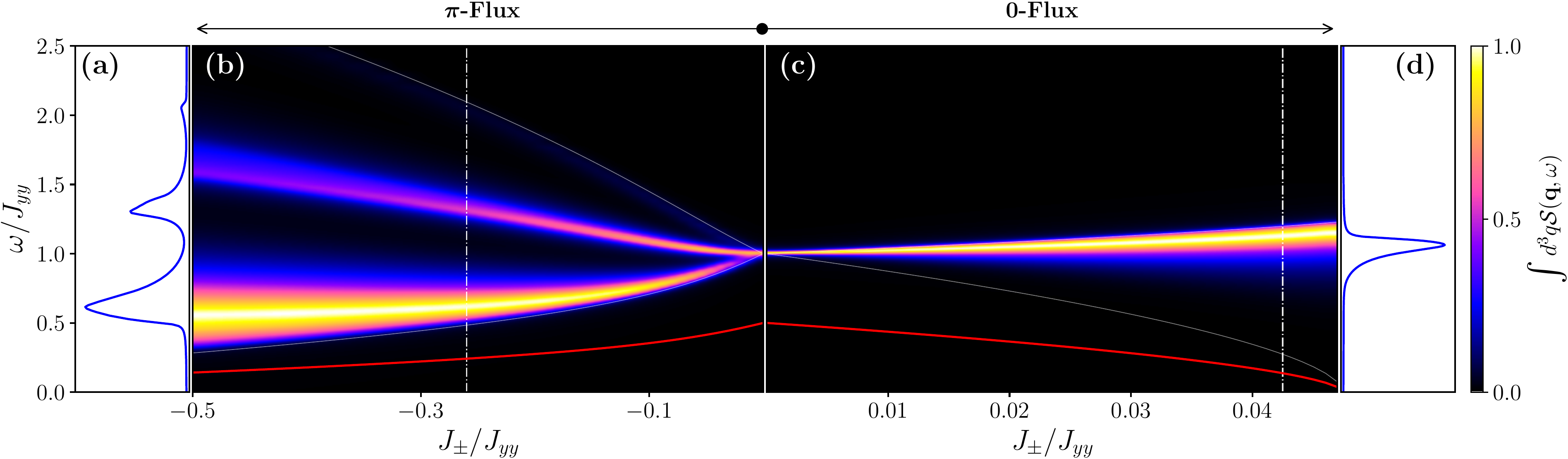}
    \caption{Momentum-integrated dynamical spin structure factor for (b) $\pi$-O-QSI and (c) $0$-O-QSI as a function of $J_{\pm}/J_{yy}$. Cuts at specific values of $J_{\pm}/J_{yy}$, indicated by the vertical white lines in (b) and (c), are presented in (a) and (d). The upper and lower edges of the two-spinon continuum are represented by solid white lines in (b) and (c), whereas the solid red lines indicate the spinon gap. }
\label{fig: 1BZ integrated DSSF}
\end{figure*}

\textit{Equal-time correlations}.--- Now that the range of stability of $0$-O-QSI and $\pi$-O-QSI has been established, we study their physical properties in detail. Since the $\mathrm{S}^{x}$ and $\mathrm{S}^{y}$ components of the pseudospins are octupolar in nature, they are not expected to linearly couple to the magnetic dipoles of neutrons. Accordingly, we assume that the only non-vanishing g-factor in the local frame is $g_{zz}$, an accurate approximation for Ce$_2$Zr$_2$O$_7$~\cite{smith2022case, bhardwaj2022sleuthing}. With $g_{xx}=g_{yy}=0$, neutron scattering probes correlations between the local $z$-components of the pseudospins associated with the spinon excitations ($\mathrm{S}^z \sim \Phi^{\dagger} e^{i {\bar A}} \Phi$), not the photons ($\mathrm{S}^y \sim E$).

We start by considering the equal-time correlations in both deconfined phases. On top of the diagonal equal-time pseudospin correlations $\mathcal{S}^{aa}_{\mathrm{LF}}$, we compute the equal-time neutron scattering structure factor
\begin{equation} \label{eq: equal-time neutron scattering}
\mathcal{S}(\boldsymbol{q})\! =\! \frac{1}{N}\! \sum_{i, j}\! \left[\hat{z}_i\! \cdot \! \hat{z}_j\! -\! \frac{\left(\hat{z}_i\! \cdot \! \boldsymbol{q}\right)\left(\hat{z}_j\! \cdot\! \boldsymbol{q}\right)}{q^2}\right]\! e^{-i \boldsymbol{q} \cdot \left(\boldsymbol{R}_i\! -\! \boldsymbol{R}_j\right)}\expval{\mathrm{S}_i^z \mathrm{S}_j^z}
\end{equation}
where $\hat{z}_i$ is the local $z$-axis at site $i$. With polarized neutrons, this total contribution can be separated into the spin-flip (SF) $\mathcal{S}_{\mathrm{SF}}(\boldsymbol{q})$, and non-spin-flip (NSF) $\mathcal{S}_{\mathrm{NSF}}(\boldsymbol{q})$ channels (neutrons are polarized perpendicular to the scattering plane). These results are presented in Fig.~\ref{fig: static correlations} for $0$-O-QSI and $\pi$-O-QSI along the $[hhl]$ plane. The qualitative features of $\mathcal{S}_{\mathrm{SF}}(\boldsymbol{q})$ are similar to $\mathcal{S}(\boldsymbol{q})$.
Thus, we do not present $\mathcal{S}_{\mathrm{SF}}(\boldsymbol{q})$ separately.

It should first be noted by looking at the equal-time correlations in the local frame and $\mathcal{S}(\mathbf{q})$ that the intensity gets reversed at the transition between $0$-O-QSI and $\pi$-O-QSI while the pattern remains similar. This can be understood by considering the Ising limit where the ground state is an equal weight superposition of 2-in-2-out configurations in the $y$-basis. Such a state corresponds to an equal weight superposition of all single tetrahedra configurations (i.e., 2-in-2-out, 3-in-1-out, 1-in-3-out, and all-in-all-out) in the $x$- or $z$-basis, leading to completely flat $\mathcal{S}^{xx}_{\text{LF}}(\mathbf{q})$, $\mathcal{S}^{zz}_{\text{LF}}(\mathbf{q})$ and $\mathcal{S}(\mathbf{q})$. Then in the $0$-O-QSI ($\pi$-O-QSI) phase close to the Ising limit, the ferromagnetic (antiferromagnetic) transverse coupling favors the all-in-all-out (2-in-2-out) configurations. As argued in Ref.~\cite{castelnovo2019rod}, decoupled tetrahedra where all-in-all-out (2-in-2-out) configurations are favored lead to $\mathcal{S}(\mathbf{q})$ with flat high-intensity (low-intensity) rod motifs as observed in panel (2.c) (panel (2.d)). Going further away from the Ising limit, the intensity of $\mathcal{S}^{xx}_{\text{LF}}(\mathbf{q})=\mathcal{S}^{zz}_{\text{LF}}(\mathbf{q})$ keeps increasing at the $(0,0,2)$ and $(0,0,0)$ points for $\pi$-O-QSI and $0$-O-QSI respectively. The previous single-tetrahedron argument outline above starts to fail, as signaled by an increasing intensity modulation along the rods in $\mathcal{S}(\mathbf{q})$. As correlations between tetrahedra in the $z$-basis increase, we interestingly find a corresponding rise of the contrast in the NSF channels. This intensity modulation along the rods and in the NSF channels is especially striking considering that such features are observed in experiments and ED~\cite{gaudet2019quantum, smith2022case, Hosoi2022Uncovering}, but all classical calculations report completely flat rods and a featureless NSF channel~\cite{Hosoi2022Uncovering, smith2022case, bhardwaj2022sleuthing}. Such features can only be obtained classically by artificially introducing next-nearest-neighbor or dipolar interactions~\cite{bhardwaj2022sleuthing, Chung2022Probing}. This observation seems to imply that the variations along the rods and in the NSF channel are due to quantum fluctuations that evade classical treatments. A detailed comparison presented in the Supplemental Material~\cite{Note1} shows that these equal-time correlations are consistent with the 32-site exact diagonalization (ED) results of Ref.~\cite{Hosoi2022Uncovering} and recent experimental measurements on Ce$_2$Zr$_2$O$_7$.


\textit{Dynamical correlations}.--- Next, we turn to the dynamical spin structure factor (DSSF). Fig.~\ref{fig: dssf path} presents the spinon dispersion and DSSF between high symmetry points for $0$-O-QSI and $\pi$-O-QSI. The intensity is concentrated near the upper edge of the two-spinon continuum for $0$-O-QSI. A detailed analysis shows that predictions from GMFT are in excellent semi-quantitative agreement with the QMC results of Ref.~\cite{huang2018dynamics} for the 0-flux phase (see Supplemental Material~\cite{Note1}). We take this highly non-trivial check as convincing evidence that GMFT does provide a reliable description of spinons dynamics in QSI. In the case of $\pi$-O-QSI, the spinon dispersion is composed of two bands (the flux enlarges the unit cell~\cite{desrochers2023symmetry, essin2014spectroscopic}) that are mostly flat (i.e., standard deviation of the two bands is much lower than their separation). This leads to a two-spinon density $\rho^{(2)}(\omega)\sim \sum_{\alpha, \beta, \mathbf{q}_{1},\mathbf{q}_{2}} \delta(\omega - \mathcal{E}_{\alpha}(\mathbf{q}_{1}) - \mathcal{E}_{\beta}(\mathbf{q}_{2}))$ with three peaks coming from processes involving two spinons in the lowest band (lowest energy peak), the two different bands (central peak), and the highest band (highest energy peak) (see Supplemental Material~\cite{Note1} for further details). Since inelastic neutron scattering probes the two-spinon continuum, these three contributions are visible in the DSSF of $\pi$-O-QSI presented in panel (2.a) of Fig.~\ref{fig: dssf path}, although the high energy peak close to the upper edges of the continuum is faint.

This continuum with three peaks is a distinctive signature. To see if it could still be measured by inelastic neutron scattering experiments on powder samples,  we present the momentum-integrated DSSF as a function of $J_{\pm}/J_{yy}$ in Fig.~\ref{fig: 1BZ integrated DSSF}. For $J_{\pm}/J_{yy}\to 0$, the DSSF collapses to a single peak at $\omega=J_{yy}$ since the spinon dispersion is entirely flat (i.e., $\mathcal{E}_{\gamma}(\mathbf{k})=J_{yy}/2$) in that limit. A single peak concentrated near the upper edge of the two-spinon continuum is observed for $0$-O-QSI from that point up to when the spinon gap vanishes and the bosons condense. For $\pi$-O-QSI, the three peaks are clearly discernible. As one moves from the Ising limit to the Heisenberg point, their separation as well as the relative intensity of the lowest energy peak compared to the second and third ones slowly increase. As a confirmation, we show in the Supplemental Material~\cite{Note1} that 32-site ED results display signatures of this multiple-peak structure despite strong finite-size limitations.

\textit{Discussion}.--- In this Letter, we used a newly introduced extension of GMFT to study octupolar QSI. We obtained a phase diagram consistent with previous studies where the deconfined $0$-O-QSI and $\pi$-O-QSI phases are separated by the $J_{\pm}=0$ line. We further showed that $0$-O-QSI and $\pi$-O-QSI have energy-integrated neutron scattering signatures that have inverted intensities in momentum space and highlighted that GMFT produces the typical rod motifs observed in experiments with intensity modulation along the rods and in the NSF channels --- features absent from classical treatments. It is then shown that $0$-O-QSI has a DSSF with a single peak close to the upper edge of the two-spinon continuum, whereas $\pi$-O-QSI has three distinctive peaks resulting from two mostly flat spinon bands.

These three peaks provide a distinctive experimentally accessible smoking-gun signature for $\pi$-flux QSI. The third peak is most likely too faint to be measured, but high-resolution inelastic neutron scattering on powder samples should be able to observe the first two. For instance, using the microscopic parameters of Refs.~\cite{smith2022case,bhardwaj2022sleuthing}, the separation of the first two peaks for Ce$_2$Zr$_2$O$_7$ should be approximately of $0.06\text{ meV}$ and could be resolved with the best available experimental apparatus~\cite{Note1}.


\begin{acknowledgments}
We thank Emily Z. Zhang, Han Yan, Andriy H. Nevidomskyy, Victor Porée, and Romain Sibille for helpful discussions.  We acknowledge support from the Natural Sciences and Engineering Research Council of Canada (NSERC) and the Centre of Quantum Materials at the University of Toronto. Computations were performed on the Niagara cluster, which SciNet hosts in partnership with the Digital Research Alliance of Canada. F.D. is further supported by the Vanier Canada Graduate Scholarship. Y.B.K. is also supported by the Guggenheim Fellowship from the John Simon Guggenheim Memorial Foundation and the Simons Fellowship from the Simons Foundation. Some parts of this work were performed at the Aspen Center for Physics, which is supported by the National Science Foundation grant PHY-1607611.
\end{acknowledgments}

\underline{\emph{Note added:}} After this manuscript was made public but before its publication, a new inelastic neutron scattering experiment on powder samples of material candidate Ce$_2$Sn$_2$O$_7$ achieved an energy resolution of 0.7~$\mu$eV and reported the presence of three peaks of decreasing intensity in the measured dynamical spin structure factor~\cite{poree2023fractional}. 



%

\end{document}


\title{Supplemental Material for \linebreak
``Spectroscopic Signatures of Fractionalization in Octupolar Quantum Spin Ice''}

\author{F\'elix Desrochers}
\email{felix.desrochers@mail.utoronto.ca}
\affiliation{%
 Department of Physics, University of Toronto, Toronto, Ontario M5S 1A7, Canada
}%
\author{Yong Baek Kim}%
\email{ybkim@physics.utoronto.ca}
\affiliation{%
 Department of Physics, University of Toronto, Toronto, Ontario M5S 1A7, Canada
}%

\date{\today}

\maketitle

\tableofcontents

\setcounter{secnumdepth}{3}
\setcounter{equation}{0}
\renewcommand{\theequation}{S\arabic{equation}}
\setcounter{table}{0}
\renewcommand{\thetable}{S\arabic{table}}
\setcounter{figure}{0}
\renewcommand{\thefigure}{S\arabic{figure}}

\section{\label{SI sec: Microscopic model of dipolar-octupolar pyrochlores} Microscopic model of dipolar-octupolar pyrochlores}

\subsection{Coordinate systems}

\subsubsection{Lattice coordinates}

The magnetically active ions in spin ice form a pyrochlore lattice, an FCC Bravais lattice with four sublattices shaping into a network of corner-sharing tetrahedra. To identify the position of a unit cell on the pyrochlore lattice, we introduce the \emph{global cartesian coordinates} (GCC), which are the standard frame coordinates of the FCC cube with edge length set to unity, and the following three basis vectors
\begin{subequations} \label{eq: basis for SIPC expressed in GCC}
    \begin{align}
        &\hat{\mathbf{e}}_1 = \frac{1}{2}\left( 0,1,1 \right)\\
        &\hat{\mathbf{e}}_2 = \frac{1}{2}\left( 1,0,1 \right)\\
        &\hat{\mathbf{e}}_3 = \frac{1}{2}\left( 1,1,0 \right).
    \end{align}
\end{subequations}
For later convenience, we also introduce $\hat{\mathbf{e}}_0 = \left( 0,0,0 \right)$. The position of the sites within the unit cell is expressed by defining $\hat{\epsilon}_i = \frac{1}{2}\hat{\mathbf{e}}_i$ ($i=1,2,3$) to be the displacement of the $i=1,2,3$ sublattices from the $i=0$ sublattice respectively (where $\hat{\epsilon}_0=\hat{\mathbf{e}}_0=\mathbf{0}$).

The diamond lattice is premedial to the pyrochlore lattice \cite{henley2010coulomb}. We refer to it as the parent diamond lattice. This parent lattice is an FCC Bravais lattice with two sublattices positioned at the center of the up and down-pointing tetrahedra. The initial pyrochlore lattice sites are at the center of the bonds on the diamond lattice. Each down tetrahedron is connected to four nearest-neighbor up tetrahedra by
\begin{subequations} \label{eq: NN vectors in parent diamond lattice}
    \begin{align}
        &\mathbf{b}_0 = \frac{-1}{4}\left( 1, 1, 1 \right)\\
        &\mathbf{b}_1 = \frac{1}{4}\left( -1,1,1 \right)\\
        &\mathbf{b}_2 = \frac{1}{4}\left( 1,-1,1 \right)\\
        &\mathbf{b}_3 = \frac{1}{4}\left( 1,1,-1 \right).
    \end{align}
\end{subequations}
Each up tetrahedron is connected to four down tetrahedra by the opposite vectors. To label the position of the sites on this parent diamond lattice, we introduce the \emph{sublattice indexed diamond coordinates} (SIDC), where the unit cell is identified by a linear combination of the three basis vectors in Eq.~\eqref{eq: basis for SIPC expressed in GCC}. The two sublattices are defined by the sublattice displacement vectors $-\eta_{\alpha} \mathbf{b}_0/2$, where $\eta_{A}=1$ and $\eta_{B}=-1$ with $\alpha$ labeling the sublattice, and $A$ ($B$) stands for down (up). This coordinate system is related to the GCC by 
\begin{align*}
\mathbf{r}_{\alpha} &= \left( r_1, r_2, r_3 \right)_{\alpha} = r_1 \hat{\mathbf{e}}_1 + r_2 \hat{\mathbf{e}}_2 + r_3 \hat{\mathbf{e}}_3 - \frac{\eta_{\alpha}}{2} \mathbf{\mathbf{b}}_{0} \hspace{2.5mm} \text{(SIDC)} \nonumber\\
&= \frac{1}{2} \left( r_2 + r_3, r_1 + r_3, r_1 + r_2 \right) - \frac{\eta_{\alpha}}{2}\mathbf{b}_{0} \hspace{12mm}\text{(GCC)}  \nonumber
\end{align*}


\subsubsection{Local coordinates}

Spins on the four different sublattices are defined in a local frame. The basis vectors of these sublattice-dependant coordinates systems are defined in table~\ref{tab: Local basis}.

\begin{table}[!ht]
\caption{\label{tab: Local basis}%
Local sublattice basis vectors
}
\begin{ruledtabular}
\begin{tabular}{ccccc}
$i$ & 0 & 1  & 2  & 3 \\
\hline
$\hat{z}_{i}$ & $\frac{1}{\sqrt{3}}\left(1,1,1\right)$ & $\frac{-1}{\sqrt{3}}\left(-1,1,1\right)$  & $\frac{-1}{\sqrt{3}}\left(1,-1,1\right)$  & $\frac{-1}{\sqrt{3}}\left(1,1,-1\right)$   \\[2mm]
$\hat{y}_{i}$ & $\frac{1}{\sqrt{2}}\left(0,-1,1\right)$  & $\frac{1}{\sqrt{2}}\left(0,1,-1\right)$  & $\frac{-1}{\sqrt{2}}\left(0,1,1\right)$ & $\frac{1}{\sqrt{2}}\left(0,1,1\right)$  \\[2mm]
$\hat{x}_{i}$ & $\frac{1}{\sqrt{6}}\left(-2,1,1\right)$ & $\frac{-1}{\sqrt{6}}\left(2,1,1\right)$  & $\frac{1}{\sqrt{6}}\left(2,1,-1\right)$  & $\frac{1}{\sqrt{6}}\left(2,-1,1\right)$    \\
\end{tabular}
\end{ruledtabular}
\end{table}

\subsection{Microscopic Hamiltonian}

A well-isolated ground state doublet can be represented by a pseudospin-$1/2$ operator~\cite{smith2022reply}. From symmetry considerations, certain multipole operators are non-vanishing in this doublet and can be used to write the pseudospin operators. For dipolar-octupolar (DO) doublets, the pseudospin operators are
\begin{subequations} \label{eq: definition of the pseudospin operators}
  \begin{align}
    \tilde{\mathrm{S}}^x &=  \mathcal{P} \left( \mathcal{C}_{0}\left( (J^{x})^{3}-\overline{J^{x} J^{y} J^{y}}\right)+\mathcal{C}_{1} J^{z} \right) \mathcal{P}  \\
    \tilde{\mathrm{S}}^y &= \mathcal{C}_{2} \mathcal{P} \left( (J^{y})^{3}-\overline{J^{y} J^{x} J^{x}} \right) \mathcal{P}   \\
    \tilde{\mathrm{S}}^z &= \mathcal{C}_3 \mathcal{P}J^{z}\mathcal{P},
  \end{align}
\end{subequations}
where $\mathcal{P}$ is a projection operator into the DO doublet, and the overline denotes symmetrized products. The constants $\mathcal{C}_0$, $\mathcal{C}_1$, $\mathcal{C}_2$ and $\mathcal{C}_3$ are determined by CEF parameters and are needed to ensure that the pseudospins form a $\mathfrak{su}(2)$ algebra (i.e., $\comm{\tilde{\mathrm{S}}^m}{\tilde{\mathrm{S}}^n}= i\sum_l \varepsilon^{mnl}\tilde{\mathrm{S}}^{l}$). $\mathcal{C}_1=0$ for $\mathrm{Ce}_2 \mathrm{Sn}_2 \mathrm{O}_7$ and $\mathrm{Ce}_2 \mathrm{Zr}_2 \mathrm{O}_7$ \cite{patri2020distinguishing}. Even though $\mathrm{S}^{x}$ is octupolar, it is referred to as a dipolar moment since it transforms identically to a dipole under the symmetry operations of the $D_{3d}$ local site symmetry group.

From the transformation properties of these pseudospin operators defined above, the most general symmetry-allowed Hamiltonian with nearest-neighbor bilinear interactions has the form 
\begin{align}\label{eq: initial XYZ+XZ Hamiltonian}
\mathcal{H}=\sum_{\langle \mathbf{R}_{i}, \mathbf{R}_{j}' \rangle}\left[\tilde{J}_{x x} \tilde{\mathrm{S}}_{\mathbf{R}_i}^x \tilde{\mathrm{S}}_{\mathbf{R}_j'}^x + \tilde{J}_{yy} \tilde{\mathrm{S}}_{\mathbf{R}_i}^{y} \tilde{\mathrm{S}}_{\mathbf{R}_j'}^{y} + \tilde{J}_{zz} \tilde{\mathrm{S}}_{\mathbf{R}_i}^{z} \tilde{\mathrm{S}}_{\mathbf{R}_j'}^{z} + \tilde{J}_{x z}\left(\tilde{\mathrm{S}}_{\mathbf{R}_i}^{x} \tilde{\mathrm{S}}_{\mathbf{R}_j'}^{z} + \tilde{\mathrm{S}}_{\mathbf{R}_i}^{z} \tilde{\mathrm{S}}_{\mathbf{R}_j'}^{x} \right)\right],
\end{align}
where $\mathbf{R}_{i}$ represent the sites of the pyrochlore lattice and the index $i\in\{0,1,2,3\}$ labels the sublattices. The $\tilde{J}_{xz}$ term can be eliminated by performing a uniform rotation around the local $\hat{y}_{i}$ axes 
\begin{subequations} \label{eq: Trf: nn pyrochlore -> XYZ}
\begin{align} 
\mathrm{S}^{x} &= \cos(\theta) \tilde{\mathrm{S}}^{x}-\sin (\theta) \tilde{\mathrm{S}}^{z} \\
\mathrm{S}^{y} &= \tilde{\mathrm{S}}^{y},\\
\mathrm{S}^{z} &= \sin(\theta) \tilde{\mathrm{S}}^{x} + \cos(\theta) \tilde{\mathrm{S}}^{z},
\end{align}
\end{subequations}
where $\tan (2 \theta)= 2 \tilde{J}_{x z}/(\tilde{J}_{z z}-\tilde{J}_{x x})$. Such a transformation yields the $XYZ$ model
\begin{align}\label{eq: XYZHamiltonian}
\mathcal{H}=\sum_{\langle \mathbf{R}_{i}, \mathbf{R}_{j}' \rangle}\left[ J_{x x} \mathrm{S}_{\mathbf{R}_i}^x \mathrm{S}_{\mathbf{R}_j'}^x + J_{yy} \mathrm{S}_{\mathbf{R}_i}^{y} \mathrm{S}_{\mathbf{R}_j'}^{y} + J_{zz} \mathrm{S}_{\mathbf{R}_i}^{z} \mathrm{S}_{\mathbf{R}_j'}^{z}\right]
\end{align}
with the renormalized couplings
\begin{subequations} \label{eq: Relation couplings nn pyrochlore -> XYZ}
\begin{align} 
&J_{xx}=\frac{\tilde{J}_{z z} + \tilde{J}_{x x}}{2}-\frac{\sqrt{\left(\tilde{J}_{z z}-\tilde{J}_{x x}\right)^{2}+4 \tilde{J}_{x z}^{2}}}{2}, \\
&J_{yy}=\tilde{J}_{y y}, \\
&J_{zz}=\frac{\tilde{J}_{z z}+\tilde{J}_{x x}}{2}+\frac{\sqrt{\left(\tilde{J}_{z z}-\tilde{J}_{x x}\right)^{2}+4 \tilde{J}_{x z}^{2}}}{2}.
\end{align}
\end{subequations}
When evaluating the equal-time and dynamical spin structure factor in the main text, it is assumed that $\tilde{J}_{xz}\approx 0$ such that $\theta\approx 0$. This appears to be true for Ce$_2$Zr$_2$O$_7$~\cite{smith2022case}.

The raising and lowering operators $\mathrm{S}^{\pm}=\mathrm{S}^z \pm i \mathrm{S}^x$ can further be introduced to rewrite the $XYZ$ Hamiltonian in the form 
\begin{align} \label{eq: XYZ with raising/lowering}
\mathcal{H}=\sum_{\langle \mathbf{R}_{i}, \mathbf{R}_{j}' \rangle}  \left[J_{y y} \mathrm{S}_{\mathbf{R}_{i}}^y \mathrm{S}_{\mathbf{R}_{j}^{\prime}}^y-J_{\pm}\left(\mathrm{S}_{\mathbf{R}_{i}}^{+} \mathrm{S}_{\mathbf{R}_{j}^{\prime}}^{-}+\mathrm{S}_{\mathbf{R}_{i}}^{-} \mathrm{S}_{\mathbf{R}_{j}^{\prime}}^{+}\right) +J_{\pm \pm}\left(\mathrm{S}_{\mathbf{R}_{i}}^{+} \mathrm{S}_{\mathbf{R}_{j}^{\prime}}^{+}+\mathrm{S}_{\mathbf{R}_{i}}^{-} \mathrm{S}_{\mathbf{R}_{j}^{\prime}}^{-}\right)\right], \end{align}
where $J_{\pm}=-\left( J_{xx} + J_{zz} \right)/4$ and $J_{\pm\pm}=\left( J_{xx} - J_{zz} \right)/4$.

\section{\label{SI sec: Gauge mean-field approximation} Gauge mean-field construction}

\subsection{Parton construction}

In GMFT, the initial spin-1/2 Hilbert space on the pyrochlore lattice $\mathscr{H}_{\text{spin}}= \otimes_N \mathscr{H}_{\text{S=1/2}}$ is augmented to a new larger one $\mathscr{H}_{\text{big}} = \mathscr{H}_{\text{spin}} \otimes \mathscr{H}_Q$ where bosonic degrees of freedom are introduced on the parent diamond lattice. This new bosonic Hilbert space $\mathscr{H}_Q$ describes the bosonic field $Q_{\mathbf{r}_{\alpha}} \in \mathbb{Z}$ defined on each site of the parent diamond lattice site. For this mapping to be exact, the discretized Gauss's law
\begin{equation}
    Q_{\mathbf{r}_{\alpha}} = \eta_{\alpha} \sum_{\mu=0}^3 \mathrm{S}^{y}_{\mathbf{r}_{\alpha}+\eta_{\alpha}\mathbf{b}_\mu/2} , \label{eq: physical constraint charge GMFT}
\end{equation}
needs to be enforced for every tetrahedron. The canonically conjugate variable to the bosonic charge is $\varphi_{\mathbf{r}_{\alpha}}$ (i.e., $\comm{\varphi_{\mathbf{r}_{\alpha}}}{Q_{\mathbf{r}_{\alpha}}}=i$). This naturally leads to the definition of raising and lowering operators $\Phi_{\mathbf{r}_{\alpha}}^\dag = e^{i\varphi_{\mathbf{r}_{\alpha}}}$ and  $\Phi_{\mathbf{r}_{\alpha}} = e^{-i\varphi_{\mathbf{r}_{\alpha}}}$ respectively. These rotors respect the constraint $|\Phi_{\mathbf{r}_{\alpha}}^{\dagger}\Phi_{\mathbf{r}_{\alpha}}|=1$ by construction. As explained in the main text, the spin variables are mapped to operators defined in this enlarged Hilbert space as in Eq. (2), where $A$ and $E$ are canonical conjugate fields that act within the $\mathscr{H}_{\text{spin}}$ subspace of $\mathscr{H}_{\text{big}}$. The local $y$-component of the spin now corresponds to the emergent electric field, and the raising/lowering operators create a pair of spinons on the parent lattice while creating/annihilating an electric field quanta to respect Eq.~\eqref{eq: physical constraint charge GMFT}.

Making this replacement directly into Eq.~\eqref{eq: XYZ with raising/lowering}, we get 
\begin{align} \label{eq: XYZ Hamiltonian with parton operators}
\mathcal{H} =&\frac{J_{y y}}{2} \sum_{\mathbf{r}_\alpha} Q_{\mathbf{r}_\alpha}^2  -\frac{J_{\pm}}{4} \sum_{\mathbf{r}_\alpha} \sum_{\mu, \nu \neq \mu} \Phi_{\mathbf{r}_\alpha+\eta_\alpha \mathbf{b}_\mu}^{\dagger} \Phi_{\mathbf{r}_\alpha+\eta_\alpha \mathbf{b}_\nu} e^{i \eta_\alpha\left(A_{\mathbf{r}_\alpha, \mathbf{r}_\alpha+\eta_\alpha \mathbf{b}_\nu}-A_{\mathbf{r}_\alpha, \mathbf{r}_\alpha+\eta_\alpha \mathbf{b}_\mu}\right)} \nonumber \\
&+\frac{J_{\pm \pm}}{8} \sum_{\mathbf{r}_\alpha} \sum_{\mu, \nu \neq \mu}\left( e^{i \eta_\alpha\left(A_{\mathbf{r}_\alpha, \mathbf{r}_\alpha+\eta_\alpha \mathbf{b}_\nu}+A_{\mathbf{r}_\alpha, \mathbf{r}_\alpha+\eta_\alpha \mathbf{b}_\mu}\right)}\Phi_{\mathbf{r}_\alpha}^{\dagger} \Phi_{\mathbf{r}_\alpha}^{\dagger}  \Phi_{\mathbf{r}_\alpha+\eta_\alpha \mathbf{b}_\mu} \Phi_{\mathbf{r}_\alpha+\eta_\alpha \mathbf{b}_\nu} + \text{ h.c.}\right).
\end{align}
The total partition function is
\begin{align}
    \mathcal{Z}=&\int \mathcal{D}[\varphi, Q,A,E,\lambda,\zeta] e^{-S_{\text {matter}} - S_{\text {EM}}},
\end{align}
where 
\begin{equation}
    S_\mathrm{EM}=\frac{U}{2} \int_{0}^{\beta}\dd{\tau} \sum_{\left\langle\mathbf{r}_\alpha \mathbf{r}_\beta^{\prime}\right\rangle} \left(E_{\mathbf{r}_{\alpha} \mathbf{r}_{\beta}^{\prime}}^{\tau }\right)^{2}  
\end{equation}
enforces the odd vacuum condition $E_{\mathbf{r}_{\alpha} \mathbf{r}_{\beta}^{\prime}}=\pm 1/2$ by taking the $U\to \infty$ limit, and $S_{\text {matter}}$ describes the quantum rotors coupled to the $U(1)$ gauge field
\begin{align}
    S_{\text{matter}}=&\int_{0}^{\beta} \dd{\tau} \left[\sum_{\mathbf{r}_{\alpha}}\left(i Q_{\mathbf{r}_{\alpha}}^{\tau} \partial_{\tau} \varphi_{\mathbf{r}_{\alpha}}^{\tau}+i \lambda_{\mathbf{r}_\alpha}^{\tau} \left(\Phi_{\mathbf{r}_{\alpha}}^{\tau *} \Phi_{\mathbf{r}_{\alpha}}^{\tau} -1\right)+ i\zeta_{\mathbf{r}_{\alpha}}^{\tau}\left(\sum_{\mu} E_{\mathbf{r}_\alpha, \mathbf{r}_\alpha + \eta_{\alpha} \mathbf{b}_{\mu}}^{\tau}-Q_{\mathbf{r}_\alpha}^{\tau} \right) \right)  + \mathcal{H}  
    \right].
\end{align}
The Lagrange multipliers $\lambda_{\mathbf{r}_\alpha}^{\tau}$ and $\zeta_{\mathbf{r}_\alpha}^{\tau}$ enforce the constraint $|\Phi_{\mathbf{r}_{\alpha}}^{\dagger}\Phi_{\mathbf{r}_{\alpha}}|=1$ and Eq.~\eqref{eq: physical constraint charge GMFT} respectively. 

\subsection{Mean-field decoupling and saddle point approximation}

To get a tractable model, we first decouple the four bosons term associated with the $J_{\pm\pm}$ coupling. To do so, we follow the prescription of Refs.~\cite{lee2012generic, savary2021quantum} and apply the following decoupling 
\begin{align}
 \Phi^{\dagger}_{1} \Phi^{\dagger}_ {2} \Phi_{3} \Phi_{4} \mathbf{s} \mathbf{s} \rightarrow &\langle\mathbf{s}\rangle\langle\mathbf{s}\rangle \left(\left\langle\Phi^{\dagger}_{1} \Phi^{\dagger}_{2}\right\rangle \Phi_{3} \Phi_{4} + \Phi^{\dagger}_{1} \Phi^{\dagger}_{2}\langle\Phi_{3} \Phi_{4}\rangle+\left\langle\Phi^{\dagger}_{1} \Phi_{3}\right\rangle \Phi^{\dagger}_{2} \Phi_4+\Phi^{\dagger}_{1} \Phi_{3} \left\langle\Phi^{\dagger}_{2} \Phi_{4}\right\rangle  + \left\langle\Phi^{\dagger}_{1} \Phi_{4}\right\rangle \Phi^{\dagger}_{2} \Phi_{3} + \Phi^{\dagger}_{1} \Phi_{4}\left\langle\Phi^{\dagger}_{2} \Phi_{3}\right\rangle \right) \nonumber \\
& +(\langle\mathbf{s}\rangle \mathbf{s}+\mathbf{s}\langle\mathbf{s}\rangle)\left(\left\langle\Phi^{\dagger}_{1} \Phi^{\dagger}_{2}\right\rangle\langle\Phi_{3} \Phi_{4}\rangle+\left\langle\Phi^{\dagger}_{1} \Phi_{3}\right\rangle\left\langle\Phi^{\dagger}_{2} \Phi_{4}\right\rangle+\left\langle\Phi^{\dagger}_{1} \Phi_{4} \right\rangle\left\langle\Phi^{\dagger}_{2} \Phi_{3} \right\rangle\right) \nonumber \\
& -2\langle\mathbf{s}\rangle\langle\mathbf{s}\rangle\left(\left\langle\Phi^{\dagger}_{1} \Phi^{\dagger}_{2}\right\rangle\langle\Phi_{3} \Phi_{4}\rangle + \left\langle\Phi^{\dagger}_{1} \Phi_{3}\right\rangle\left\langle\Phi^{\dagger}_{2} \Phi_{4}\right\rangle + \left\langle\Phi^{\dagger}_{1} \Phi_{4}\right\rangle\left\langle\Phi^{\dagger}_{2} \Phi_{3}\right\rangle\right),
\end{align}
where we use the shorthand notation $\mathbf{s}=e^{iA}/2$, and taking the expectation value $\expval{\mathbf{s}}=e^{i\overline{A}}/2$ correspond to fixing the gauge field to a constant background. If we further fix all gauge connections to a constant background (i.e., $A\to \bar{A}$), the decoupling simplifies to
\begin{align}
 \Phi^{\dagger}_{1} \Phi^{\dagger}_ {2} \Phi_{3} \Phi_{4} \mathbf{s} \mathbf{s} \rightarrow \langle\mathbf{s}\rangle\langle\mathbf{s}\rangle & \left(\left\langle\Phi^{\dagger}_{1} \Phi^{\dagger}_{2}\right\rangle \Phi_{3} \Phi_{4} + \Phi^{\dagger}_{1} \Phi^{\dagger}_{2}\langle\Phi_{3} \Phi_{4}\rangle + \left\langle\Phi^{\dagger}_{1} \Phi_{3}\right\rangle \Phi^{\dagger}_{2} \Phi_4 \right.\nonumber\\
 &\left. + \Phi^{\dagger}_{1} \Phi_{3}\left\langle\Phi^{\dagger}_{2} \Phi_{4}\right\rangle  + \left\langle\Phi^{\dagger}_{1} \Phi_{4}\right\rangle \Phi^{\dagger}_{2} \Phi_{3} + \Phi^{\dagger}_{1} \Phi_{4}\left\langle\Phi^{\dagger}_{2} \Phi_{3}\right\rangle \right).
\end{align}
We can then introduce the inter-site pairing $\chi$, on-site pairing $\chi^{0}$, and inter-sublattice hopping $\xi$ MF parameters to rewrite the MF Hamiltonian as
\begin{align} \label{eq: GMFT Hamiltonian}
\mathcal{H}_{\mathrm{MF}}=&\frac{J_{y y}}{2} \sum_{\mathbf{r}_\alpha} Q_{\mathbf{r}_\alpha}^2 -\frac{J_{\pm}}{4} \sum_{\mathbf{r}_\alpha} \sum_{\mu, \nu \neq \mu} \Phi_{\mathbf{r}_\alpha+\eta_\alpha \mathbf{b}_\mu}^{\dagger} \Phi_{\mathbf{r}_\alpha+\eta_\alpha \mathbf{b}_\nu} e^{i \eta_\alpha\left(\overline{A}_{\mathbf{r}_\alpha, \mathbf{r}_\alpha+\eta_\alpha \mathbf{b}_\nu}-\overline{A}_{\mathbf{r}_\alpha, \mathbf{r}_\alpha+\eta_\alpha \mathbf{b}_\mu}\right)} \nonumber \\
&+\frac{J_{\pm \pm}}{8} \sum_{\mathbf{r}_\alpha} \sum_{\mu, \nu \neq \mu}\left[e^{i \eta_\alpha\left(\overline{A}_{\mathbf{r}_\alpha, \mathbf{r}_\alpha+\eta_\alpha \mathbf{b}_\nu}+\overline{A}_{\mathbf{r}_\alpha, \mathbf{r}_\alpha+\eta_\alpha \mathbf{b}_\mu}\right)}\right. \left(\Phi_{\mathbf{r}_\alpha}^{\dagger} \Phi_{\mathbf{r}_\alpha}^{\dagger} \chi_{\mathbf{r}_\alpha+\eta_\alpha \mathbf{b}_\mu, \mathbf{r}_\alpha+\eta_\alpha \mathbf{b}_\nu}+\bar{\chi}^{0}_{\mathbf{r}_\alpha, \mathbf{r}_\alpha} \Phi_{\mathbf{r}_\alpha+\eta_\alpha \mathbf{b}_\mu} \Phi_{\mathbf{r}_\alpha+\eta_\alpha \mathbf{b}_\nu}\right.  \nonumber \\
&\hspace{2.8cm}\left.\left. +2 \Phi_{\mathbf{r}_\alpha}^{\dagger} \Phi_{\mathbf{r}_\alpha+\eta_\alpha \mathbf{b}_\mu} \xi_{\mathbf{r}_\alpha, \mathbf{r}_\alpha+\eta_\alpha \mathbf{b}_\nu}+2 \Phi_{\mathbf{r}_\alpha}^{\dagger} \Phi_{\mathbf{r}_\alpha+\eta_\alpha \mathbf{b}_\nu} \xi_{\mathbf{r}_\alpha, \mathbf{r}_\alpha+\eta_\alpha \mathbf{b}_\mu}\right)+\text { h.c. }\right],
\end{align}
where the MF parameters have to satisfy the self-consistency conditions
\begin{subequations} \label{eq: self-consistency conditions}
\begin{align}
    \chi_{\mathbf{r}_{\alpha}+\eta_{\alpha}\mathbf{b}_\mu, \mathbf{r}_{\alpha}+\eta_{\alpha}\mathbf{b}_{\nu}} &= \expval{\Phi_{\mathbf{r}_{\alpha}+\eta_{\alpha}\mathbf{b}_\mu} \Phi_{\mathbf{r}_{\alpha}+\eta_{\alpha}\mathbf{b}_{\nu}}} \\
    \overline{\chi}^{0}_{\mathbf{r}_{\alpha}, \mathbf{r}_{\alpha}} &= \expval{\Phi_{\mathbf{r}_{\alpha}}^{\dagger} \Phi_{\mathbf{r}_{\alpha}}^{\dagger} }\\
    \xi_{\mathbf{r}_{\alpha},\mathbf{r}_{\alpha}+\eta_{\alpha}\mathbf{b}_{\mu}} &= \expval{\Phi_{\mathbf{r}_{\alpha}}^{\dagger} \Phi_{\mathbf{r}_{\alpha}+\eta_{\alpha}\mathbf{b}_{\mu}}}.
\end{align}
\end{subequations}
We see that the first term of $\mathcal{H}_{\text{MF}}$ corresponds to the energy cost for the existence of spinons. The second describes intra-sublattice hopping of the spinon while coupled to the fixed constant background, whereas the $J_{\pm\pm}$ term describes both inter-sublattice hopping and pairing. Turning back to the total partition function, we further allow the gauge charges to take on any integer value $Q_{\mathbf{r}_{\alpha}}\in(-\infty,\infty)$ instead of being constrained to $|Q_{\mathbf{r}_{\alpha}}|\le 4S$ such that we can integrate them out and get
\begin{align}
    \mathcal{Z}_{\text{MF}}= \int \mathcal{D}[\Phi^{*}, \Phi]  e^{-S_{\text{GMFT}}},
\end{align}
with the saddle point action
\begin{align}
    S_{\text {GMFT}} = \int_{0}^{\beta} \dd{\tau} &\left(  \sum_{\mathbf{r}_{\alpha}} \left(\frac{1}{2 J_{yy}}  \partial_{\tau}\Phi_{\mathbf{r}_\alpha}^{\tau *} \partial_{\tau} \Phi_{\mathbf{r}_\alpha}^{\tau} + i \lambda_{\mathbf{r}_{\alpha}}^{\tau} \left(\Phi_{\mathbf{r}_{\alpha}}^{\tau*} \Phi_{\mathbf{r}_{\alpha}}^{\tau} -1\right) \right) + \mathcal{H}_{\text {GMFT}}  \right),
\end{align}
and $\mathcal{H}_{\text{GMFT}}$ contains both the $J_{\pm}$ and $J_{\pm\pm}$ terms of Eq.~\eqref{eq: GMFT Hamiltonian}.

\section{\label{SI sec: Transformation of the parton operators} Transformation of the parton operators}

\subsection{\label{SI subsec: Space group} Space group}

The space group (SG) of the diamond lattice ($F d \overline{3} m$) is minimally generated by five operators: three translations $T_{i}$ ($i=1,2,3$), a rotoreflection $\overline{C}_6$  (i.e., $\overline{C}_6 = IC_3$ where $C_3$ is a threefold rotation around $\left[111\right]$ and $I$ is the inversion), and a non-symmorphic screw operation $S$. These space-group generators act on the position vector written in the SIDC as
\begin{subequations}
\begin{align}
T_{i}:& \mathbf{r}_{\alpha} \mapsto \left(r_{1}+\delta_{i, 1}, r_{2}+\delta_{i, 2}, r_{3}+\delta_{i, 3}\right)_{\alpha} \\
\overline{C}_{6}:& \mathbf{r}_{\alpha} \mapsto \left(-r_{3}, -r_{1}, -r_{2} \right)_{\pi_{A,B}(\alpha)}\\
S:& \mathbf{r}_{\alpha} \mapsto \left(-r_{1},-r_{2}, r_{1}+r_{2}+r_{3}+\delta_{\alpha,A}\right)_{\pi_{A, B}(\alpha)},
\end{align}
\end{subequations}
where $\pi_{A,B}(\alpha)$  are cyclic permutations of the $A$ and $B$ sublattices.

\subsection{Space group transformations}

For dipolar-octupolar doublets \cite{rau2019frustrated}, the pseudospins transform under the space group generators as
\begin{subequations} 
  \begin{align}
     T_i :  \qty\Big{ \mathrm{S}^{+}_{\mathbf{R}_{i}}, \mathrm{S}^{-}_{\mathbf{R}_{i}}, \mathrm{S}^{z}_{\mathbf{R}_{i}} } \mapsto &  \qty\Big{  \mathrm{S}^{+}_{T_i(\mathbf{R}_{i})}, \mathrm{S}^{-}_{T_i(\mathbf{R}_{i})}, \mathrm{S}^{z}_{T_i(\mathbf{R}_{i})} }\\
    \overline{C}_6 :  \qty\Big{  \mathrm{S}^{+}_{\mathbf{R}_{i}}, \mathrm{S}^{-}_{\mathbf{R}_{i}}, \mathrm{S}^{z}_{\mathbf{R}_{i}} } \mapsto &  \qty\Big{   \mathrm{S}^{+}_{\overline{C}_6(\mathbf{R}_{i})},  \mathrm{S}^{-}_{\overline{C}_6(\mathbf{R}_{i})}, \mathrm{S}^z_{\overline{C}_6(\mathbf{R}_{i})} } \\
    S :  \qty\Big{ \mathrm{S}^{+}_{\mathbf{R}_{i}}, \mathrm{S}^{-}_{\mathbf{R}_{i}}, \mathrm{S}^{z}_{\mathbf{R}_{i}} } \mapsto &  \qty\Big{  - \mathrm{S}^{+}_{S(\mathbf{R}_{i})} , -  \mathrm{S}^{-}_{S(\mathbf{R}_{i})} , \mathrm{S}^{z}_{S(\mathbf{R}_{i})} }.
  \end{align}
\end{subequations}
In terms of the GMFT parton construction, these transformations translate to
\begin{widetext}
\begin{subequations} 
  \begin{align}
    T_{i} &:  \qty\Big{ \frac{1}{2} \Phi^{\dag}_{\mathbf{r}_A} e^{i A_{\mathbf{r}_{A},\mathbf{r}_{A}+ \mathbf{b}_\mu}} \Phi_{\mathbf{r}_{A}+\mathbf{b}_\mu},
    \frac{1}{2} \Phi^{\dag}_{\mathbf{r}_A+\mathbf{b}_\mu} e^{-iA_{\mathbf{r}_{A},\mathbf{r}_{A}+ \mathbf{b}_\mu}}  \Phi_{\mathbf{r}_{A}},
    E_{\mathbf{r}_{A},\mathbf{r}_{A}+\mathbf{b}_\mu} }\nonumber \\ 
    &\mapsto  \qty\Big{ \frac{1}{2} \Phi^{\dag}_{T_{i} (\mathbf{r}_A)} e^{i A_{T_{i} (\mathbf{r}_{A}),T_{i} (\mathbf{r}_{A}+ \mathbf{b}_\mu)}}  \Phi_{T_{i} (\mathbf{r}_{A}+\mathbf{b}_\mu)}, \frac{1}{2}
    \Phi^{\dag}_{T_{i} (\mathbf{r}_A+\mathbf{b}_\mu)} e^{-iA_{T_{i} (\mathbf{r}_{A}),T_{i} (\mathbf{r}_{A}+ \mathbf{b}_\mu)}}  \Phi_{T_{i} (\mathbf{r}_{A})},
    E_{T_{i} (\mathbf{r}_{A}),T_{i} (\mathbf{r}_{A}+\mathbf{b}_\mu)} 
    } \\ 
    \overline{C}_{6} &:  \qty\Big{ \frac{1}{2} \Phi^{\dag}_{\mathbf{r}_A} e^{i A_{\mathbf{r}_{A},\mathbf{r}_{A}+ \mathbf{b}_\mu} } \Phi_{\mathbf{r}_{A}+\mathbf{b}_\mu},
    \frac{1}{2} \Phi^{\dag}_{\mathbf{r}_A+\mathbf{b}_\mu} e^{-iA_{\mathbf{r}_{A},\mathbf{r}_{A}+ \mathbf{b}_\mu}}  \Phi_{\mathbf{r}_{A}},
    E_{\mathbf{r}_{A},\mathbf{r}_{A}+\mathbf{b}_\mu} }\nonumber \\ 
    &\mapsto   \qty\Big{ \frac{1}{2}   \Phi^{\dag}_{\overline{C}_{6}(\mathbf{r}_A)} e^{i A_{\overline{C}_{6}(\mathbf{r}_{A}), \overline{C}_{6}(\mathbf{r}_{A}+\mathbf{b}_\mu)} } \Phi_{\overline{C}_{6}(\mathbf{r}_{A}+\mathbf{b}_\mu)} ,  \frac{1}{2} \Phi^{\dag}_{\overline{C}_{6}(\mathbf{r}_A+\mathbf{b}_\mu)} e^{-iA_{\overline{C}_{6}(\mathbf{r}_{A}),\overline{C}_{6}(\mathbf{r}_{A}+ \mathbf{b}_\mu)}} \Phi_{\overline{C}_{6}(\mathbf{r}_{A})} , E_{\overline{C}_{6}(\mathbf{r}_{A}),\overline{C}_{6}(\mathbf{r}_{A}+\mathbf{b}_\mu)}  } \\
    S &: \qty\Big{ \frac{1}{2} \Phi^{\dag}_{\mathbf{r}_A} e^{i A_{\mathbf{r}_{A},\mathbf{r}_{A}+ \mathbf{b}_\mu}}  \Phi_{\mathbf{r}_{A}+\mathbf{b}_\mu},
    \frac{1}{2} \Phi^{\dag}_{\mathbf{r}_A+\mathbf{b}_\mu} e^{-iA_{\mathbf{r}_{A},\mathbf{r}_{A}+ \mathbf{b}_\mu}}  \Phi_{\mathbf{r}_{A}},
    E_{\mathbf{r}_{A},\mathbf{r}_{A}+\mathbf{b}_\mu} }\nonumber \\ 
    &\mapsto   \qty\Big{ -\frac{1}{2}   \Phi^{\dag}_{S(\mathbf{r}_A)} e^{i A_{S(\mathbf{r}_{A}), S(\mathbf{r}_{A}+\mathbf{b}_\mu)} } \Phi_{S(\mathbf{r}_{A}+\mathbf{b}_\mu)} ,  -\frac{1}{2} \Phi^{\dag}_{S(\mathbf{r}_A+\mathbf{b}_\mu)} e^{-iA_{S(\mathbf{r}_{A}), S(\mathbf{r}_{A}+ \mathbf{b}_\mu)}} \Phi_{S(\mathbf{r}_{A})} , E_{S(\mathbf{r}_{A}),S(\mathbf{r}_{A}+\mathbf{b}_\mu)}  } .
  \end{align}
\end{subequations}
\end{widetext}

\subsection{Local \texorpdfstring{$U(1)$}{U(1)} transformations}

Before performing a saddle point approximation and fixing the gauge connection to a constant background, the Hamiltonian has the following $U(1)$ gauge structure 
\begin{subequations}\label{eq: gauge transformations}
    \begin{align} 
        \Phi_{\mathbf{r}_{\alpha}} &\rightarrow \Phi_{\mathbf{r}_{\alpha}} e^{i \theta_{\mathbf{r}_{\alpha}}} \\
         A_{\mathbf{r}_{\alpha} \mathbf{r}_{\beta}^{\prime}}&\rightarrow          A_{\mathbf{r}_{\alpha} \mathbf{r}_{\beta}^{\prime}} - \theta_{\mathbf{r}^{\prime}_{\beta}} + \theta_{\mathbf{r}_{\alpha}} 
    \end{align}
\end{subequations}
as a direct consequence of the physical constraint~\eqref{eq: physical constraint charge GMFT}. However, at the MF level, $\mathcal{H}_{\text{GMFT}}$ does not have a $U(1)$ gauge structure anymore. Let us then study how the GMFT Hamiltonian transforms under an arbitrary local $U(1)$ transformation of the form 
\begin{align}
    G: \Phi_{\mathbf{r}_{\alpha}} \mapsto \Phi_{\mathbf{r}_{\alpha}} e^{i\theta_{\mathbf{r}_{\alpha}}}.
\end{align}
First, for the $J_{\pm}$ term describing intra-sublattice spinon hopping transforms as
\begin{align}
    G&: \Phi_{\mathbf{r}_{\alpha}+\eta_{\alpha}\mathbf{b}_{\mu}}^{\dagger} \Phi_{\mathbf{r}_{\alpha}+\eta_{\alpha}\mathbf{b}_{\nu}} e^{i\eta_{\alpha}\left(\overline{A}_{\mathbf{r}_{\alpha},\mathbf{r}_{\alpha}+\eta_{\alpha}\mathbf{b}_{\nu}} - \overline{A}_{\mathbf{r}_{\alpha},\mathbf{r}_{\alpha}+\eta_{\alpha}\mathbf{b}_{\mu}} \right)} \nonumber  \\
    &\mapsto  \Phi_{\mathbf{r}_{\alpha}+\eta_{\alpha}\mathbf{b}_{\mu}}^{\dagger} \Phi_{\mathbf{r}_{\alpha}+\eta_{\alpha}\mathbf{b}_{\nu}} e^{i\eta_{\alpha}\left[\left(\overline{A}_{\mathbf{r}_{\alpha},\mathbf{r}_{\alpha}+\eta_{\alpha}\mathbf{b}_{\nu}} + \eta_{\alpha}\left(\theta_{\mathbf{r}_{\alpha}+\eta_{\alpha}\mathbf{b}_{\nu}} - \theta_{\mathbf{r}_{\alpha}}\right)  \right) - \left( \overline{A}_{\mathbf{r}_{\alpha},\mathbf{r}_{\alpha}+\eta_{\alpha}\mathbf{b}_{\mu}} + \eta_{\alpha}\left(\theta_{\mathbf{r}_{\alpha}+\eta_{\alpha}\mathbf{b}_{\mu}} - \theta_{\mathbf{r}_{\alpha}}\right) \right) \right]}. 
\end{align}
The local $U(1)$ transformation can be absorbed in the gauge field background by mapping it to
\begin{align} \label{eq: gauge trf gauge field}
    G: \overline{A}_{\mathbf{r}_{\alpha},\mathbf{r}_{\alpha}+\eta_{\alpha}\mathbf{b}_{\mu}} \mapsto \overline{A}_{\mathbf{r}_{\alpha},\mathbf{r}_{\alpha}+\eta_{\alpha}\mathbf{b}_{\mu}} + \eta_{\alpha} \left(\theta_{\mathbf{r}_{\alpha}+ \eta_{\alpha} \mathbf{b}_{\mu}} - \theta_{\mathbf{r}_{\alpha}}\right)  = G_{\theta}\left( \overline{A}_{\mathbf{r}_{\alpha},\mathbf{r}_{\alpha}+\eta_{\alpha}\mathbf{b}_{\mu}} \right).
\end{align}
Next, for the inter-site pairing coupling associated with the $J_{\pm\pm}$ term, we get the mapping
\begin{align}
    G&: e^{i \eta_\alpha\left(\overline{A}_{\mathbf{r}_\alpha, \mathbf{r}_\alpha+\eta_\alpha \mathbf{b}_\nu}+\overline{A}_{\mathbf{r}_\alpha, \mathbf{r}_\alpha+\eta_\alpha \mathbf{b}_\mu}\right)}\Phi_{\mathbf{r}_\alpha}^{\dagger} \Phi_{\mathbf{r}_\alpha}^{\dagger} \chi_{\mathbf{r}_\alpha+\eta_\alpha \mathbf{b}_\mu, \mathbf{r}_\alpha+\eta_\alpha \mathbf{b}_\nu} \nonumber \\
    &\mapsto  e^{i\eta_{\alpha}\left[\left(\overline{A}_{\mathbf{r}_{\alpha},\mathbf{r}_{\alpha}+\eta_{\alpha}\mathbf{b}_{\nu}} + \eta_{\alpha}\left(\theta_{\mathbf{r}_{\alpha}+\eta_{\alpha}\mathbf{b}_{\nu}} - \theta_{\mathbf{r}_{\alpha}}\right)  \right) + \left( \overline{A}_{\mathbf{r}_{\alpha},\mathbf{r}_{\alpha}+\eta_{\alpha}\mathbf{b}_{\mu}} + \eta_{\alpha}\left(\theta_{\mathbf{r}_{\alpha}+\eta_{\alpha}\mathbf{b}_{\mu}} - \theta_{\mathbf{r}_{\alpha}}\right) \right) \right]}  \nonumber \\
    &\hspace{2cm} \times \Phi_{\mathbf{r}_\alpha}^{\dagger} \Phi_{\mathbf{r}_\alpha}^{\dagger} \chi_{\mathbf{r}_\alpha+\eta_\alpha \mathbf{b}_\mu, \mathbf{r}_\alpha+\eta_\alpha \mathbf{b}_\nu} e^{-i \left( \theta_{\mathbf{r}_\alpha+\eta_\alpha \mathbf{b}_\mu} + \theta_{\mathbf{r}_\alpha+\eta_\alpha \mathbf{b}_\nu} \right)},
\end{align}
such that the inter-site pairing field transforms as 
\begin{align} \label{eq: gauge trf inter-site pairing}
    G:  \chi_{\mathbf{r}_\alpha+\eta_\alpha \mathbf{b}_\mu, \mathbf{r}_\alpha+\eta_\alpha \mathbf{b}_\nu} \mapsto \chi_{\mathbf{r}_\alpha+\eta_\alpha \mathbf{b}_\mu, \mathbf{r}_\alpha+\eta_\alpha \mathbf{b}_\nu} e^{-i \left( \theta_{\mathbf{r}_\alpha+\eta_\alpha \mathbf{b}_\mu} + \theta_{\mathbf{r}_\alpha+\eta_\alpha \mathbf{b}_\nu} \right)} = G_{\theta}\left( \chi_{\mathbf{r}_\alpha+\eta_\alpha \mathbf{b}_\mu, \mathbf{r}_\alpha+\eta_\alpha \mathbf{b}_\nu}\right). 
\end{align}
For the on-site pairing coupling associated with the $J_{\pm\pm}$ term,
\begin{align}
    G &: e^{i \eta_\alpha\left(\overline{A}_{\mathbf{r}_\alpha, \mathbf{r}_\alpha+\eta_\alpha \mathbf{b}_\nu}+\overline{A}_{\mathbf{r}_\alpha, \mathbf{r}_\alpha+\eta_\alpha \mathbf{b}_\mu}\right)} \bar{\chi}^{0}_{\mathbf{r}_\alpha, \mathbf{r}_\alpha} \Phi_{\mathbf{r}_\alpha+\eta_\alpha \mathbf{b}_\mu} \Phi_{\mathbf{r}_\alpha+\eta_\alpha \mathbf{b}_\nu} \nonumber \\
    &\mapsto  e^{i\eta_{\alpha}\left[\left(\overline{A}_{\mathbf{r}_{\alpha},\mathbf{r}_{\alpha}+\eta_{\alpha}\mathbf{b}_{\nu}} + \eta_{\alpha}\left(\theta_{\mathbf{r}_{\alpha}+\eta_{\alpha}\mathbf{b}_{\nu}} - \theta_{\mathbf{r}_{\alpha}}\right)  \right) + \left( \overline{A}_{\mathbf{r}_{\alpha},\mathbf{r}_{\alpha}+\eta_{\alpha}\mathbf{b}_{\mu}} + \eta_{\alpha}\left(\theta_{\mathbf{r}_{\alpha}+\eta_{\alpha}\mathbf{b}_{\mu}} - \theta_{\mathbf{r}_{\alpha}}\right) \right) \right]} \bar{\chi}^{0}_{\mathbf{r}_\alpha, \mathbf{r}_\alpha} e^{i2\theta_{\mathbf{r}_{\alpha}}} \Phi_{\mathbf{r}_\alpha+\eta_\alpha \mathbf{b}_\mu} \Phi_{\mathbf{r}_\alpha+\eta_\alpha \mathbf{b}_\nu} , 
\end{align}
such that the on-site pairing field transforms as
\begin{align}\label{eq: gauge trf on-site pairing}
    G:  \overline{\chi}_{\mathbf{r}_\alpha, \mathbf{r}_\alpha}^{0} \mapsto \overline{\chi}_{\mathbf{r}_\alpha, \mathbf{r}_\alpha}^{0} e^{i 2 \theta_{\mathbf{r}_\alpha} } = G_{\theta}\left( \overline{\chi}_{\mathbf{r}_\alpha, \mathbf{r}_\alpha}^{0} \right). 
\end{align}
Finally, the inter-site hopping term transforms under a local $U(1)$ transformation as
\begin{align}
    G &: e^{i \eta_\alpha\left(\overline{A}_{\mathbf{r}_\alpha, \mathbf{r}_\alpha+\eta_\alpha \mathbf{b}_\nu}+\overline{A}_{\mathbf{r}_\alpha, \mathbf{r}_\alpha+\eta_\alpha \mathbf{b}_\mu}\right)} \xi_{\mathbf{r}_\alpha, \mathbf{r}_\alpha+\eta_{\alpha}\mathbf{b}_{\nu}} \Phi_{\mathbf{r}_\alpha}^\dagger  \Phi_{\mathbf{r}_\alpha+\eta_\alpha \mathbf{b}_\mu} \nonumber \\
    &\mapsto  e^{i\eta_{\alpha}\left[\left(\overline{A}_{\mathbf{r}_{\alpha},\mathbf{r}_{\alpha}+\eta_{\alpha}\mathbf{b}_{\nu}} + \eta_{\alpha}\left(\theta_{\mathbf{r}_{\alpha}+\eta_{\alpha}\mathbf{b}_{\nu}} - \theta_{\mathbf{r}_{\alpha}}\right)  \right) + \left( \overline{A}_{\mathbf{r}_{\alpha},\mathbf{r}_{\alpha}+\eta_{\alpha}\mathbf{b}_{\mu}} + \eta_{\alpha}\left(\theta_{\mathbf{r}_{\alpha}+\eta_{\alpha}\mathbf{b}_{\mu}} - \theta_{\mathbf{r}_{\alpha}}\right) \right) \right]} \xi_{\mathbf{r}_\alpha, \mathbf{r}_\alpha+\eta_{\alpha}\mathbf{b}_{\nu}} e^{i (\theta_{\mathbf{r}_\alpha} - \theta_{\mathbf{r}_\alpha + \eta_{\alpha} \mathbf{b}_{\nu}})} \Phi_{\mathbf{r}_\alpha}^\dagger  \Phi_{\mathbf{r}_\alpha+\eta_\alpha \mathbf{b}_\mu}, 
\end{align}
such that we have
\begin{align} \label{eq: gauge trf inter-sublattice hopping}
    G: \xi_{\mathbf{r}_{\alpha},\mathbf{r}_{\alpha}+\eta_{\alpha}\mathbf{b}_{\nu}} \mapsto \xi_{\mathbf{r}_{\alpha},\mathbf{r}_{\alpha}+\eta_{\alpha}\mathbf{b}_{\nu}} e^{i(\theta_{\mathbf{r}_{\alpha}} - \theta_{\mathbf{r}_{\alpha}+\eta_{\alpha}\mathbf{b}_{\nu}})} = G_{\theta}\left( \xi_{\mathbf{r}_{\alpha},\mathbf{r}_{\alpha}+\eta_{\alpha}\mathbf{b}_{\nu}} \right).
\end{align}
To summarize, the GMFT Hamiltonian transforms under a local $U(1)$ gauge transformation as
\begin{align} \label{eq: transformation GMFT Hamiltonian local U(1) trsf}
    G: \mathcal{H}_{\text{GMFT}}\left(\left\{ \overline{A},\chi,\chi^{0},\xi \right\}\right) \mapsto \mathcal{H}_{\text{GMFT}}\left(\left\{ G_{\theta}(\overline{A}), G_{\theta}(\chi), G_{\theta}(\chi^{0}), G_{\theta}(\xi) \right\}\right),
\end{align}
where the gauge field background and MF parameters transform as in Eqs.~\eqref{eq: gauge trf gauge field},~\eqref{eq: gauge trf inter-site pairing},~\eqref{eq: gauge trf on-site pairing}, and~\eqref{eq: gauge trf inter-sublattice hopping}. This defines an equivalence relation between different configurations $\left\{ \overline{A},\chi,\chi^{0},\xi \right\}$. 

\section{\label{SI sec: Classification of symmetric spin liquids} Classification of \texorpdfstring{$U(1)$}{U(1)} symmetric spin liquids}

\subsection{Generalities}

\subsubsection{Projective construction}

We now discuss the general ideas behind the classification of symmetry fractionalization classes within GMFT. We briefly summarize the argument outlined in Ref.~\cite{desrochers2023symmetry} to which we refer the interested reader for a more thorough exposition.

It should first be noted that the general gauge transformation of Eq. \eqref{eq: gauge transformations} is generated by 
\begin{align}
    U(\{\theta\})&=\prod_{\mathbf{r}_\alpha} \exp\left(i\theta_{\mathbf{r}_\alpha} \left(Q_{\mathbf{r}_\alpha}- \sum_{\mu} E_{\mathbf{r}_{\alpha},\mathbf{r}_{\alpha}+\eta_{\alpha}\mathbf{b}_\mu}\right)\right).
\end{align}
Under such a transformation, the GMFT Hamiltonian transforms as in Eq.~\eqref{eq: transformation GMFT Hamiltonian local U(1) trsf}. Therefore, starting from the ground state $\ket{\Psi_{\text{GS}} \left( \left\{ \overline{A},\chi,\chi^{0},\xi \right\} \right) }$ of a GMFT Hamiltonian $\mathcal{H}_{\text{GMFT}}\left( \left\{ \overline{A},\chi,\chi^{0},\xi \right\} \right)$, one can think of using a projector-like transformation $\mathcal{P}_{\text{Gauss}}$ to recover a physical spin wavefunction. Such a projection operator should remove charge configurations that do not respect $|Q_{\mathbf{r}_{\alpha}}|\le 4S$ and acts on the $\mathscr{H}_{\text{spin}}$ part of the state to enforce Eq.~\eqref{eq: physical constraint charge GMFT}. However, since $\comm{U}{\mathcal{P}_{\text{Gauss}}}=0$ because $U$ acts trivially on any state that respects the lattice Gauss's law, we get the crucial observation that all GMFT eigenstates that only differ by a gauge transformation yield the same physical spin wavefunction
\begin{align*}
   &\mathcal{P}_{\text{Gauss}} U \ket{\Psi_{\text{GS}} \left( \left\{ \overline{A},\chi,\chi^0,\xi \right\} \right) } = \mathcal{P}_{\text{Gauss}} \ket{\Psi_{\text{GS}} \left( \left\{ G_{\theta}(\overline{A}), G_{\theta}(\chi), G_{\theta}(\chi^0), G_{\theta}(\xi) \right\} \right) } \\\
    =& U \mathcal{P}_{\text{Gauss}}  \ket{\Psi_{\text{GS}} \left(\left\{ \overline{A},\chi,\chi^0,\xi \right\} \right)} \\
    =& \mathcal{P}_{\text{Gauss}}  \ket{\Psi_{\text{GS}} \left(\left\{ \overline{A},\chi,\chi^0,\xi \right\} \right)}.
\end{align*}
This argument is independent of the way one chooses to implement the projection back to the physical spin space $\mathcal{P}_{\text{Gauss}}$. We can thus conclude that although the gauge structure is not explicitly present, the MF theory still has redundancies in its description. 

This redundancy has important consequences when considering how to implement symmetries in GMFT. It implies that a GMFT eigenstate for a given Ansatz $\{\overline{A},\chi,\chi^0,\xi\}$ corresponds to a physical spin wavefunction symmetric under a given transformation $\mathcal{O}$ if it is symmetric under $\mathcal{O}$ \emph{up to a gauge transformation}, in complete analogy to the projective symmetry group (PSG) construction for Abrikosov fermions and Schwinger bosons~\cite{wen2002quantum, wang2006spin, messio2013time, bieri2016projective}. Equivalently stated, a GMFT wavefunction is symmetric under $\mathcal{O}$ if there exist a gauge transformation $G_{\mathcal{O}}$ such that 
\begin{align}
    G_{\mathcal{O}}\circ\mathcal{O}: \mathcal{H}_{\text{GMFT}}\left( \left\{ \overline{A},\chi,\chi^0,\xi \right\} \right) \mapsto  \mathcal{H}_{\text{GMFT}}\left( \left\{ \overline{A},\chi,\chi^0,\xi \right\} \right).
\end{align}
As a result, all Ansätze corresponding to physical states invariant under a given set of symmetries $\{\mathcal{O}_1,\mathcal{O}_2,...\}$ can be classified by identifying the associated gauge-enriched operations $\{\tilde{\mathcal{O}}_1, \tilde{\mathcal{O}}_2,...\} = \{G_{\mathcal{O}_1}\circ\mathcal{O}_1,G_{\mathcal{O}_2}\circ\mathcal{O}_2,...\}$ that leave $\mathcal{H}_{\text{GMFT}}$ invariant.

\begin{figure}
\includegraphics[width=1.00\linewidth]{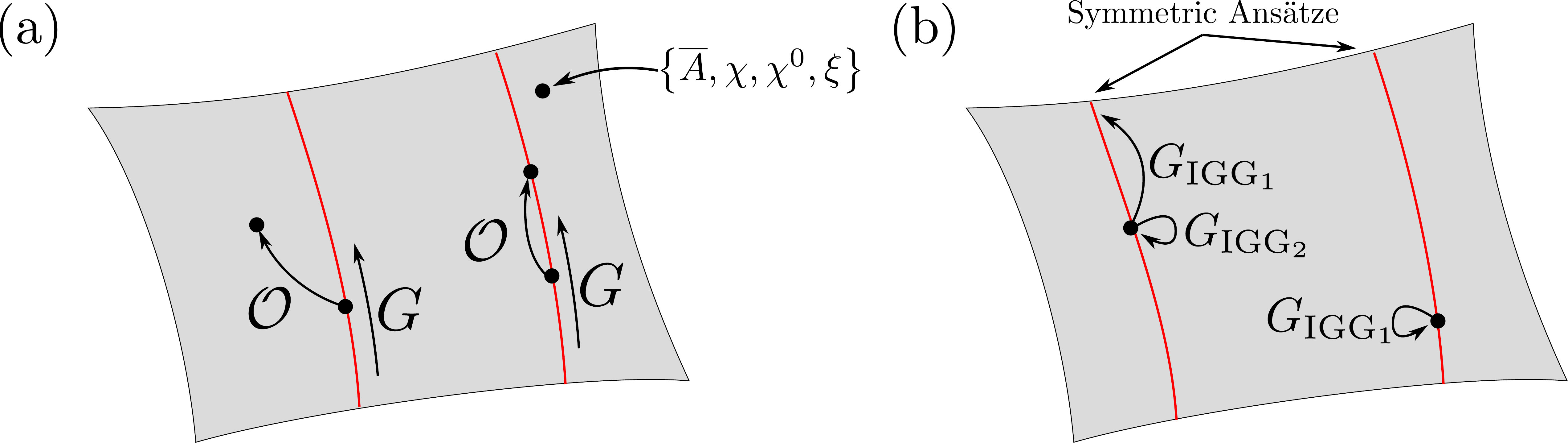}
\caption{Graphical representation of the projective construction. (a) The space represents all possible configurations $\{\overline{A},\chi,\chi^{0},\xi\}$, and the red curves correspond to configurations related by a gauge transformation $G$. A GMFT eigenstate corresponding to a certain equivalence class of configuration will yield a symmetric spin wavefunction under a transformation $\mathcal{O}$ if this operation maps a representative configuration to a gauge equivalent one, as in the right part of the figure. On the other hand, the GMFT eigenstate corresponds to a spin wavefunction that is not symmetric under $\mathcal{O}$ if it maps the configuration to another one that is not related by a local $U(1)$ transformation as in the left part of the figure. (b) All Ansätze symmetric under given transformations are invariant under a subgroup of pure gauge transformation known as the invariant gauge group (IGG). The family of Ansätze to the right of the figure is invariant under $\text{IGG}_{1}$ whereas the one to the left is not but rather has $\text{IGG}_{2}\subset \text{IGG}_{1}$ as invariant gauge group. \label{fig: projective symmetry}}
\end{figure}

The subgroup of pure gauge transformations $G_{\text{IGG}}\circ\mathds{1}$ that leave the GMFT Hamiltonian invariant is known as the invariant gauge group (IGG). The IGG corresponds to the emergent low-energy gauge structure of the model. The ideas behind the classification of symmetry fractionalization classes with GMFT are summarized in Fig.~\ref{fig: projective symmetry}.

\subsubsection{Consistency conditions}

To classify symmetry classes, one starts from all algebraic constraints of the form
\begin{equation} \label{eq: algebraic SG relations}
\mathcal{O}_{1} \circ \mathcal{O}_{2} \circ \cdots=1
\end{equation}
which translates directly to the gauge-enriched relations
\begin{equation} \label{eq: algebraic SG relations gauge-enriched}
\widetilde{\mathcal{O}}_{1} \circ \widetilde{\mathcal{O}}_{2} \circ \cdots=\left(G_{\mathcal{O}_{1}} \circ \mathcal{O}_{1}\right) \circ\left(G_{\mathcal{O}_{2}} \circ \mathcal{O}_{2}\right) \circ \cdots  = e^{i\psi} \in \text{IGG},
\end{equation}
with $\psi\in\left[0, 2\pi \right)$ if IGG$=U(1)$ and $\psi\in\left\{0, \pi \right\}$ if IGG$=\mathbb{Z}_{2}$. We can use the following conjugation relation
\begin{widetext}
\begin{align}
\mathcal{O}_{i} \circ G_{\mathcal{O}_{j}} \circ \mathcal{O}_{i}^{-1}: \Phi_{\mathbf{r}_{\alpha}} \mapsto e^{i \phi_{\mathcal{O}_{j}}\left[\mathcal{O}_{i}^{-1}\left(\mathbf{r}_{\alpha}\right)\right]} \Phi_{\mathbf{r}_{\alpha}},
\end{align}
\end{widetext}
to map all these gauge-enriched constraints to phase relations of the form
\begin{equation}
\begin{aligned}\label{eq: phase equations for PSG}
&\phi_{\mathcal{O}_{1}}\left(\mathbf{r}_{\alpha}\right) + \phi_{\mathcal{O}_{2}}\left[\mathcal{O}_{1}^{-1}\left(\mathbf{r}_{\alpha}\right)\right] + \phi_{\mathcal{O}_{3}}\left[\mathcal{O}_{2}^{-1}\circ\mathcal{O}_{1}^{-1}\left(\mathbf{r}_{\alpha}\right)\right]+\cdots= \psi\hspace{3mm} \text{mod }2\pi.
\end{aligned}
\end{equation}
The GMFT classes for a given IGG are then obtained by listing the gauge inequivalent solutions of all phase equations of the form~\eqref{eq: phase equations for PSG}. It must be impossible to relate two distinct GMFT classes by a general gauge transformation $G$ that maps the phase factor to 
\begin{align}
    \phi_{\mathcal{O}}(\mathbf{r}_{\alpha}) &\to \phi_{\mathcal{O}}(\mathbf{r}_{\alpha}) + \phi_{G}(\mathbf{r}_{\alpha}) -  \phi_{G}(\mathcal{O}^{-1}(\mathbf{r}_{\alpha})).
\end{align}

\subsection{\label{SI subsec: PSG solution - Algebraic constraints} Algebraic constraints}

For the parent diamond lattice, the algebraic constraints are
\begin{subequations} \label{eq: SG generators constraints}
  \begin{align}
    T_{i} T_{i+1} T_{i}^{-1} T_{i+1}^{-1} &=1,  i=1,2,3 \\
    \bar{C}_{6}^{6} &=1, \\
    S^{2} T_{3}^{-1} &=1,  \\
    \bar{C}_{6} T_{i} \bar{C}_{6}^{-1} T_{i+1} &=1, i=1,2,3 \\
    S T_{i} S^{-1} T_{3}^{-1} T_{i} &=1,  i=1,2, \\
    S T_{3} S^{-1} T_{3}^{-1} &=1  \\
    \left(\bar{C}_{6} S\right)^{4} &=1 \\
    \left(\bar{C}_{6}^{3} S\right)^{2} &=1.  
  \end{align}
\end{subequations}
The corresponding gauge-enriched operations are
\begin{widetext}
\begin{subequations}
\begin{align}
\left(G_{T_{i}} T_{i}\right)\left(G_{T_{i+1}} T_{i+1}\right)\left(G_{T_{i}} T_{i}\right)^{-1}\left(G_{T_{i+1}} T_{i+1}\right)^{-1}&\in IGG, \\
\left(G_{\bar{C}_{6}} \bar{C}_{6}\right)^{6} &\in IGG, \\
\left(G_{S} S\right)^{2}\left(G_{T_{3}} T_{3}\right)^{-1} &\in IGG,\\
\left(G_{\bar{C}_{6}} \bar{C}_{6}\right)\left(G_{T_{i}} T_{i}\right)\left(G_{\bar{C}_{6}} \bar{C}_{6}\right)^{-1}\left(G_{T_{i+1}} T_{i+1}\right) &\in IGG, \\
\left(G_{S} S\right)\left(G_{T_{i}} T_{i}\right)\left(G_{S} S\right)^{-1}\left(G_{T_{3}} T_{3}\right)^{-1}\left(G_{T_{i}} T_{i}\right) &\in IGG, \\
\left(G_{S} S\right)\left(G_{T_{3}} T_{3}\right)\left(G_{S} S\right)^{-1}\left(G_{T_{3}} T_{3}\right)^{-1} &\in IGG, \\
\left[\left(G_{\bar{C}_{6}} \bar{C}_{6}\right)\left(G_{S} S\right)\right]^{4} &\in IGG,\\
\left[\left(G_{\bar{C}_{6}} \bar{C}_{6}\right)^{3}\left(G_{S} S\right)\right]^{2} &\in IGG .
\end{align}
\end{subequations}
These constraints are explicitly 
\begin{subequations} 
\begin{empheq}[]{align}
\phi_{T_{i}}\left(\mathbf{r}_{\alpha}\right)+\phi_{T_{i+1}}\left[T_{i}^{-1}\left(\mathbf{r}_{\alpha}\right)\right]-\phi_{T_{i}}\left[T_{i+1}^{-1}\left(\mathbf{r}_{\alpha}\right)\right]-\phi_{T_{i+1}}\left(\mathbf{r}_{\alpha}\right) &=\psi_{T_i},\label{eq: psg classification T_i}\\
\phi_{\bar{C}_{6}}\left(\mathbf{r}_{\alpha}\right)+\phi_{\bar{C}_{6}}\left[\bar{C}_{6}^{-1}\left(\mathbf{r}_{\alpha}\right)\right]+\phi_{\bar{C}_{6}}\left[\bar{C}_{6}^{-2}\left(\mathbf{r}_{\alpha}\right)\right]+\phi_{\bar{C}_{6}}\left[\bar{C}_{6}^{-3}\left(\mathbf{r}_{\alpha}\right)\right]+\phi_{\bar{C}_{6}}\left[\bar{C}_{6}^{-4}\left(\mathbf{r}_{\alpha}\right)\right]+\phi_{\bar{C}_{6}}\left[\bar{C}_{6}^{-5}\left(\mathbf{r}_{\alpha}\right)\right] &=\psi_{\bar{C}_{6}} \label{eq: psg classification C} \\
\phi_{S}\left(\mathbf{r}_{\alpha}\right) + \phi_{S}\left[S^{-1}\left(\mathbf{r}_{\alpha}\right)\right]-\phi_{T_{3}}\left(\mathbf{r}_{\alpha}\right) &=\psi_{S} \label{eq: psg classification S}  \\
\phi_{\bar{C}_{6}}\left(\mathbf{r}_{\alpha}\right)+\phi_{T_{i}}\left[\bar{C}_{6}^{-1}\left(\mathbf{r}_{\alpha}\right)\right]-\phi_{\bar{C}_{6}}\left[T_{i+1}\left(\mathbf{r}_{\alpha}\right)\right]+\phi_{T_{i+1}}\left[T_{i+1}\left(\mathbf{r}_{\alpha}\right)\right] &=\psi_{\bar{C}_{6} T_{i}} \label{eq: psg classification CT_i}\\
\phi_{S}\left(\mathbf{r}_{\alpha}\right) + \phi_{T_{i}}\left[S^{-1}\left(\mathbf{r}_{\alpha}\right)\right]-\phi_{S}\left[T_{3}^{-1} T_{i}\left(\mathbf{r}_{\alpha}\right)\right]-\phi_{T_{3}}\left[T_{i}\left(\mathbf{r}_{\alpha}\right)\right]+\phi_{T_{i}}\left[T_{i}\left(\mathbf{r}_{\alpha}\right)\right]&=\psi_{S T_{i}} \label{eq: psg classification S T_i}  \\
\phi_{S}\left(\mathbf{r}_{\alpha}\right) + \phi_{T_{3}}\left[S^{-1}\left(\mathbf{r}_{\alpha}\right)\right]-\phi_{S}\left[T_{3}^{-1}\left(\mathbf{r}_{\alpha}\right)\right]-\phi_{T_{3}}\left(\mathbf{r}_{\alpha}\right) &=\psi_{S T_{3}} \label{eq: psg classification S T_3} \\
\phi_{\overline{C}_{6}}\left(\mathbf{r}_{\alpha}\right)+\phi_{S}\left[\bar{C}_{6}^{-1}\left(\mathbf{r}_{\alpha}\right)\right] + \phi_{\bar{C}_{6}}\left[\left(\bar{C}_{6} S\right)^{-1}\left(\mathbf{r}_{\alpha}\right)\right] + \phi_{S}\left[\left(\bar{C}_{6} S \bar{C}_{6}\right)^{-1}\left(\mathbf{r}_{\alpha}\right)\right]+\phi_{\bar{C}_{6}}\left[\left(\bar{C}_{6} S \bar{C}_{6} S\right)^{-1}\left(\mathbf{r}_{\alpha}\right)\right] & \nonumber\\
+\phi_{S}\left[\left(\bar{C}_{6} S \bar{C}_{6} S \bar{C}_{6}\right)^{-1}\left(\mathbf{r}_{\alpha}\right)\right] +\phi_{\bar{C}_{6}}\left[\left(\bar{C}_{6} S \bar{C}_{6} S \bar{C}_{6} S\right)^{-1}\left(\mathbf{r}_{\alpha}\right)\right] + \phi_{S}\left[\left(\bar{C}_{6} S \bar{C}_{6} S \bar{C}_{6} S \bar{C}_{6}\right)^{-1}\left(\mathbf{r}_{\alpha}\right)\right] &=\psi_{\bar{C}_{6} S} \label{eq: psg classification CS} \\
\phi_{\bar{C}_{6}}\left(\mathbf{r}_{\alpha}\right)+\phi_{\bar{C}_{6}}\left[\bar{C}_{6}^{-1}\left(\mathbf{r}_{\alpha}\right)\right]+\phi_{\bar{C}_{6}}\left[\bar{C}_{6}^{-2}\left(\mathbf{r}_{\alpha}\right)\right]+\phi_{S}\left[\bar{C}_{6}^{-3}\left(\mathbf{r}_{\alpha}\right)\right] + \phi_{\bar{C}_{6}}\left[\left(\bar{C}_{6}^{3} S\right)^{-1}\left(\mathbf{r}_{\alpha}\right)\right] \hspace{2.3cm} & \nonumber\\
+ \phi_{\bar{C}_{6}}\left[\left(\bar{C}_{6}^{3} S \bar{C}_{6}\right)^{-1}\left(\mathbf{r}_{\alpha}\right)\right] + \phi_{\bar{C}_{6}}\left[\left(\bar{C}_{6}^{3} S \bar{C}_{6}^{2}\right)^{-1}\left(\mathbf{r}_{\alpha}\right)\right] + \phi_{S}\left[S\left(\mathbf{r}_{\alpha}\right)\right] &=\psi_{S \bar{C}_{6}}  \label{eq: psg classification SC}
\end{empheq}
\end{subequations}
\end{widetext}
where all $\psi\in\left[0,2\pi\right)$ if IGG=$U(1)$ or $\psi\in\left\{0,\pi\right\}$ if IGG=$\mathbb{Z}_2$, $i=1,2,3$ for Eqs.~\eqref{eq: psg classification T_i} and~\eqref{eq: psg classification CT_i} and $i=1,2$ for Eq.~\eqref{eq: psg classification S T_i}. All phase equations are defined modulo $2\pi$. We will not indicate that subtlety explicitly for the sake of simplicity. We emphasize that these constraints are distinct from the effective spin-1/2 case discussed in Ref.~\cite{desrochers2023symmetry} as a consequence of the different ways the pseudospins transform.

\subsection{\label{SI subsec: PSG solution - Solution of the PSG constraints} Solution of the constraints}

The IGG of the GMFT Hamiltonian of Eq. (3) in the main text is $\mathbb{Z}_2$ in the presence of inter-site and on-site pairing field. Here we will arbitrarily set these pairing fields to zero to classify $U(1)$ QSLs. 

\subsubsection{\label{SI subsubsec: PSG solution - Solution of the PSG constraints: inter-unit cell part} Inter-unit cell part}

Let us first consider the constraints coming from the commutativity of the translation operators given in Eq.~\eqref{eq: psg classification T_i}. Using our gauge freedom, we can set $\phi_{T_1}(r_1,r_2,r_3)_{\alpha}=\phi_{T_2}(0,r_2,r_3)_{\alpha}=\phi_{T1}(0,0,r_3)_{\alpha}=0$, which then leads to 
\begin{subequations} \label{eq U(1) classification: T1, T2, T3 first equation}
  \begin{align}
    \phi_{T_1}(\mathbf{r}_{\alpha}) &= 0 \\
    \phi_{T_2}(\mathbf{r}_{\alpha}) &= -\psi_{T_1} r_1\\
    \phi_{T_3}(\mathbf{r}_{\alpha}) &= \psi_{T_3} r_1 - \psi_{T_2} r_2.
  \end{align}
\end{subequations}
Then using Eq.~\eqref{eq: psg classification CT_i}, we get the equations
\begin{subequations}
\begin{align}
     \psi_{\overline{C}_6 T_1} =& \phi_{\overline{C}_6}(r_1,r_2,r_3)_{\alpha} -\phi_{\overline{C}_6}(r_1,r_2+1,r_3)_{\alpha} -r_1 \psi_{T_1} \\
    \psi_{\overline{C}_6 T_2} =&\phi_{\overline{C}_6}(r_1,r_2,r_3)_{\alpha} -\phi_{\overline{C}_6}(r_1,r_2,r_3+1)_{\alpha}   + \psi_{T_1} r_2 - \psi_{T_2} r_2 + \psi_{T_3} r_1 \\
    \psi_{\overline{C}_6 T_3}=&\phi_{\overline{C}_6}(r_1,r_2,r_3)_{\alpha} -\phi_{\overline{C}_6}(r_1+1,r_2,r_3)_{\alpha}  + \psi_{T_2} r_3 - \psi_{T_3} r_2.
\end{align}
\end{subequations}
that yield $\psi_{T_1}=\psi_{T_2}=\psi_{T_3}$ and
\begin{align} \label{eq U(1) classification: first equation for phi C6}
    \phi_{\overline{C}_6}(\mathbf{r}_{\alpha}) =& \phi_{\overline{C}_6}(\mathbf{0}_{\alpha}) - r_2 \psi_{\overline{C}_6 T_1} -r_3 \psi_{\overline{C}_6 T_2} - r_1 \psi_{\overline{C}_6 T_{3}} - \psi_{T_1}(r_1 r_2 - r_1 r_3).
\end{align}
Replacing the translation phase factors in the constraints~\eqref{eq: psg classification S T_i} and~\eqref{eq: psg classification S T_3} leads to
\begin{subequations}
\begin{align}
    \psi_{S T_1} =&\phi_{S}(r_1,r_2,r_3)_{\alpha} -\phi_{S}(r_1+1,r_2,r_3-1)_{\alpha} + (-1-r_1+r_2)\psi_{T_1}  \\
    \psi_{S T_2} =&\phi_{S}(r_1,r_2,r_3)_{\alpha} -\phi_{S}(r_1,r_2+1,r_3-1)_{\alpha}  + (1-r_1+r_2) \psi_{T_1} \\
    \psi_{S T_3} =&\phi_{S}(r_1,r_2,r_3)_{\alpha} -\phi_{S}(r_1,r_2,r_3-1)_{\alpha} + (r_2 - r_1) \psi_{T_1},
\end{align}
\end{subequations}
which impose $\psi_{T_1}=n_1 \pi$ with $n_1 \in\left\{0, 1 \right\}$ and 
\begin{align}\label{eq U(1) classification: first equation for phi S}
    \phi_{S}(\mathbf{r}_{\alpha}) =& \phi_{S}(\mathbf{0}_{\alpha}) -r_1 \psi_{ST_1} - r_2 \psi_{S T_2} + \frac{1}{2} n_1 \pi \left( -r_1 + r_2 + 2 r_1 r_2 - r_1^2 + r_2^2  \right) + (r_1 + r_2 + r_3)  \psi_{ST_3} .
\end{align}
We can find all other constraints by replacing the phase factors for $T_{i}$, $S$, and $\overline{C}_{6}$ in the remaining equations. Eq.~\eqref{eq: psg classification C} expressing the finite order of the rotoreflection leads to
\begin{subequations}
  \begin{align}
    3\phi_{\overline{C}_6}(\mathbf{0}_A) + 3\phi_{\overline{C}_6}(\mathbf{0}_B)  &= \psi_{\overline{C}_6}.
    \end{align}
\end{subequations}
Eq.~\eqref{eq: psg classification S} associated with $\psi_{S}$ yields
\begin{subequations}
\begin{align}
    \psi_{S} =&  (r_1 + r_2 + 2 r_3) \psi_{S T_3}  + \phi_S(\mathbf{0}_A) + \phi_S(\mathbf{0}_B) \\
    \psi_S =& ( r_1 + r_2 + 2 r_3 - 1) \psi_{S T_3} +  \phi_S(\mathbf{0}_A) + \phi_S(\mathbf{0}_B) 
\end{align}
\end{subequations}
which leads to the constraints
\begin{subequations}
  \begin{align}
    \psi_{ST_3} &= 0 \\
    \phi_{S}(\mathbf{0}_A) + \phi_{S}(\mathbf{0}_B) &= \psi_{S}.
  \end{align}
\end{subequations}
Eq.~\eqref{eq: psg classification CS} gives
\begin{subequations}
\begin{align}
     \psi_{\overline{C}_6 S} &= \psi_{\overline{C}_6 T_1} + \psi_{\overline{C}_6 T_2} + \psi_{\overline{C}_6 T_3} - \psi_{ST_1} - \psi_{ST_2} + 4 (\phi_{\overline{C}_{6}}(\mathbf{0}_A) + \phi_{S}(\mathbf{0}_B)) \\
    \psi_{\overline{C}_6 S} &= 4 (\phi_{\overline{C}_{6}}(\mathbf{0}_B) + \phi_{\overline{C}_{6}}(\mathbf{0}_A)).  
\end{align}
\end{subequations}
At last, Eq.~\eqref{eq: psg classification SC} gives the constraints
\begin{subequations}
  \begin{align}
    -3 \psi_{\overline{C}_6 T_1} + 3 \psi_{\overline{C}_6 T_2} - \psi_{\overline{C}_6 T_3} + 2 \psi_{ST_1} &= 0 \\
    -3 \psi_{\overline{C}_6 T_1} - \psi_{\overline{C}_6 T_2} + 3 \psi_{\overline{C}_6 T_3} + 2 \psi_{ST_2} &= 0 \\
    - \psi_{\overline{C}_6 T_1} + \psi_{\overline{C}_6 T_2} + \psi_{\overline{C}_6 T_3} + 4 \phi_{\overline{C}_{6}}(\mathbf{0}_A) + 2 \phi_{\overline{C}_{6}}(\mathbf{0}_B) + 2 \phi_{S}(\mathbf{0}_B) &= \psi_{S \overline{C}_6} \\
    2 \phi_{\overline{C}_{6}}(\mathbf{0}_A) + 4 \phi_{\overline{C}_{6}}(\mathbf{0}_B) + 2 \phi_{S}(\mathbf{0}_A) &= \psi_{S \overline{C}_6}.
  \end{align}
\end{subequations}

\subsubsection{\label{SI subsubsec: PSG solution - Solution of the PSG constraints: intra-unit cell part} Gauge fixing and intra-unit cell part}

We must fix all remaining gauge degrees of freedom and solve the intra-unit cell constraints. Let us briefly summarize the results we have determined thus far. We obtained the phase equations~\eqref{eq U(1) classification: T1, T2, T3 first equation},~\eqref{eq U(1) classification: first equation for phi C6} and~\eqref{eq U(1) classification: first equation for phi S}, and the constraints
\begin{subequations} \label{eq: intra-cell condition} 
\begin{align}
    3\phi_{\overline{C}_6}(\mathbf{0}_A) + 3\phi_{\overline{C}_6}(\mathbf{0}_B)  &= \psi_{\overline{C}_6} \label{eq: intra-cell condition Eq.1} \\
    \phi_{S}(\mathbf{0}_A) +  \phi_{S}(\mathbf{0}_B) &=\psi_{S}  \label{eq: intra-cell condition Eq.2} \\
    \psi_{\overline{C}_6 T_1} + \psi_{\overline{C}_6 T_2} + \psi_{\overline{C}_6 T_3} - \psi_{ST_1} - \psi_{ST_2} + 4 (\phi_{\overline{C}_{6}}(\mathbf{0}_A) + \phi_{S}(\mathbf{0}_B)) &= \psi_{\overline{C}_6 S} \label{eq: intra-cell condition Eq.3} \\
    4 (\phi_{\overline{C}_{6}}(\mathbf{0}_B) + \phi_{\overline{C}_{6}}(\mathbf{0}_A)) &= \psi_{\overline{C}_6 S} \label{eq: intra-cell condition Eq.4} \\
    -3 \psi_{\overline{C}_6 T_1} + 3 \psi_{\overline{C}_6 T_2} - \psi_{\overline{C}_6 T_3} + 2 \psi_{ST_1} &= 0 \label{eq: intra-cell condition Eq.5} \\
    -3 \psi_{\overline{C}_6 T_1} - \psi_{\overline{C}_6 T_2} + 3 \psi_{\overline{C}_6 T_3} + 2 \psi_{ST_2} &= 0 \label{eq: intra-cell condition Eq.6} \\
    - \psi_{\overline{C}_6 T_1} + \psi_{\overline{C}_6 T_2} + \psi_{\overline{C}_6 T_3} + 4 \phi_{\overline{C}_{6}}(\mathbf{0}_A) + 2 \phi_{\overline{C}_{6}}(\mathbf{0}_B) + 2 \phi_{S}(\mathbf{0}_B) &= \psi_{S \overline{C}_6} \label{eq: intra-cell condition Eq.7} \\
    2 \phi_{\overline{C}_{6}}(\mathbf{0}_A) + 4 \phi_{\overline{C}_{6}}(\mathbf{0}_B) + 2 \phi_{S}(\mathbf{0}_A) &= \psi_{S \overline{C}_6} \label{eq: intra-cell condition Eq.8}
\end{align}
\end{subequations}
These constraints can be simplified by fixing some gauge degrees of freedom to remove redundant solutions.

We can fix $\psi_{\overline{C}_6 T_1}=\psi_{\overline{C}_6 T_2}=0$ by using our IGG structure ($\phi_{\mathcal{O}}\to\phi_{\mathcal{O}}+\theta$, where $\theta \in\left[ 0,2\pi\right)$) for $\phi_{T_1}$ and $\phi_{T_2}$ since the phase associated with $T_1$, $T_2$, and $T_3$ appear an odd number of times in Eq.~\eqref{eq: psg classification CT_i}. Similarly, we can fix $\psi_{ST_1}=0$ because $T_3$ is also present an odd number of times in Eq.~\eqref{eq: psg classification S T_i}. This elimination of redundant degrees of freedom directly implies $\psi_{\overline{C}_6 T_3}=\psi_{S T_2}=0$ from Eqs.~\eqref{eq: intra-cell condition Eq.5} and~\eqref{eq: intra-cell condition Eq.6}. In the same vein, we can use the remaining IGG degree of freedom of $\overline{C}_6$ and $S$ to fix $\phi_{\overline{C}_6}(\mathbf{0}_{A})=\phi_{S}(\mathbf{0}_{B})$ which directly implies $\psi_{\overline{C}_{6}S}=0$ from Eq.~\eqref{eq: intra-cell condition Eq.3}. Next, we can use a constant sublattice-dependent gauge transformation 
\begin{equation} \label{eq: intra-cell condition sublattice gauge transformation}
    \phi(\mathbf{r}_{\alpha}) = \phi_{\alpha}, \hspace{5mm} \text{where }\alpha\in\{\text{A},\text{B}\}.
\end{equation}
Given that the phase factor generally transform as $\phi_{\mathcal{O}}(\mathbf{r}_{\alpha})\to \phi_{\mathcal{O}}(\mathbf{r}_{\alpha}) + \phi(\mathbf{r}_{\alpha}) - \phi\left[ \mathcal{O}^{-1}(\mathbf{r}_{\alpha}) \right]$, our initial gauge fixing for $\phi_{T_1}$, $\phi_{T_{2}}$ and $\phi_{T_{3}}$ remains unaffected by such a gauge transformation while $\phi_{\overline{C}_6}$ and $\phi_{S}$ are mapped to
\begin{subequations}
  \begin{align}
    \phi_{\overline{C}_6}(\mathbf{0}_{\alpha}) &\to \eta_{\alpha}  (\phi_{A} - \phi_{B}) +  \phi_{\overline{C}_6}(\mathbf{0}_{\alpha}) \\
    \phi_{S}(\mathbf{0}_{\alpha}) &\to  \eta_{\alpha}  (\phi_{A} - \phi_{B}) +  \phi_{S}(\mathbf{0}_{\alpha}) 
  \end{align}
\end{subequations}
We can then choose $\phi_{\alpha}$ to fix
\begin{equation}
    \phi_{\overline{C}_{6}} (\mathbf{0}_B) = 0. \label{eq U(1) classification: gauge fixing phi intra sublattice C 1}
\end{equation}  
We directly get from Eqs.~\eqref{eq: intra-cell condition Eq.1},~\eqref{eq: intra-cell condition Eq.2}, and~\eqref{eq: intra-cell condition Eq.7} that $\phi_{S}(\mathbf{0}_A) = \psi_{\overline{C}_{6}}=\psi_{S \overline{C}_{6}}=0$. Finally, Eqs.~\eqref{eq: intra-cell condition Eq.4} and~\eqref{eq: intra-cell condition Eq.4} imply $\psi_{S}=n_{S}\pi$ with $n_{S}\in\{0,1\}$ and $\psi_{\overline{C}_{6} S}=0$. 

We conclude that there are four GMFT classes given by the phase factors
\begin{subequations} \label{eq: phase factors U(1) PSG classification}
\begin{align}
\phi_{T_{1}}\left(\mathbf{r}_{\alpha}\right)=& 0 \\
\phi_{T_{2}}\left(\mathbf{r}_{\alpha}\right)=& n_{1} \pi r_{1} \\
\phi_{T_{3}}\left(\mathbf{r}_{\alpha}\right)=& n_{1} \pi\left(r_{1}+r_{2}\right) \\
\phi_{\bar{C}_{6}}\left(\mathbf{r}_{\alpha}\right) =& n_1 \pi r_1 (r_2 + r_3) \\
\phi_{S}\left(\mathbf{r}_{\alpha}\right) =&  n_{S}\pi\delta_{\alpha,A} + \frac{n_{1} \pi}{2} \left(-r_{1}(r_{1}+1) + r_{2}(r_{2}+1) + 2r_{1}r_{2} \right).
\end{align}
\end{subequations}
Only two fully symmetric fractionalization classes (corresponding to $n_S=0$ in the current classification) are present in the effective spin-1/2 case~\cite{desrochers2023symmetry}. The presence of novel classes in the octupolar dominant regime of the dipolar-octupolar doublet is a direct consequence of the non-trivial pseudospin transformation properties, as highlighted previously.

\section{\label{SI sec: From symmetry classification to saddle point action} From symmetry classification to saddle point action}

\subsection{\label{SI subsec: Relating fields on different bonds} Relating fields on different bonds}

The value of the gauge field and MF parameters on every bond needs to be determined to build the GMFT action of every classified symmetry fractionalization class. To do so, one arbitrarily fixes the gauge field and solves the self-consistency conditions for the MF parameters on a subset of representative bonds and sites that are not symmetry-related. Since all bonds and sites can be related by space group operations for the diamond lattice, we only need to fix the gauge field and know the MF parameter on a single bond/site. The configuration on the rest of the lattice is then deduced by using the transformation properties of the gauge field and various MF parameters. 
 
The transformation properties of the GMFT parameters can be deduced by using the spinon transformations and requiring that the Hamiltonian is invariant under the gauge-enriched symmetry operations. Indeed, the gauge-enriched operators $\widetilde{\mathcal{O}}$ must be symmetries of the GMFT Hamiltonian
\begin{align}
    \widetilde{\mathcal{O}}:\mathcal{H}_{\text{GMFT}}\mapsto \mathcal{H}_{\text{GMFT}}.
\end{align}
For the GMFT Hamiltonian to be invariant under the projective operation $\widetilde{\mathcal{O}}=G_{\mathcal{O}}\circ \mathcal{O}$, the requirements
\begin{subequations} \label{eq: mapping of the fields under SG operations}
\begin{align} 
    \overline{A}_{\mathcal{O}(\mathbf{r}_\alpha),\mathcal{O}(\mathbf{r}_\alpha+\eta_{\alpha}\mathbf{b}_{\mu})} &= \overline{A}_{\mathbf{r}_\alpha, \mathbf{r}_\alpha+\eta_{\alpha}\mathbf{b}_{\mu}} + \eta_{\alpha} \left( \phi_{\mathcal{O}}\left(\mathcal{O}\left(\mathbf{r}_{\alpha} + \eta_{\alpha} \mathbf{b}_{\mu}\right)\right) - \phi_{\mathcal{O}}\left(\mathcal{O}\left(\mathbf{r}_{\alpha} \right) \right) \right) \label{eq: gauge field background} \\
    \chi_{\mathcal{O}\left(\mathbf{r}_\alpha+\eta_{\alpha}\mathbf{b}_{\mu} \right), \mathcal{O}\left(\mathbf{r}_\alpha+\eta_{\alpha}\mathbf{b}_{\nu} \right)} &= \chi_{\mathbf{r}_\alpha+\eta_{\alpha}\mathbf{b}_{\mu} , \mathbf{r}_\alpha+\eta_{\alpha}\mathbf{b}_{\nu}} \exp\left[ -i \left( \phi_{\mathcal{O}}\left(\mathcal{O}\left(\mathbf{r}_{\alpha} + \eta_{\alpha} \mathbf{b}_{\mu}\right)\right) + \phi_{\mathcal{O}}\left(\mathcal{O}\left(\mathbf{r}_{\alpha} + \eta_{\alpha} \mathbf{b}_{\nu} \right) \right)  \right) \right] \label{eq: inter-site pairing} \\
    \overline{\chi}_{\mathcal{O}\left(\mathbf{r}_\alpha\right), \mathcal{O}\left(\mathbf{r}_\alpha \right)}^{0} &= \overline{\chi}_{\mathbf{r}_\alpha, \mathbf{r}_\alpha}^{0} \exp\left[ 2 i  \phi_{\mathcal{O}}\left(\mathcal{O}\left(\mathbf{r}_{\alpha} \right)\right) \right] \label{eq: transformation on-site pairing field} \\
    \xi_{\mathcal{O}(\mathbf{r}_\alpha),\mathcal{O}(\mathbf{r}_\alpha+\eta_{\alpha}\mathbf{b}_{\mu})} &= \xi_{\mathbf{r}_\alpha, \mathbf{r}_\alpha+\eta_{\alpha}\mathbf{b}_{\mu}} \exp\left[ i \left( \phi_{\mathcal{O}}\left(\mathcal{O}\left(\mathbf{r}_{\alpha} \right)\right) - \phi_{\mathcal{O}}\left(\mathcal{O}\left(\mathbf{r}_{\alpha} + \eta_{\alpha} \mathbf{b}_{\mu} \right) \right)  \right) \right] \label{eq: inter-sublattice hopping}
\end{align}
\end{subequations}
must be satisfied.

\subsection{\label{SI subsec: Relating different bonds} Relating different bonds}

From the above transformation properties, we see that after fixing all fields on a representative bond/point, we can obtain the complete configurations if we know how to relate a single point to all other points (because of the on-site paring $\chi^{0}$), a nearest-neighbor bond to all other nearest-neighbor bonds (because of the gauge field background $\overline{A}$ and inter-sublattice hopping $\xi$), and a the second-nearest-neighbor bond to all other second-nearest-neighbor bonds (because of the inter-site pairing $\chi$). Consequently, let us find these transformations in terms of the space group generators. 

For the on-site pairing field $\chi^{0}$, if we pick $\mathbf{0}_A$ to be the representative point, then it can be related to all other points of the lattice by:
\begin{subequations}
    \begin{align}
        T_{1}^{r_{1}}\circ T_{2}^{r_2}\circ T_{3}^{r_{3}}: \mathbf{0}_{A}\mapsto (r_1,r_2,r_3)_{A} \\
        T_{1}^{r_{1}}\circ T_{2}^{r_2}\circ T_{3}^{r_{3}}\circ \overline{C}_6: \mathbf{0}_{A}\mapsto (r_1,r_2,r_3)_{B} 
    \end{align}
\end{subequations}

For the gauge field background $\overline{A}$ and inter-sublattice inter-sublattice hopping $\xi$, we can pick $\mathbf{0}_A\to\mathbf{0}_B$ as the representative nearest-neighbor bond. All other nearest-neighbor bonds of the lattice are related to it by the transformations
\begin{subequations}
    \begin{align}
        T_{1}^{r_{1}}\circ T_{2}^{r_2}\circ T_{3}^{r_{3}}:& (\mathbf{0}_A \to \mathbf{0}_B) \mapsto ((r_1,r_2,r_3)_A \to (r_1,r_2,r_3)_B) \\
        T_{1}^{r_{1}}\circ T_{2}^{r_2}\circ T_{3}^{r_{3}}\circ\overline{C}_6^{4}\circ S\circ\overline{C}_6:& (\mathbf{0}_A \to \mathbf{0}_B) \mapsto ((r_1,r_2,r_3)_A \to (r_1 + 1,r_2,r_3)_B)  \\
       T_{1}^{r_{1}}\circ T_{2}^{r_2}\circ T_{3}^{r_{3}} \circ\overline{C}_6^{2} \circ S \circ\overline{C}_6 :& (\mathbf{0}_A \to \mathbf{0}_B) \mapsto ((r_1,r_2,r_3)_A \to (r_1,r_2+1,r_3)_B)  \\
        T_{1}^{r_{1}}\circ T_{2}^{r_2}\circ T_{3}^{r_{3}}\circ S \circ\overline{C}_6:& (\mathbf{0}_A \to \mathbf{0}_B) \mapsto ((r_1,r_2,r_3)_A \to (r_1,r_2,r_3+1)_B).
    \end{align}
\end{subequations}
We can further map the representative bond to bonds with the opposite direction by first inverting it with $\overline{C}_6$ (i.e., $\overline{C}_6:(\mathbf{0}_A \to \mathbf{0}_B) \mapsto (\mathbf{0}_B \to \mathbf{0}_A)$) and then use the same transformations as above.

For the inter-site pairing $\chi$, we can pick $\mathbf{0}_B\to(1,0,0)_B$ as the representative second-nearest-neighbor bond. All other second-nearest-neighbor bonds of the parent diamond lattice can be obtained by
\begin{subequations}
    \begin{align}
        T_{1}^{r_{1}}\circ T_{2}^{r_2}\circ T_{3}^{r_{3}}:& (\mathbf{0}_B\to(1,0,0)_B) \mapsto ((r_1,r_2,r_3)_B \to (r_1+1,r_2,r_3)_B) \\
        T_{1}^{r_{1}}\circ T_{2}^{r_2}\circ T_{3}^{r_{3}} \circ \overline{C}_6^{4}:& (\mathbf{0}_B\to(1,0,0)_B) \mapsto ((r_1,r_2,r_3)_B \to (r_1,r_2+1,r_3)_B) \\
        T_{1}^{r_{1}}\circ T_{2}^{r_2}\circ T_{3}^{r_{3}} \circ \overline{C}_6^{2}:& (\mathbf{0}_B\to(1,0,0)_B) \mapsto ((r_1,r_2,r_3)_B \to (r_1,r_2,r_3+1)_B) \\
        T_{1}^{r_{1}}\circ T_{2}^{r_2}\circ T_{3}^{r_{3}} \circ \overline{C}_6^{4}\circ S \circ \overline{C}_6^{3}:& (\mathbf{0}_B\to(1,0,0)_B) \mapsto ((r_1+1,r_2,r_3)_B \to (r_1,r_2+1,r_3)_B) \\
        T_{1}^{r_{1}}\circ T_{2}^{r_2}\circ T_{3}^{r_{3}} \circ \overline{C}_6^{4}\circ S \circ \overline{C}_6:& (\mathbf{0}_B\to(1,0,0)_B) \mapsto ((r_1+1,r_2,r_3)_B \to (r_1,r_2,r_3+1)_B) \\
        T_{1}^{r_{1}}\circ T_{2}^{r_2}\circ T_{3}^{r_{3}} \circ \overline{C}_6^{2}\circ S \circ \overline{C}_6^{3} :& (\mathbf{0}_B\to(0,1,0)_B) \mapsto ((r_1,r_2+1,r_3)_B \to (r_1,r_2,r_3+1)_B) \\
        T_{1}^{r_{1}}\circ T_{2}^{r_2}\circ T_{3}^{r_{3}}\circ \overline{C}_6^{3} :& (\mathbf{0}_B\to(1,0,0)_B) \mapsto ((r_1,r_2,r_3)_A \to (r_1-1,r_2,r_3)_A) \\
        T_{1}^{r_{1}}\circ T_{2}^{r_2}\circ T_{3}^{r_{3}} \circ \overline{C}_6:& (\mathbf{0}_B\to(1,0,0)_B) \mapsto ((r_1,r_2,r_3)_A \to (r_1,r_2-1,r_3)_A) \\
        T_{1}^{r_{1}}\circ T_{2}^{r_2}\circ T_{3}^{r_{3}} \circ \overline{C}_6^{5} :& (\mathbf{0}_B\to(1,0,0)_B) \mapsto ((r_1,r_2,r_3)_A \to (r_1,r_2,r_3-1)_A) \\
        T_{1}^{r_{1}}\circ T_{2}^{r_2}\circ T_{3}^{r_{3}} \circ \overline{C}_6 \circ S \circ \overline{C}_6^{3}:& (\mathbf{0}_B\to(1,0,0)_A) \mapsto ((r_1-1,r_2,r_3)_A \to (r_1,r_2-1,r_3)_A) \\
        T_{1}^{r_{1}}\circ T_{2}^{r_2}\circ T_{3}^{r_{3}} \circ \overline{C}_6 \circ S \circ \overline{C}_6 :& (\mathbf{0}_B\to(1,0,0)_B) \mapsto ((r_1-1,r_2,r_3)_A \to (r_1,r_2,r_3-1)_A) \\
        T_{1}^{r_{1}}\circ T_{2}^{r_2}\circ T_{3}^{r_{3}} \circ \overline{C}_6^{5} \circ S \circ \overline{C}_6^{3}:& (\mathbf{0}_B\to(0,1,0)_B) \mapsto ((r_1,r_2-1,r_3)_A \to (r_1,r_2,r_3-1)_A)
    \end{align}
\end{subequations}
We can further map the representative bond to bonds with the opposite direction by first inverting it with $S\circ\overline{C}_6\circ S \circ \overline{C}_6$ (i.e., $S\circ\overline{C}_6\circ S \circ \overline{C}_6:(\mathbf{0}_B\to(1,0,0)_B)) \mapsto ((1,0,0)_B\to\mathbf{0}_B)$) and then use the same transformations as above.

\subsection{\label{SI subsec: Saddle point field configuration} Saddle point field configuration of the \texorpdfstring{$U(1)$}{U(1)} symmetric Ansätze}

Now that the transformation properties of the GMFT parameters and the relation between bonds of the lattice in terms of space group generators are known, we can build the saddle point action for every symmetry fractionalization class. However, before doing so, we make a few crucial comments regarding our procedure to build the GMFT phase diagram of Fig. 1 in the main text. In our classification of symmetric $U(1)$ Ansätze, we have assumed that all pairing terms vanish (i.e., $\chi=\chi^0=0$). However, all MF parameters are nonzero within the long-range ordered phases where the bosonic spinons are condensed (i.e., $\expval{\Phi}\ne 0$). Therefore, a rigorous construction of the GMFT phase diagram would require a complete classification of both $U(1)$ and $\mathbb{Z}_2$ symmetric Ansätze before solving the self-consistency of all classes at different points to compare their ground state energy. Only the $\mathbb{Z}_2$ Ansätze could describe ordered phases since they do not assume that any MF parameter vanish. We have classified symmetric $\mathbb{Z}_2$ Ansätze for the octupolar regime and will present these results in an upcoming publication. In this analysis, we have concluded that the phase factors of Eq.~\eqref{eq: phase factors U(1) PSG classification} also describe a subset of four symmetric $\mathbb{Z}_2$ Ansätze that are obtained by allowing the pairing fields to be nonzero. For simplicity's sake, we have restricted our attention to the four symmetric Ansätze described by the phase factor of Eq.~\eqref{eq: phase factors U(1) PSG classification} that can describe symmetric $U(1)$ and $\mathbb{Z}_2$ QSLs as well as ordered phases to build the phase diagram. Of course, the phase diagram reported in the main text does not rule out the possibility of a stable symmetric $\mathbb{Z}_2$ QSL since we have only restricted ourselves to a subset of possible Ansätze. In light of these remarks, we allow the pairing fields to be nonzero in the rest of the analysis. 

Now for the saddle point field configuration, we first conclude that the on-site pairing field is constant over the whole lattice since it transforms as in Eq.~\eqref{eq: transformation on-site pairing field} and the phase factors \eqref{eq: phase factors U(1) PSG classification} are all always either 0 or $\pi$. Therefore, if we fix $\chi^{0}_{\mathbf{0}_{A},\mathbf{0}_{A}} = \chi^{0}$ we have $\chi^{0}_{\mathbf{r}_{\alpha},\mathbf{r}_{\alpha}} = \chi^{0}$. 

Fixing the gauge field background on the representative bond $\overline{A}_{\mathbf{0}_{A},\mathbf{0}_B}=\overline{A}$, we have for all other bonds 
\begin{subequations}
\begin{align}
\overline{A}_{(r_{1},r_{2},r_{3})_{A},(r_{1},r_{2},r_{3})_{B}} &= \overline{A} + n_{s}\pi \\
\overline{A}_{(r_{1},r_{2},r_{3})_{A},(r_{1}+1,r_{2},r_{3})_{B}} &= \overline{A} + n_{s}\pi + n_{1} \pi (r_{2} + r_{3}) \\
\overline{A}_{(r_{1},r_{2},r_{3})_{A},(r_{1},r_{2}+1,r_{3})_{B}} &= \overline{A} + n_{s}\pi + n_{1} \pi r_{3} \\
\overline{A}_{(r_{1},r_{2},r_{3})_{A},(r_{1},r_{2},r_{3}+1)_{B}} &= \overline{A} + n_{s}\pi.
\end{align}
\end{subequations}

For the inter-sublattice hopping field, its value on other bonds is related to the one on the representative bond $\xi_{\mathbf{0}_{A},\mathbf{0}_B}=\xi$ by
\begin{subequations}
\begin{align}
\xi_{(r_{1},r_{2},r_{3})_{A},(r_{1},r_{2},r_{3})_{B}} &= \xi \\
\xi_{(r_{1},r_{2},r_{3})_{A},(r_{1}+1,r_{2},r_{3})_{B}} &= \xi \exp\left[ i\pi(n_{s} + n_{1} (r_{2} + r_{3})) \right]  \\
\xi_{(r_{1},r_{2},r_{3})_{A},(r_{1},r_{2}+1,r_{3})_{B}} &= \xi \exp\left[ i\pi(n_{s} + n_{1} r_{3}) \right] \\
\xi_{(r_{1},r_{2},r_{3})_{A},(r_{1},r_{2},r_{3}+1)_{B}} &= \xi \exp\left[ i\pi n_{s} \right] .
\end{align}
\end{subequations}

Lastly, for the inter-site pairing $\chi_{\mathbf{0}_{B},(1,0,0)_{B}}=\chi$, we have 
\begin{subequations}
\begin{align}
        \chi_{(r_1,r_2,r_3)_B , (r_1+1,r_2,r_3)_B}   &= \chi \exp\left[ i n_1 \pi ( r_{2} + r_{3} ) \right] \\
        \chi_{(r_1,r_2,r_3)_B , (r_1,r_2+1,r_3)_B}   &= \chi \exp\left[ i n_1 \pi  r_{3} \right] \\
        \chi_{(r_1,r_2,r_3)_B , (r_1,r_2,r_3+1)_B}   &= \chi  \\
        \chi_{(r_1+1,r_2,r_3)_B , (r_1,r_2+1,r_3)_B} &= \chi \exp\left[ i n_1 \pi ( r_2 + 1 ) \right] \\
        \chi_{(r_1+1,r_2,r_3)_B , (r_1,r_2,r_3+1)_B} &= \chi \exp\left[ i n_1 \pi ( r_{2} + r_{3} +1 ) \right] \\
        \chi_{(r_1,r_2+1,r_3)_B , (r_1,r_2,r_3+1)_B} &= \chi \exp\left[ i n_1 \pi ( r_{3} + 1 ) \right] \\
        \chi_{(r_1,r_2,r_3)_A , (r_1-1,r_2,r_3)_A}   &= \chi \exp\left[ i n_1 \pi ( r_{2} + r_{3} ) \right] \\
        \chi_{(r_1,r_2,r_3)_A , (r_1,r_2-1,r_3)_A}   &= \chi \exp\left[ i n_1 \pi r_{3} \right] \\
        \chi_{(r_1,r_2,r_3)_A , (r_1,r_2,r_3-1)_A}   &= \chi  \\
        \chi_{(r_1-1,r_2,r_3)_A , (r_1,r_2-1,r_3)_A} &= \chi \exp\left[ i n_1 \pi ( r_{2} + 1 ) \right] \\
        \chi_{(r_1-1,r_2,r_3)_A , (r_1,r_2,r_3-1)_A} &= \chi \exp\left[ i n_1 \pi ( r_{2} + r_{3} +1 ) \right] \\
        \chi_{(r_1,r_2-1,r_3)_A , (r_1,r_2,r_3-1)_A} &= \chi \exp\left[ i n_1 \pi ( r_{3} + 1  ) \right]
\end{align}
\end{subequations}

\begin{figure}
\includegraphics[width=0.75\linewidth]{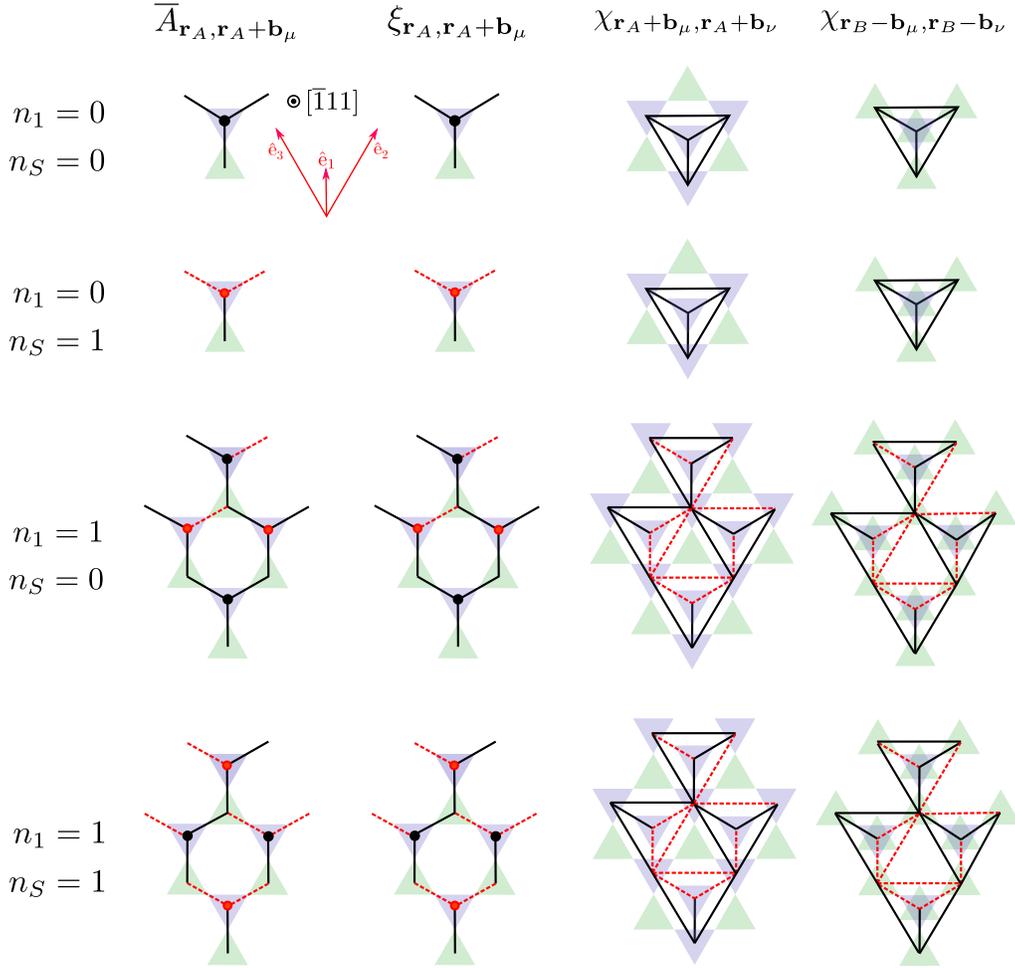}
\caption{Configuration of the gauge field background and MF parameters within the unit cell for the fully symmetric fractionalization classes. The full black line is zero for the gauge field background, and the dashed red line is $\pi$. For the other MF parameters, the full black line is the value on the representative bond (i.e., $\xi$ and $\chi$), and the dashed red line is the opposite (i.e., $-\xi$ and $-\chi$). The full circles represent bonds coming out
of the plane. The triangles are tetrahedra of the A (purple) and B sublattices (green) as seen from above. 
\label{fig: unit cell}}
\end{figure}

We can arbitrarily fix $\overline{A}=0$ for simplicity's sake, whereas the parameters $\chi^{0}$, $\chi$, and $\xi$ are determined by solving the self-consistency conditions \eqref{eq: self-consistency conditions}. Some restrictions can further be determined on the different MF parameters. For the pairing field, by acting with $S\circ\overline{C}_6\circ S\circ\overline{C}_6$, we get the conclusion that $\chi_{\mathbf{r}_1,\mathbf{r}_2}=\chi_{\mathbf{r}_2,\mathbf{r}_1}$, which could have also been deduced from the self-consistency condition. The definition of the inter-sublattice hopping field implies that $\xi_{\mathbf{r}_1,\mathbf{r}_2}=\xi_{\mathbf{r}_2,\mathbf{r}_1}^{*}$. By acting with $\overline{C}_{6}$ on the representative bond, we also conclude that $\xi_{\mathbf{0}_{B},\mathbf{0}_{A}} = \xi = \xi_{\mathbf{0}_{A},\mathbf{0}_{B}}^*$ which implies that $\xi$ is real.

As a last remark, we note that if the inter-site pairing field vanishes, the field configuration for the symmetry classes with $n_{S}=0$ and $n_{S}=1$ can be related by the gauge transformation
\begin{align}
    G: \Phi_{\mathbf{r}_\alpha} \mapsto \Phi_{(r_1,r_2,r_3)_\alpha} e^{i \pi \text{mod}(r_{1}+r_{2}+r_{3},2)}. 
\end{align}
Therefore, in the deconfined phases with nearest-neighbor interactions, symmetry classes with $n_{S}=0$ and $n_{S}=1$ are identical. There are only two distinct deconfined phases $n_{1}=0$ and $n_{1}=1$ that we label as the 0- and $\pi$-flux state, respectively, since translation acts projectively in the later phase~\cite{desrochers2023symmetry}. However, the $n_{S}=0$ and $n_{S}=1$ classes correspond to distinct states in the presence of spinon pair condensation (i.e., distinct $\mathbb{Z}_2$ QSL) and the confined phase (i.e., distinct LRO phases). This mapping between the $n_{S}=0$ and $n_{S}=1$ classes may break down in the presence of interactions beyond the nearest-neighbor level which would introduce other MF parameters. The above results regarding the field configuration for every class are summarized in Fig.~\ref{fig: unit cell}.

\section{\label{SI sec: Evaluating observables} Evaluating observables}

\subsection{Saddle point action in the large-\texorpdfstring{$N$}{N} approximation}

To be able to evaluate various observables, we apply the usual prescription and perform a large-$N$ approximation by replacing the hard constraint on the rotor length at every site $|\Phi^{\dagger}_{\mathbf{r}_{\alpha}}\Phi_{\mathbf{r}_{\alpha}}|=1$ by an average one 
\begin{equation}
    \frac{1}{N}\sum_{\mathbf{r}_{\alpha}}\expval{\Phi_{\mathbf{r}_{\alpha}}^{\dagger}\Phi_{\mathbf{r}_{\alpha}}}=\kappa
\end{equation}
for $\alpha\in\{A,B\}$ where $N$ is the number of diamond lattice primitive unit cells and $\kappa$ is a real parameter. This amounts to replacing the site- and time-dependent Lagrange multiplier $i\lambda_{\mathbf{r}_{\alpha}}^{\tau}$ by sublattice-dependent global ones $\lambda^{\alpha}$. 

We can then use the translation symmetry of the problem and define the Fourier transform of the spinon field operator as
\begin{align}
    \Phi_{\mathbf{r}_{\alpha}}^{\tau} &= \frac{1}{\sqrt{\beta N_{u.c.}}}\sum_{\mathbf{k}, i\omega_n} \Phi_{\mathbf{k},i\omega_n,\mathbf{r}_{s},\alpha}e^{-i\left(\omega_n \tau - \mathbf{k}\cdot\mathbf{r}_{\alpha} \right)},
\end{align}
where $N_{u.c.}$ is the number of unit cells, $\beta=1/k_{B} T$ is the inverse temperature, and the position on the diamond lattice is
\begin{align}
    \mathbf{r}_{\alpha} &= \mathbf{r}_{u.c.} + \mathbf{r}_{s} - \frac{\eta_{\alpha}}{2}\mathbf{b}_{0}
\end{align}
with $\mathbf{r}_{u.c.}$ and $\mathbf{r}_s$ labeling the position of the GMFT Ansatz unit cell and sublattice respectively. The wavevector sum is performed over the reduced first Brillouin zone associated with a GMFT Ansatz. The GMFT action then takes the form
\begin{align} \label{eq: GMFT action after FT}
    S_{\text{GMFT}} &= \sum_{\mathbf{k},i\omega_n} \vec{\Gamma}^{\dagger}_{\mathbf{k},i\omega_{n}} \left[\mathscr{G}(\mathbf{k},i\omega_{n})\right]^{-1} \vec{\Gamma}_{\mathbf{k},i\omega_{n}},
\end{align}
where the spinon vector field is 
\begin{align} \label{eq: spinon vector field after FT}
    \vec{\Gamma}_{\mathbf{k},i\omega_n}^{\dagger} =&\left( \Phi_{\mathbf{k},i\omega_n,1,A}^{*}, ...,  \Phi_{\mathbf{k},i\omega_n,N_{sl},A}^{*},\Phi_{\mathbf{k},i\omega_n,1,B}^{*}, ...,  \Phi_{\mathbf{k},i\omega_n,N_{sl},B}^{*},\right.\\
    &\hspace{1.5mm}\left.\Phi_{-\mathbf{k},-i\omega_n,1,A}, ...,  \Phi_{-\mathbf{k},-i\omega_n,N_{sl},A},\Phi_{-\mathbf{k},-i\omega_n,1,B}, ...,  \Phi_{-\mathbf{k},-i\omega_n,N_{sl},B} \right) 
\end{align}
with the indices labeling all sites of either the $A$ or $B$ sublattices inside the unit cell of a specific GMFT Ansatz, and the inverse spinon Matsubara Green's function is
\begin{align} 
    \left[\mathscr{G}(\mathbf{k},i\omega_{n})\right]^{-1} &=  \frac{\omega_n^{2}}{2J_{yy}} \mathds{1}_{4N_{sl}\times 4N_{sl}} + M(\mathbf{k}).  \label{eq: definition Matsubara Green's function}
\end{align}
$M^{\alpha}(\mathbf{k})$ is a $4N_{sl}\times 4N_{sl}$ matrix, with $N_{sl}$ being the number of primitive diamond lattice unit cells within the unit cell of a specific Ansatz (i.e., $N_{sl}=1$ and $N_{sl}=4$ if $n_{1}=0$ and $n_{1}=1$ respectively). The spinon dispersion is of the form
\begin{align}
    \mathcal{E}_{\gamma}(\mathbf{k})=\sqrt{2J_{yy} \varepsilon_{\gamma}(\mathbf{k})},
\end{align}
where $\varepsilon_{\gamma}(\mathbf{k})$ are the eigenvalues of the $M(\mathbf{k})$ matrix.

\subsection{\label{subsec: Important remarks on the large-N approximation} Important remarks on the large-\texorpdfstring{$N$}{N} approximation}

When interpreting the real and imaginary parts of $\Phi_{\mathbf{r}_\alpha}=q_{\mathbf{r}_{\alpha},1}+i q_{\mathbf{r}_{\alpha}, 2}$ as two-dimensional coordinates, the large-$N$ approximation amounts to releasing the particle from the unit circle $q_{\mathbf{r}_{\alpha},1}^2+q_{\mathbf{r}_{\alpha},2}^2=1$ and allowing it to move on the entire two-dimensional plane instead. As a direct consequence, the momentum of the particle $Q_{\mathbf{r}_{\alpha}}=p_{\mathbf{r}_{\alpha}, 1}+i p_{\mathbf{r}_{\alpha}, 2}$, where $p_{\mathbf{r}_{\alpha}, i}$ is the conjugate momentum of $q_{\mathbf{r}_{\alpha}, i}$, becomes continuous instead of taking on discrete values. The global constraint imposed by $\lambda^{\alpha}$ only impacts its average displacement. The crucial point is determining what this average displacement should be to reproduce results from the initial model accurately. In the existing GMFT literature, $\kappa=1$ is always chosen. However, we would like to argue that there are \emph{a priori} no reasons why such a choice should be made. Indeed, in analogy to how the average boson occupancy can be tuned to interpolate between the quantum and classical regime in Schwinger boson mean-field theory~\cite{wang2006spin, sachdev1992kagome, messio2010schwinger}, the $\kappa$ parameter should be tuned to reproduce results in a given limit without consideration for the initial hard constraint.

\begin{figure}
\includegraphics[width=0.85\linewidth]{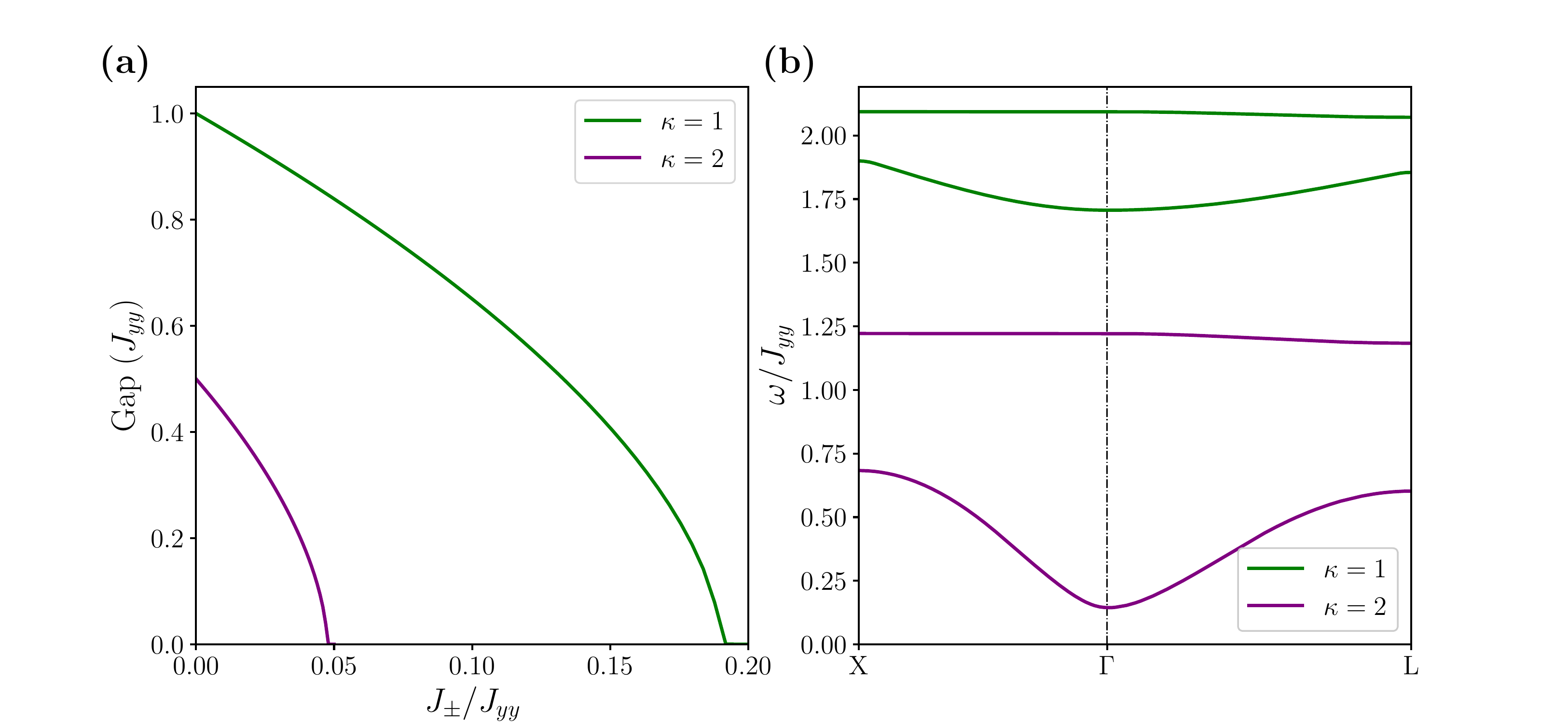}
\caption{(a) Spinon gap of the 0-flux state as a function of $J_{\pm}/J_{yy}$ with $J_{\pm\pm}=0$ for $\kappa=1$ and $\kappa=2$. (b) Lower and upper edges of the two-spinon continuum of the 0-flux state along high symmetry lines in the first Brillouin zone for $\kappa=1$ and $\kappa=2$ with $J_{\pm}/J_{yy}=0.046$ and $J_{\pm\pm}=0$. The position of the continuum can be directly compared with the QMC results of Ref.~\cite{huang2018dynamics}.
\label{fig: comparison scc}}
\end{figure}

A regime where GMFT should be expected to reproduce known results is the Ising limit. In such a limit, the spinon dispersion becomes completely flat at an energy of $J_{yy}/2$~\cite{lacroix2011introduction}. 
In the Ising limit, the rotor length self-consistency equation of GMFT reduces to
\begin{align}
    \lambda^{\alpha} = \frac{ J_{yy}}{2\kappa^{2}},
\end{align}
and the spinon dispersion 
\begin{align}
    \mathcal{E}_{\gamma}(\mathbf{k}) = \sqrt{J_{yy}^{2}/ \kappa^{2}}.
\end{align}
In order to respect the classical limit $\mathcal{E}_{\gamma}(\mathbf{k})=J_{yy}/2$, the parameter $\kappa=2$ needs to be chosen.

As briefly mentioned in the main text and argued in Ref.~\cite{desrochers2023symmetry}, the choice $\kappa=2$ further improves the accuracy of GMFT. First, as illustrated in Fig.~\ref{fig: comparison scc}(a), we find a critical value of $J_{\pm}/J_{yy} \approx 0.048$  with $\kappa=2$ for the transition between 0-flux QSI and the ordered state --- in excellent agreement with QMC~\cite{banerjee2008unusual, shannon2012quantum, kato2015numerical, huang2020extended}. In contrast, GMFT with $\kappa=1$ widely overestimates the stability of the 0-flux state by predicting a transition at $J_{\pm}/J_{yy}\approx 0.192$. Next, the position of the lower and upper edges of the two-spinon continuum obtained by GMFT with $\kappa=2$ also agrees with QMC. The lower and upper edges of the two-spinon continuum for $\kappa=2$ presented in Fig.~\ref{fig: comparison scc}(b) are in excellent agreement with Ref.~\cite{huang2018dynamics}. It should be contrasted with the results obtained for $\kappa=1$, which has a much smaller width and is located at an average energy of about twice as large.

\subsection{\label{subsec: Construction of the phase diagram} Construction of the phase diagram}

The phase diagram is constructed by solving the self-consistency equations for every symmetric Ansatz at different points in phase space and comparing their energies. If the spinons are deconfined and gapped, the phase is labeled as 0-flux and $\pi$-flux QSI if $n_{1}=0$ and $n_{1}=1$ respectively. If the spinon gap vanishes at $\mathbf{k}_c$, the bosons condense, the emerged photon is removed through the Anderson-Higgs mechanism, and we get an ordered magnet with ordering wavevector $\mathbf{k}_c$ (see Refs.~\cite{savary2012coulombic, lee2012generic, savary2021quantum} for more details). We finally note that no $\mathbb{Z}_2$ QSL where spinons pairs are condensed (i.e., $\chi\ne 0$) is observed. It does not exclude the possibility of $\mathbb{Z}_2$ QSLs since we did not study every possible symmetric GMFT Ansatz with a $\mathbb{Z}_2$ gauge structure as explained in section~\ref{SI subsec: Saddle point field configuration}.


\section{\label{SI sec: Comparison of GMFT with numerical results of the literature } Comparison of GMFT with numerical results from the literature }

\subsection{\label{SI subsec: 32-site exact diagonalization results of Hosoi, Zhang, Patri, and Kim} 32-site exact diagonalization results of Hosoi, Zhang, Patri, and Kim}

\begin{figure}
\centering
\includegraphics[width=0.6\linewidth]{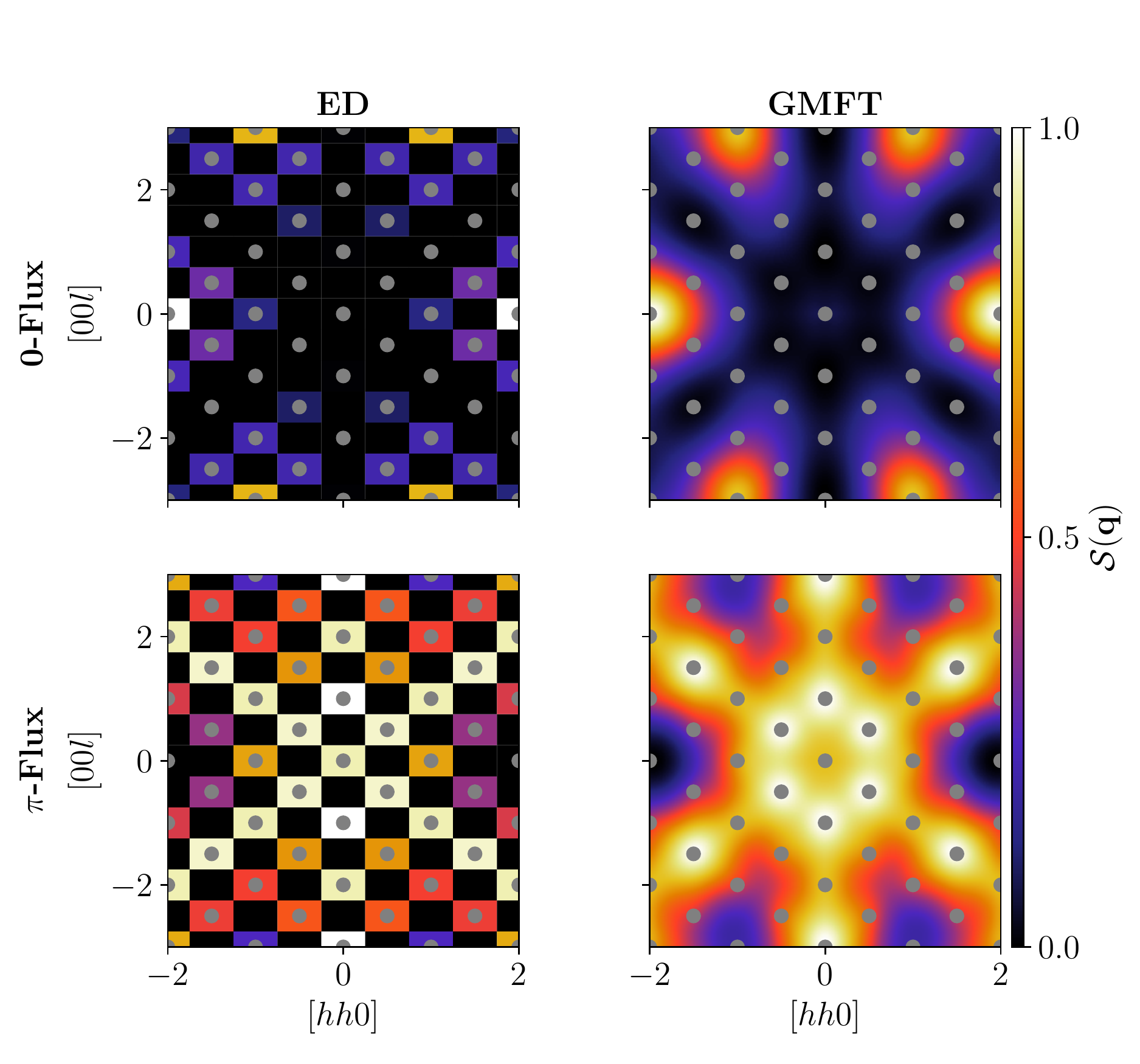}
\caption{Comparison of the total equal-time neutron scattering amplitude from the 32-site exact diagonalization (ED) results of Ref.~\cite{Hosoi2022Uncovering} (left) and GMFT (right) for the 0-flux (first row) and $\pi$-flux (second row) octupolar QSI state. The raw ED data is presented with no interpolation and with missing data points colored black. Momentum points accessible by ED are also indicated in the GMFT calculation to help the comparison. \label{fig: comparison with ED static correlations}}
\end{figure}

\subsubsection{Equal-time correlations}

The main text discusses at length how static correlations obtained from GMFT are in great agreement with the most recent 32-site exact diagonalization estimates of Ref.~\cite{Hosoi2022Uncovering} for the dipolar-octupolar case. To make direct comparison straightforward, Fig.~\ref{fig: comparison with ED static correlations} compares the total equal-time neutron scattering amplitude of the 0- and $\pi$-flux octupolar QSI state predicted from GMFT to the ED numerical results before interpolation. The two are in excellent agreement and do not show any clear qualitative differences. 

\subsubsection{Dynamical correlations}

It would be desirable to test our prediction for the presence of a 3-peak structure in the DSSF of the $\pi$-flux phase against available numerical results. However, testing the predictions from GMFT for the $\pi$-flux state is a challenging task since QMC suffers from a sign problem in this regime. The only currently available numerical results from quantum simulations for the dynamical properties of the $\pi$-flux state are the 32-site ED results of Ref.~\cite{Hosoi2022Uncovering}. Of course, these results should be taken with a grain of salt, considering that 32-site ED for the pyrochlore will suffer from strong finite-size effects and only provides information for a very limited set of points in reciprocal space. Nonetheless, since it is the only available way to test our prediction, we here try to see if these also exhibit multiple peaks in the DSSF of the $\pi$-flux state.

\begin{figure}
\centering
\includegraphics[width=0.7\linewidth]{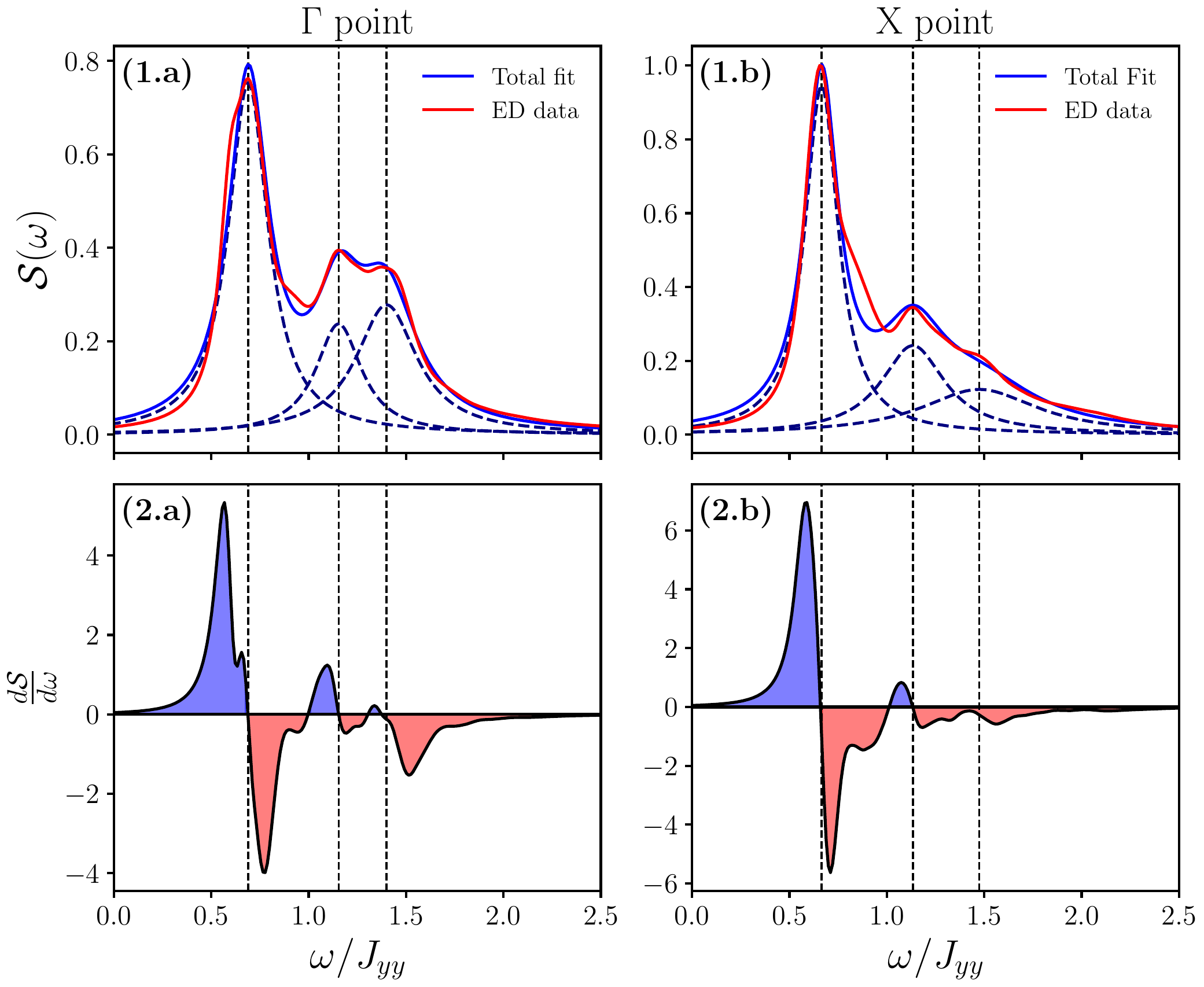}
\caption{Analysis of the 32-site exact diagonalization results of Ref.~\cite{Hosoi2022Uncovering} for the dynamical spin structure factor of the $\pi$-flux octupolar quantum spin ice phase with $\left(J_{xx}, J_{yy}, J_{zz}\right)=(0.5,1.0,0.25)$ at the (a) $\Gamma$ and (b) X point. (1) The raw data compared to a phenomenological fit with the sum of three Lorentzian functions. (2) Derivative of the DSSF as a function of energy to highlight local maxima. Vertical dashed lines denote the mean position of the three Lorentzian functions obtained from the fitting procedure. \label{fig: ED results for the DSSF}}
\end{figure}

\begin{figure}
\centering
\includegraphics[width=0.7\linewidth]{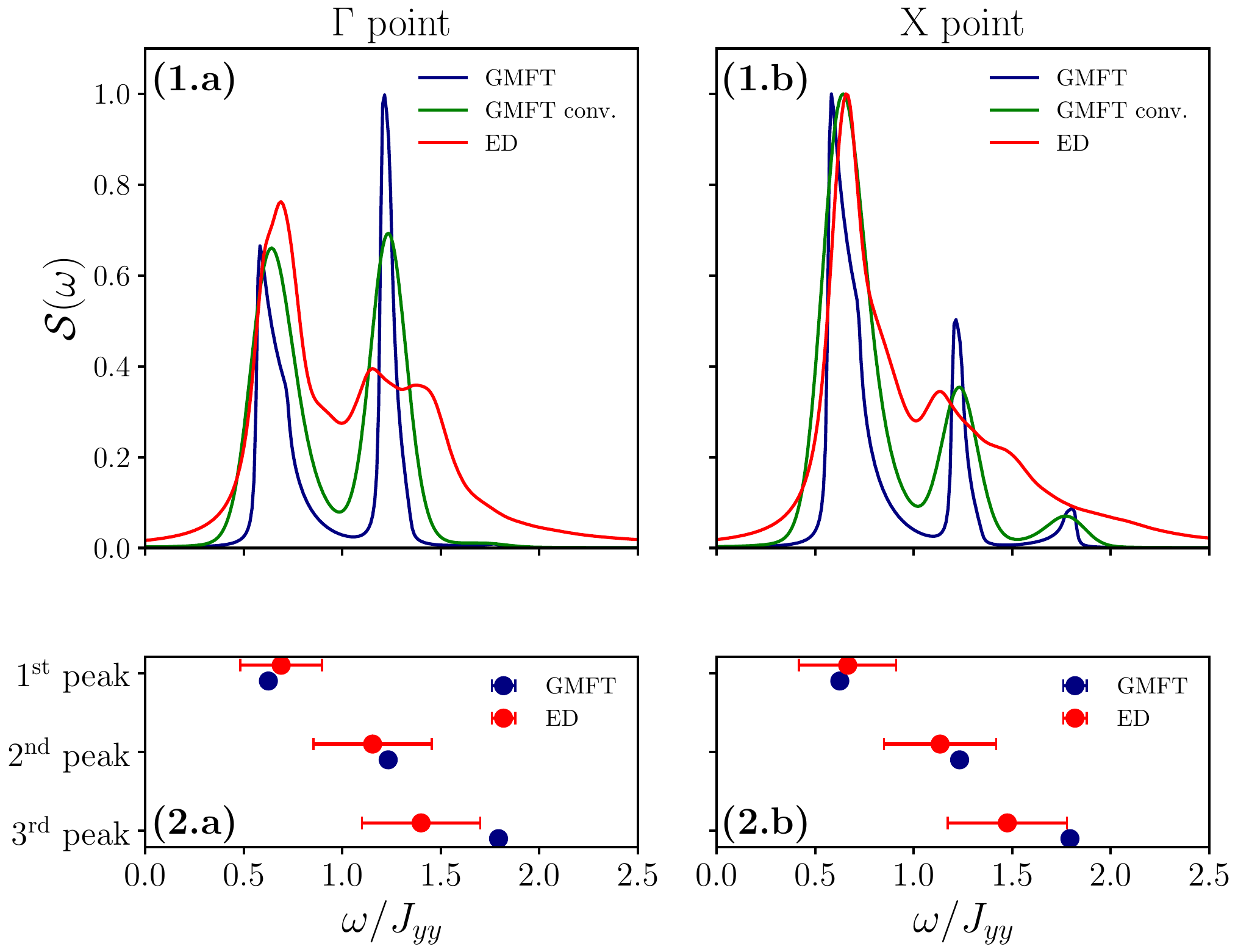}
\caption{(1) Comparison for the dynamical spin structure factor of the $\pi$-flux octupolar quantum spin ice phase predicted by GMFT and 32-site ED at the (a) $\Gamma$ and (b) X point. The GMFT results convolved with a Gaussian (labeled GMFT conv.) are also presented to mimic finite resolution effects. (2) Comparison of the position of the three peaks for both methods. For GMFT, the peaks correspond to the local maxima. The peaks are extracted from the ED data using a phenomenological fit with a sum of three Lorentzian functions (see Fig.~\ref{fig: ED results for the DSSF}).}  \label{fig: comparison of ED and GMFT results for the DSSF}
\end{figure}

The ED results for the dynamical spin structure factor of the $\pi$-flux octupolar QSI phase at the $\Gamma$ and X points (the only point in reciprocal space where the DSSF was calculated in Ref.~\cite{Hosoi2022Uncovering}) are presented in Fig.~\ref{fig: ED results for the DSSF}. We first show in panels (2.a) and (2.b) the derivative of the DSSF as a function of energy. The derivative indicates that the DSSF has three local maxima at the $\Gamma$ point and two local maxima at the $X$ point (there is a third point where the derivative almost goes through zero around 1.5 $J_{yy}$). Then in panels (1.a) and (1.b), we present a phenomenological fit of the ED results using a sum of three Lorentzian (where the position, width, and height of each Lorentzian is unconstrained). The residual sum of squares is minimized with three Lorentzian of non-zero height centered around the three local maxima at the $\Gamma$ point and around comparable positions for the X point. Therefore, without any comparison with GMFT, it can already be concluded that the ED results for the DSSF of $\pi$-flux QSI has three local maxima as a function of energy and is qualitatively different from prediction for the $0$-flux state where there is only one peak.

Moving to a more quantitative comparison of the ED results with GMFT, we present in Fig.~\ref{fig: comparison of ED and GMFT results for the DSSF} a direct comparison of the predictions for the two methods. It should first be noted that the position of the first peak in the two methods are in great agreement at both the $\Gamma$ and X point, a remarkable result considering the limitations of both methods. Now for the higher energy peaks, their relative intensities are in disagreement, and the second and third peak overlap in ED, whereas they are well separated in GMFT. We extract the peak position in ED by fitting the DSSF to a phenomenological sum of three Lorentzian and plot the results for the three peaks in panels (2.a) and (2.b). It can be noted that althought the position of the first and second peaks are very similar, the position of the third is lower in ED than in GMFT. It is hard to argue whether these differences could be explained by strong finite-size effects in ED or effects beyond GMFT, like gauge fluctuation and finite spinon lifetime. We do not speculate on the origin of the observed discrepancies, and such discussions are beside the goal of the comparison. Our main point from this comparison with ED is that the only available numerical results for the $\pi$-flux QSI show multiple peaks in its DSSF (despite some quantitative differences that can reasonably be attributed to finite-size effects) and thus support our claim and physical interpretation.

\subsection{\label{SI subsec: Comparison with QMC} Quantum Monte Carlo results of Huang, Deng, Wan, and Meng}

\begin{figure}
\centering
\includegraphics[width=0.5\linewidth]{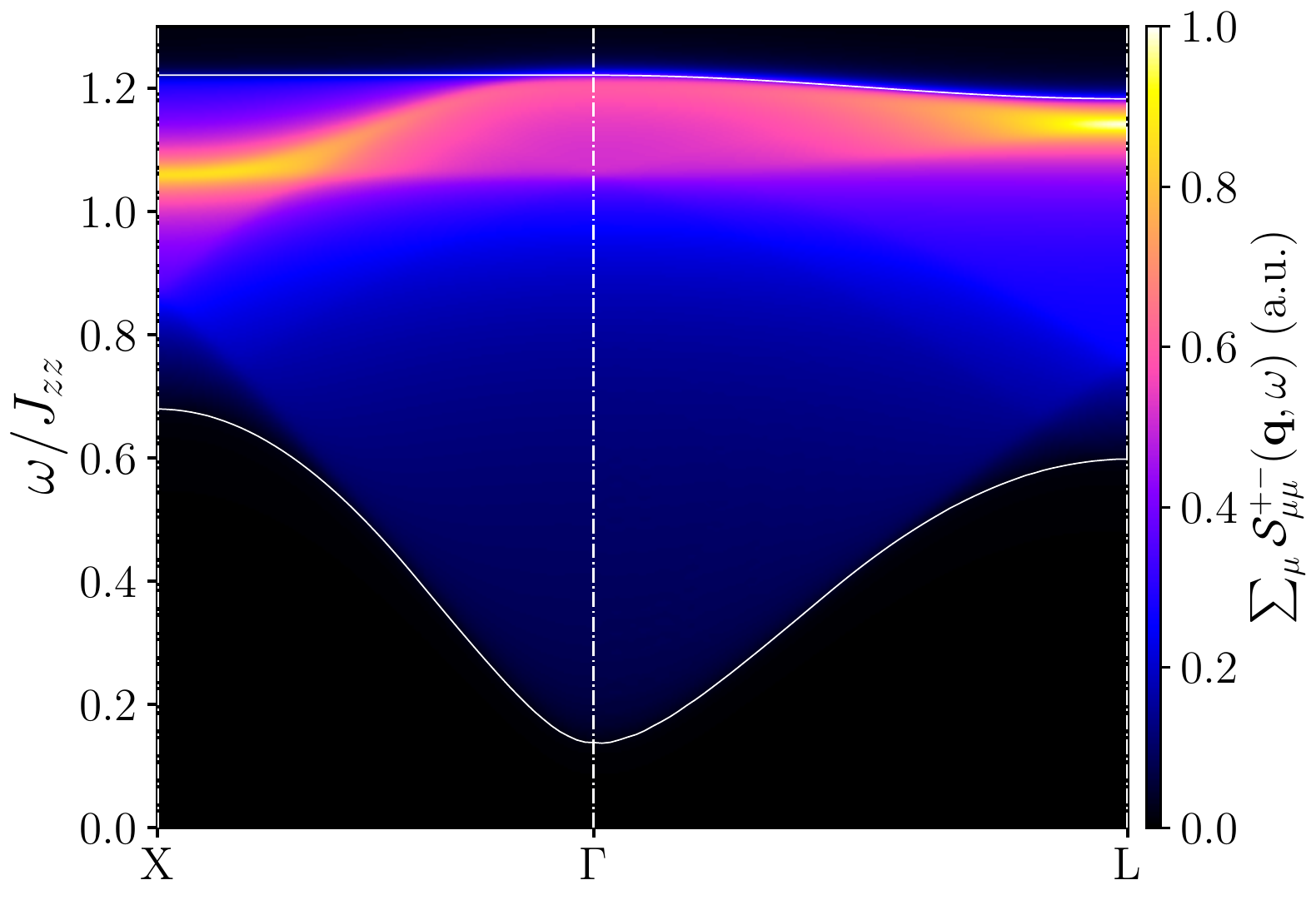}
\caption{GMFT prediction for the diagonal part of the dynamical spin structure factor in the local frame $\sum_{\mu}\mathcal{S}^{+-}_{\mu\mu}(\mathbf{q},\omega)$ for the 0-flux state with $J_{\pm}/J_{zz}=0.046$. The white lines denote the positions of the upper and lower edges of the two-spinon continuum. These results can be directly compared with the QMC results presented in Fig. 2 (c)-(d) of Ref.~\cite{huang2018dynamics}. \label{fig: comparison with QMC dynamical correlations}}
\end{figure}

The investigation conducted in Ref.\cite{huang2018dynamics} employed QMC to examine sublattice-dependent dynamical correlations. These correlations are defined as
\begin{align}
\mathcal{S}^{+-}_{\mu \nu}(\mathbf{q},\omega)=&\frac{1}{N_{\text{u.c.}}} \sum_{\mathbf{R}_{\mu}, \mathbf{R}_{\nu}'} e^{i \mathbf{q} \cdot\left(\mathbf{R}_{\mu} - \mathbf{R}_{\nu}' \right)} \int \dd{t} e^{i \omega t} \left\langle \mathrm{S}_{\mathbf{R}_{\mu}}^{+}(t) \mathrm{S}_{\mathbf{R}_{\nu}'}^{-}(0)\right\rangle,
\end{align}
where $\mathbf{R}{\mu}$ and $\mathbf{R}{\nu}$ label all sites of one of the four pyrochlore sublattices, denoted by $\mu,\nu\in{ 0,1,2,3}$, spanning the entire lattice. The spins are expressed in the local frame. The investigation focused on the diagonal component of the dynamical spin structure factor, namely $\sum_{\mu}\mathcal{S}^{+-}{\mu\mu}(\mathbf{q},\omega)$, for the XXZ model with a ratio of $J{\pm}/J_{zz}=0.046$. In this context, $J_{zz}$ is the dominant coupling analogous to $J_{yy}$ in the octupolar case. The study examined the $\Gamma\to\text{X}$ and $\Gamma\to\text{L}$ directions. GMFT results are presented in Fig.\ref{fig: comparison with QMC dynamical correlations}. Strikingly, these results exhibit exceptional agreement with the QMC results of Ref.~\cite{huang2018dynamics}. Notably, the upper and lower boundaries of the two-spinon continuum coincide. Moreover, important characteristics are captured, such as the flat upper edge of the two-spinon continuum along the $\Gamma\to\text{X}$ path, a slight decrease along the $\Gamma\to\text{L}$ path, and a slightly lower position of the lower edge at the L point compared to the X point. The behavior of the spectral weight is also consistent, with both calculations demonstrating a broad continuum and the majority of spectral weight concentrated near the upper edge of the two-spinon continuum. Furthermore, both methods reveal an increase in spectral intensity along the $\Gamma\to\text{L}$ and $\Gamma\to\text{X}$ paths, featuring local maxima at the X and L points.

In light of this comparison, we emphasize that GMFT gives outstandingly good agreement with the best numerical estimates for the stability of 0-flux QSI and its static and dynamic properties. Such an agreement is achieved even though gauge-field fluctuations are ignored. Since GMFT provides a highly accurate description of the dynamical properties of the 0-flux QSI state, GMFT predictions should also provide a reliable description of dynamical correlations for the $\pi$-flux state. Indeed, there are no reasons to believe that the method should be less accurate for the $\pi$-flux state since gauge fluctuations should be as important in the $\pi$-flux phase as in the 0-flux phase near the Ising point.

\section{\label{SI sec: Physical origin of the three peaks} Physical origin of the three peaks}

\begin{figure}
\centering
\includegraphics[width=0.9\linewidth]{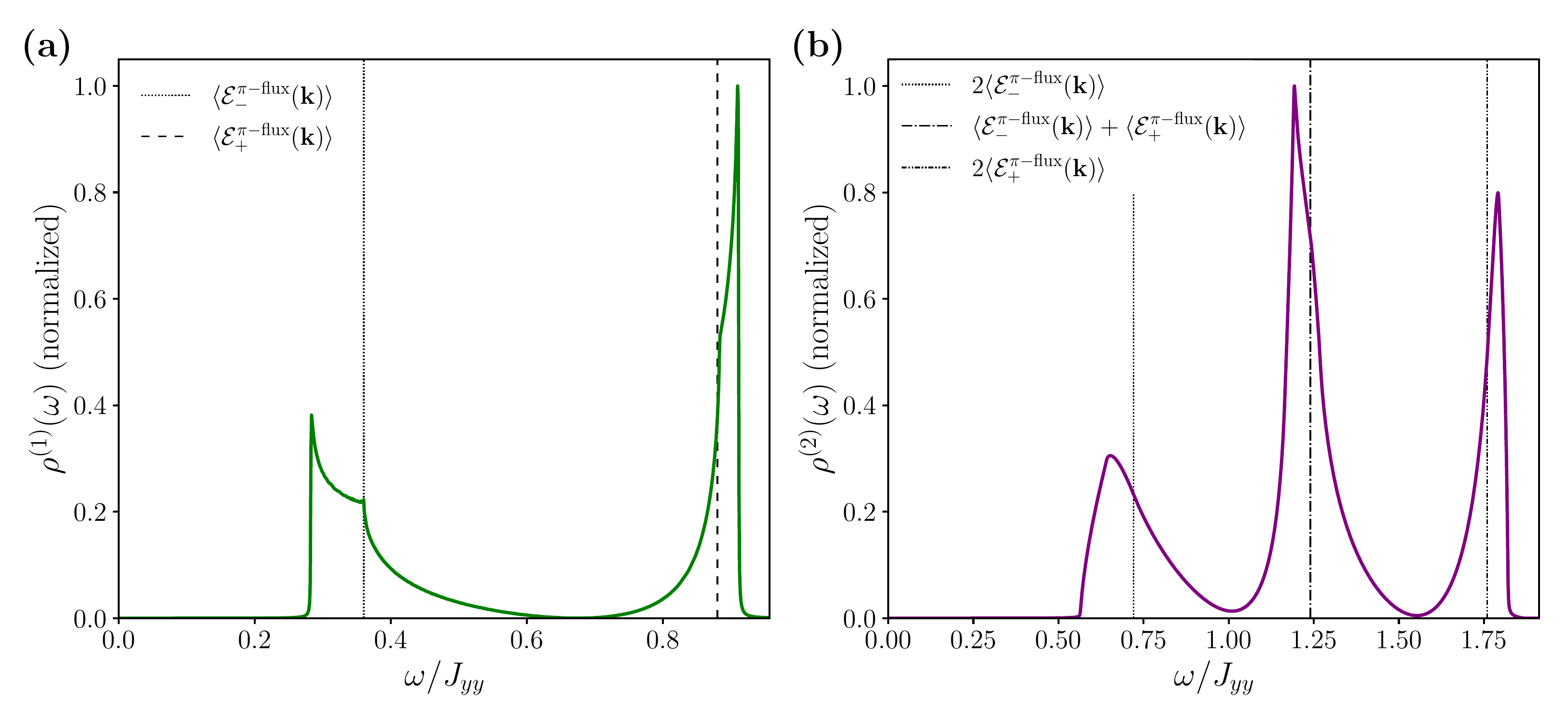}
\caption{(a) Normalized one-spinon density of states $\rho^{(1)}(\omega)$ and (b) normalized two-spinon density of states $\rho^{(2)}(\omega)$ for $\pi$-flux quantum spin ice with $J_{\pm}/J_{yy}=-0.1875$. \label{fig: dos pi}}
\end{figure}

The main text discussed the physical origin of the three inelastic peaks observed in the dynamical spin structure factor of $\pi$-flux QSI. It was explained that the three peaks originate from the two mostly flat spinon bands. We here expand on this further for clarity.

Using the gauge fixing introduced in Ref.~\cite{desrochers2023symmetry}, the two non-degenerate spinon bands of $\pi$-flux QSI are
\begin{align}
    \mathcal{E}_{\pm}^{\pi\text{-flux}}(\mathbf{k}) &= \sqrt{2J_{yy}\left( \lambda \pm \frac{1}{2}\sqrt{J_{\pm}^2 \left(3 -\sin{k_x}\sin{k_y} - \sin{k_x}\sin{k_z} - \sin{k_y}\sin{k_z} \right)}\right)} 
\end{align}
where, in the deconfined phase, $\lambda=\lambda^{A}=\lambda^{B}$ and the notation $\mathbf{k}=k_x\hat{x}+k_{y}\hat{y}+k_z\hat{z}$ was used. The minima (maxima) of this dispersion occur for the $\mathcal{E}_{-}^{\pi\text{-flux}}(\mathbf{k})$ ($\mathcal{E}_{+}^{\pi\text{-flux}}(\mathbf{k})$) band at momenta satisfying
\begin{align}
    \sin \left(k_x\right) \sin \left(k_y\right)+\sin \left(k_x\right) \sin \left(k_z\right)+\sin \left(k_y\right) \sin \left(k_z\right)=-1.
\end{align}
Solving the above constraint, one can see that the $\pi$-flux state has degenerate lines of dispersion minima (maxima) at
\begin{subequations}
\begin{align}
    & k_a=\left(n+\frac{1}{2}\right) \pi \\
    & k_b=\left(n+ 2m -\frac{1}{2}\right) \pi \\
    & k_c \in \mathbb{R},
\end{align}
\end{subequations}
where $n,m \in \mathbb{Z}$, and $(a, b, c)$ stands for any permutation of the $(x, y, z)$ triplet.

These extended lines of maxima and minima make the bands mostly flat. In other words, the variation of each band is smaller than the average separation between the bands (i.e., $\Delta\mathcal{E}_{-}^{\pi\text{-flux}}(\mathbf{k}), \Delta\mathcal{E}_{+}^{\pi\text{-flux}}(\mathbf{k})< \expval{\mathcal{E}_{+}^{\pi\text{-flux}}(\mathbf{k})-\mathcal{E}_{-}^{\pi\text{-flux}}(\mathbf{k})}$). This also leads to a one-spinon density of states $\rho^{(1)}(\omega) = \sum_{\mathbf{k},\alpha}\delta(\omega-\mathcal{E}_{\alpha}^{\pi\text{-flux}}(\mathbf{k}))/N_{u.c.}$ with two sharp peaks as seen in Fig.~\ref{fig: dos pi} (a). It should further be noted that the one-spinon density of states vanishes to zero between the two peaks since, at this intermediate energy, there are only Dirac cone band crossings, as seen in Fig. 3 of the main text. Then, the two-spinon density of states $\rho^{(2)}(\omega) = \sum_{\mathbf{k},\mathbf{q},\alpha,\beta}\delta(\omega - \mathcal{E}_{\alpha}^{\pi\text{-flux}}(\mathbf{k}) - \mathcal{E}_{\beta}^{\pi\text{-flux}}(\mathbf{q}))/N_{u.c.}^2$ will naturally have three peaks as seen in Fig.~\ref{fig: dos pi} (b). To understand this, one should recall that the two-spinon density of states can be obtained by the convolution of the one-spinon density of states $\rho^{(2)}(\omega) = (\rho^{(1)}*\rho^{(1)})(\omega)=\int d\Omega \rho^{(1)}(\Omega)\rho^{(1)}(\omega-\Omega)$. If $\rho^{(1)}(\omega)$ has two peaks then naturally its autocorrelation $\rho^{(2)}(\omega)$ will have three peaks: the first one coming from the overlap of the first $\rho^{(1)}(\omega)$ peak, the second from the overlap of the two, and finally the third one from the overlap of the second peak of $\rho^{(1)}(\omega)$. Because the first peak of $\rho^{(1)}(\omega)$ has a lower intensity than the second one, this argument further naturally explains the relative intensity of the three $\rho^{(2)}(\omega)$ peaks seen in Fig.~\ref{fig: dos pi} (b). Since the dynamical spin structure factor is weighted by the two-spinon density of states, the above argument explains the three peaks of $\pi$-flux QSI reported in the main text. 

The explanation for the origin of the three bands can further be simplified by considering a system with only two momentum-independent bands $\epsilon_1$ and $\epsilon_2$. In this case, the one-spinon density of states is naturally only non-zero exactly at $\epsilon_1$ and $\epsilon_2$, and the two-spinon density at $2\epsilon_1$, $\epsilon_1+\epsilon_2$ and $2\epsilon_2$. When allowing for the bands to vary slightly with momentum, one should instead get sharp peaks that are still centered at $\expval{\epsilon_1(\mathbf{k})}$ and $\expval{\epsilon_2(\mathbf{k})}$ for $\rho^{(1)}(\omega)$ and $2\expval{\epsilon_1(\mathbf{k})}$, $\expval{\epsilon_1(\mathbf{k})} + \expval{\epsilon_2(\mathbf{k})}$ and $2\expval{\epsilon_2(\mathbf{k})}$ for $\rho^{(2)}(\omega)$. As seen in Fig.~\ref{fig: dos pi}, This conceptual argument holds for $\pi$-flux QSI where the two peaks of $\rho^{(1)}(\omega)$ are more or less centered around $\left\langle\mathcal{E}_{-}^{\pi\text{-flux}}(\mathbf{k})\right\rangle$ and $\left\langle\mathcal{E}_{+}^{\pi\text{-flux}}(\mathbf{k})\right\rangle$, and the three peaks of $\rho^{(2)}(\omega)$ around $2\left\langle\mathcal{E}_{-}^{\pi\text{-flux}}(\mathbf{k})\right\rangle$, $\left\langle\mathcal{E}_{-}^{\pi\text{-flux}}(\mathbf{k})\right\rangle + \left\langle\mathcal{E}_{+}^{\pi\text{-flux}}(\mathbf{k})\right\rangle$ and $2\left\langle\mathcal{E}_{+}^{\pi\text{-flux}}(\mathbf{k})\right\rangle$.

\begin{figure}
\centering
\includegraphics[width=0.9\linewidth]{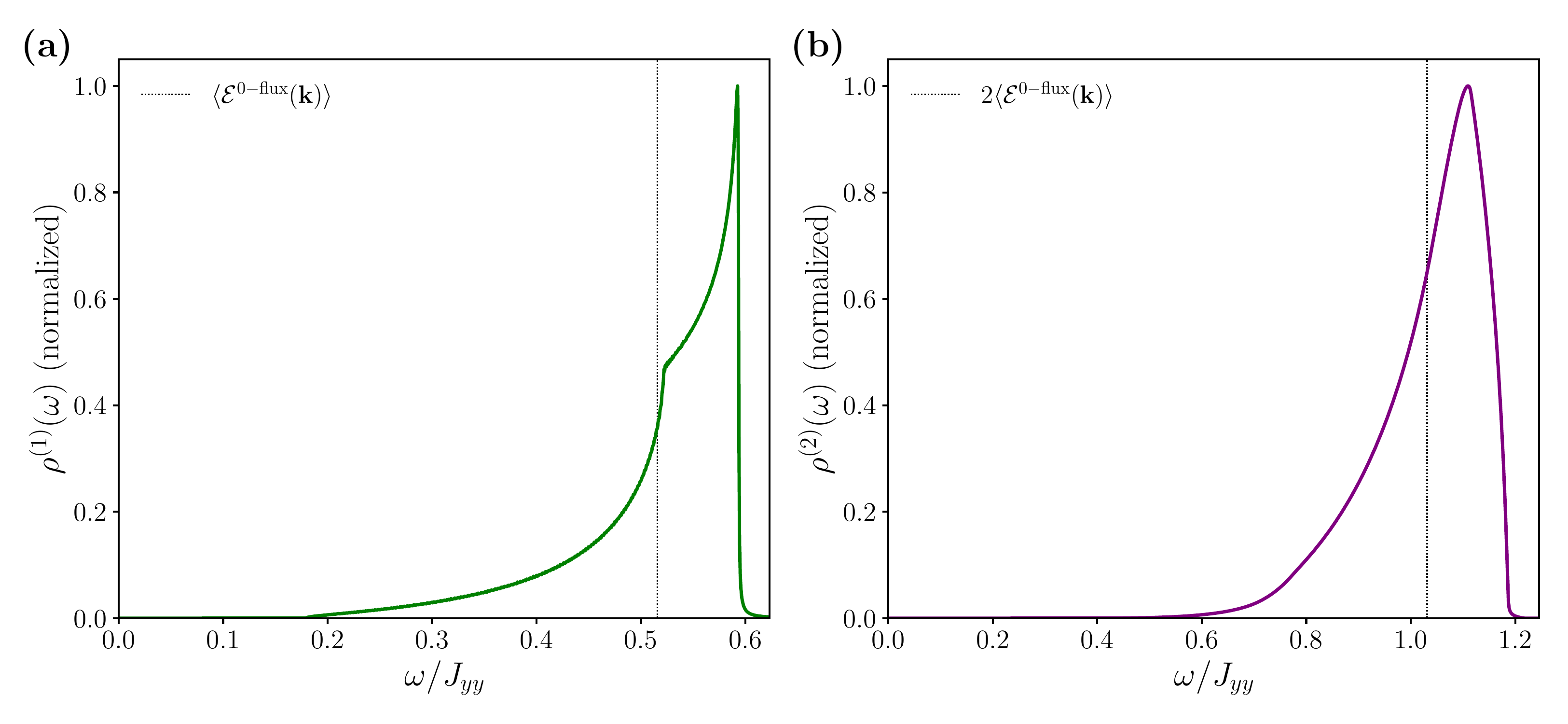}
\caption{(a) Normalized one-spinon density of states $\rho^{(1)}(\omega)$ and (b) normalized two-spinon density of states $\rho^{(2)}(\omega)$ for 0-flux quantum spin ice with $J_{\pm}/J_{yy}=0.04$. \label{fig: dos 0}}
\end{figure}

This situation should be contrasted to the 0-flux case where the spinon dispersion is 
\begin{align}
    \mathcal{E}^{0\text{-flux}}(\mathbf{k}) &= \sqrt{2J_{yy} \left( \lambda - J_{\pm} \left( \cos \left(\frac{k_x}{2}\right) \cos \left(\frac{k_y}{2}\right)+\cos \left(\frac{k_x}{2}\right) \cos \left(\frac{k_z}{2}\right)+\cos \left(\frac{k_y}{2}\right) \cos \left(\frac{k_z}{2}\right) \right)\right)}.
\end{align}
Such a dispersion leads to a single peak in both its one-spinon and two-spinon density of states, as illustrated in Fig.~\ref{fig: dos 0}. This naturally explains why the dynamical spin structure factor of 0-flux QSI reported in the main text only has a single local maximum as a function of energy. 

\begin{figure}
\centering
\includegraphics[width=0.9\linewidth]{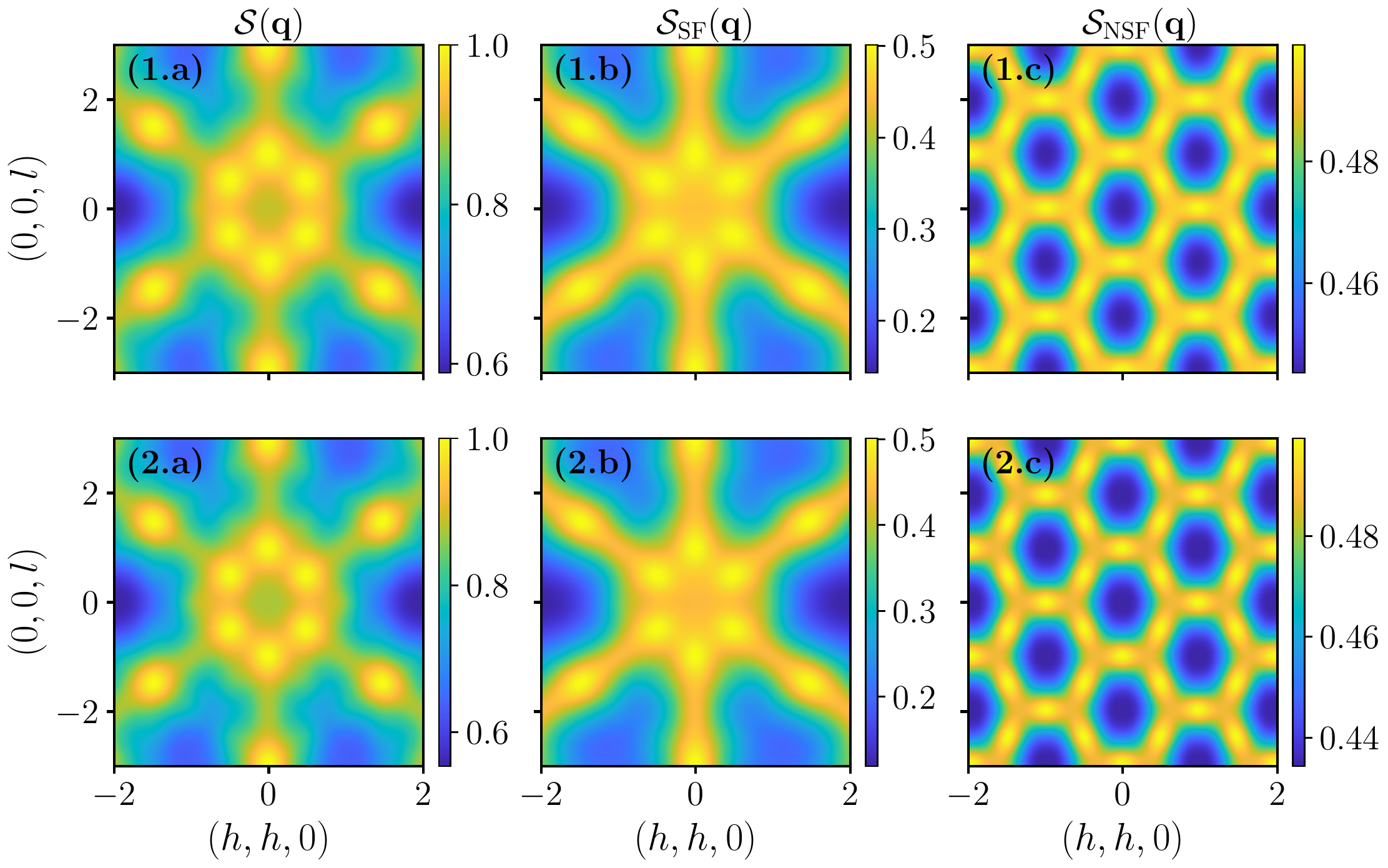}
\caption{GMFT predictions for Ce$_2$Zr$_2$O$_7$ based on the exchange coupling parameters determined in (1) Ref.~\cite{smith2022case} and (2) Ref.~\cite{bhardwaj2022sleuthing} for the (a) total neutron scattering equal-time structure factor $\mathcal{S}(\mathbf{q})$ and its contribution to the (b) spin flip $\mathcal{S}_{\mathrm{SF}}(\mathbf{q})$ and (c) non-spin-flip $\mathcal{S}_{\mathrm{NSF}}(\mathbf{q})$ channel in the [hhl] plane. \label{fig: prediction for CZO equal-time}}
\end{figure}

\section{\label{SI sec: Prediction for the position of the peaks in CZO} Predictions for C\MakeLowercase{e}$_2$Z\MakeLowercase{r}$_2$O$_7$}

\subsection{\label{SI subsec: Equal-time correlations} Equal-time correlations}

Fig~\ref{fig: prediction for CZO equal-time} shows the prediction from GMFT for the equal-time correlations in the local frame for Ce$_2$Zr$_2$O$_7$ using the parameter sets determined in Ref.~\cite{smith2022case} ($J_{yy}=0.064~\mathrm{meV}, J_{xx}=0.063~\mathrm{meV}, J_{zz}=0.011~\mathrm{meV}$) and Ref.~\cite{bhardwaj2022sleuthing} ($J_{yy}=0.08~\mathrm{meV}, J_{xx}=0.05~\mathrm{meV}, J_{zz}=0.02~\mathrm{meV}$). The predictions using both parameter sets are consistent and do not display any clear quantitative differences. 

The total equal-time structure factor $\mathcal{S}(\mathbf{q})$ can be directly compared with the energy-integrated neutron scattering measurements reported in Refs.~\cite{gao2019experimental, gaudet2019quantum}. It can be noted that the predictions of GMFT show some striking similarities with experimental measurements. Both display broad rod-like motifs with intensity modulation along the rods. We can also note further corresponding micro-details like the presence of a local minimum at the zone centered surrounded by a ring of high intensity with a maximum at the X point (i.e., [0,0,1]) and local maxima at the [0,0,3] point. The corresponding contributions to the spin-flip $\mathcal{S}_{\text{SF}}(\mathbf{q})$ and to the non-spin-flip channels $\mathcal{S}_{\text{NSF}}(\mathbf{q})$ can be also contrasted with the measurements of Ref.~\cite{smith2022case}. The predictions from GMFT for the $\pi$-flux state are again consistent with these measurements. The spin-flip channel displays the same overall pattern as the total equal-time intensity but with an intensity modulation along the rods of lesser importance. As reported in Ref.~\cite{smith2022case}, the non-spin flip channels show maxima and minima at the zone boundaries and zone center, respectively. Such similarities are striking and support the identification of Ce$_2$Zr$_2$O$_7$ as a realization of $\pi$-flux quantum spin ice.

\begin{figure}[h!]
\centering
\includegraphics[width=0.7\linewidth]{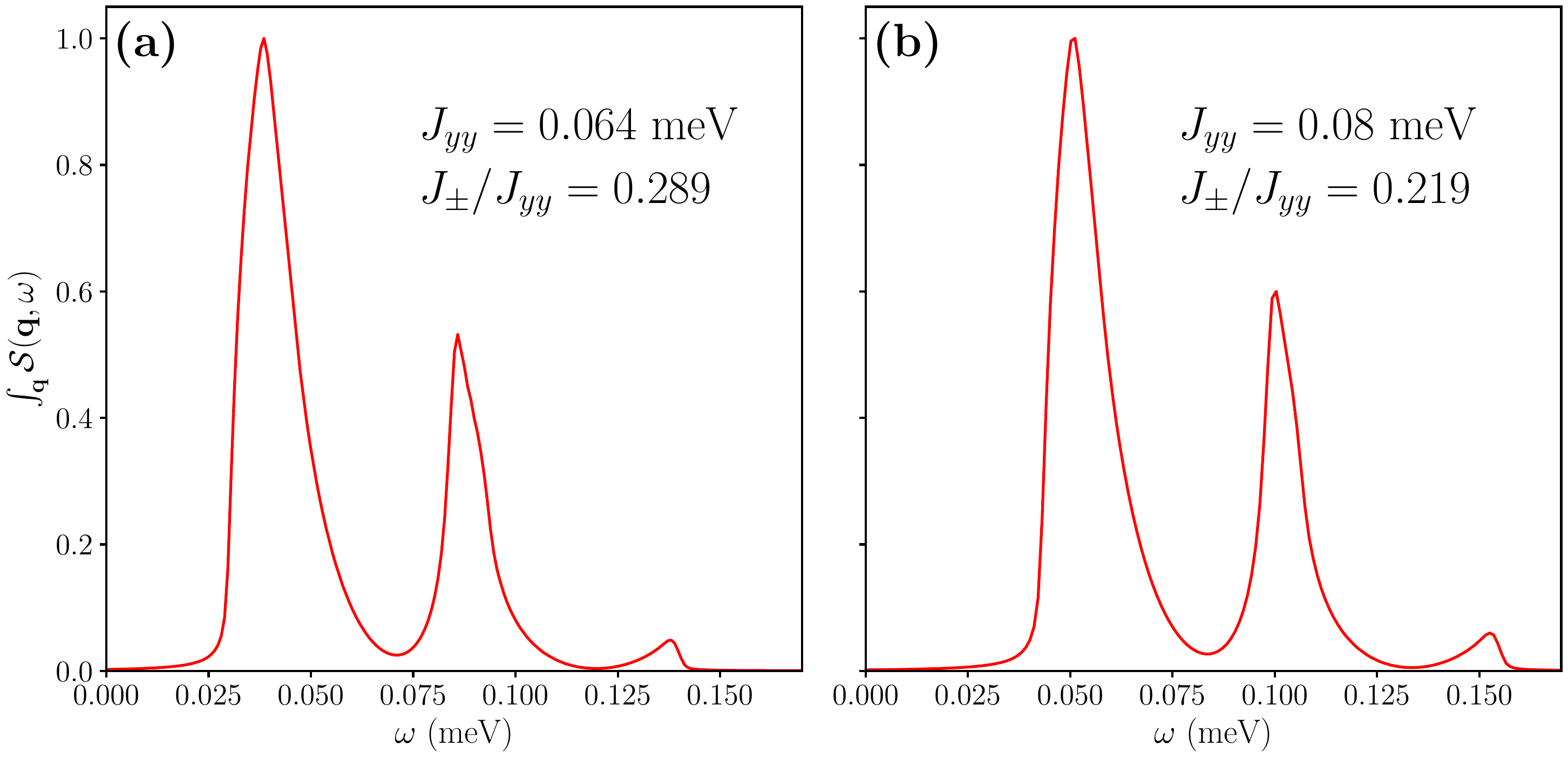}
\caption{Prediction for the momentum-integrated dynamical spin structure factor of Ce$_2$Zr$_2$O$_7$ based on the values of the exchange coupling parameters determined in (a) Ref.~\cite{smith2022case} and (b) Ref.~\cite{bhardwaj2022sleuthing}.   \label{fig: prediction for CZO}}
\end{figure}

\subsection{\label{SI subsec: Prediction for the position of the peaks in CZO} Position of the three peaks}

To provide an estimate of the experimental resolution necessary to observe the peaks, Fig.~\ref{fig: prediction for CZO} presents the prediction from GMFT for the momentum-integrated DSSF of Ce$_2$Zr$_2$O$_7$ using the exchange coupling constants of Refs.~\cite{smith2022case, bhardwaj2022sleuthing}. The peaks are separated by approximately 0.05~meV. This could be resolved with currently available experimental techniques. For instance, the resolution of 0.7~$\mu$eV achieved through backscattering neutron spectroscopy in Ref.~\cite{poree2023fractional} is well above what is necessary to observe the peaks.


%
